\documentclass[preprintnumbers,amsmath,amssymb,floatfix,10pt,prd,onecolumn,superscriptaddress,nofootinbib]{revtex4-2}
\usepackage{latexsym}
\usepackage{epsfig}
\usepackage{epstopdf}
\usepackage{graphicx}
\usepackage{amssymb}
\usepackage{amsmath}
\usepackage{amsfonts}
\usepackage{subfigure}
\usepackage{dcolumn}
\usepackage{bm}
\usepackage{color}
\usepackage{comment}
\usepackage[utf8]{inputenc}
\usepackage[dvipsnames]{xcolor}
\usepackage{hyperref}
\hypersetup{colorlinks=true,linkcolor=blue,citecolor=red,urlcolor=magenta}

\begin{document}
\title{\bf $4D$ Einstein-Gauss-Bonnet Black Hole in Power-Yang-Mills Field: A Shadow Study}
\author{M. Zubair}
\email{mzubairkk@gmail.com;drmzubair@cuilahore.edu.pk}\affiliation{Department of Mathematics, COMSATS University Islamabad, Lahore Campus, Lahore, Pakistan}
\author{Muhammad Ali Raza}
\email{maliraza01234@gmail.com}\affiliation{Department of Mathematics, COMSATS University Islamabad, Lahore Campus, Lahore, Pakistan}
\author{Furkat Sarikulov} \email{furqatsariquloff@gmail.com}
\affiliation{School of Mathematics and Natural Sciences, New Uzbekistan University, Mustaqillik Ave. 54, Tashkent 100007, Uzbekistan}
\affiliation{Ulugh Beg Astronomical Institute, Astronomicheskaya 33, Tashkent 100052, Uzbekistan} 
\author{Javlon Rayimbaev}
\email{javlon@astrin.uz}
\affiliation{Institute of Fundamental and Applied Research, National Research University TIIAME, Kori Niyoziy 39, Tashkent 100000, Uzbekistan}
\affiliation{Akfa University, Kichik Halqa Yuli Street 17,  Tashkent 100095, Uzbekistan}
\affiliation{National University of Uzbekistan, Tashkent 100174, Uzbekistan}
\affiliation{Tashkent State Technical University, Tashkent 100095, Uzbekistan}

\begin{abstract}
We consider a static black hole immersed in the Power-Yang-Mills field in four dimensional Einstein-Gauss-Bonnet gravity and investigate the effect of various parameters on the radius of the photon sphere. The modified form of the Newman-Janis algorithm is used for obtaining a rotating black hole solution in this gravity. Further, we try to explore the influence of the Yang-Mills magnetic charge $Q$ with power $q$, Gauss-Bonnet parameter $\alpha$ and spin $a$ on the horizon radius. The geodesic equations are constructed by incorporating the Hamilton-Jacobi formalism. The radial component of the geodesic equations gives the effective potential which is further used in deriving the mathematical structure for the shadows by using Bardeen's procedure for a fixed observer at infinity. The shadows are calculated and plotted in terms of two celestial coordinates for an equatorial observer. It is observed that all the parameters have a very significant effect on the shadow and related physical observables. {We also obtain the constraint values for the spin, magnetic charge and Gauss-Bonnet parameters, using the shadow size of supermassive black holes Sagittarius A$^*$ and M$87$* from the EHT observations for the cases of $q=0.6$ and $0.9$. It is shown that there are upper and lower bounds for the charge and spin of M$87$* at $q=0.6$, while only the upper bounds for charge and spin of Sagittarius A$^*$. Finally, we investigate the energy emission rate in the Hawking radiation around the $4D$ Einstein-Gauss-Bonnet black hole in the Power-Yang-Mills field.} 
\end{abstract}
\maketitle
\date{\today}

\section{Introduction}
As we know General Relativity (GR) may not explain some observational problems in the strong field region near black holes (BHs) and thus GR can not be considered a complete theory of gravity. However, we can construct modified gravitational theories beyond GR by changing the Einstein-Hilbert (EH) action, specifically, the EH action comprising the higher-order derivative terms. Hence, under the low energy limit of String Theory, we get the effective action supplemented by the higher order curvature terms \cite{1986NuPhB.277....1G,1987NuPhB.291...41G,1987PhLB..191..354M,1985PhLB..156..315Z,1987NuPhB.293..385M}. In this context, the generalization of GR is known as Lovelock theory which is a higher derivative gravity theory \cite{1971JMP....12..498L,1972JMP....13..874L,1990PhRvD..41.3696D}. When the curvature terms of second order are included in the EH action, we get Gauss-Bonnet (GB) theory as a special case of Lovelock theory that works only for dimensions $D\geq5$ \cite{10.2307/1968467}. However, in the past few years, GB theory has been studied for the case $D=4$. As we know, the integral of GB terms being a topological invariant is unaltered when inserted into the $4D$ Einstein field equations (EFE). However, it is observed that the higher order terms contribute in four dimensions by dilaton coupling \cite{1996PhRvD..54.5049K,2009PhRvD..80j4032M,2014PhRvL.112y1102S,2015PhRvD..92h3524K}. Recently, according to Glavan and Lin \cite{2020PhRvL.124h1301G}, without considering the dilaton coupling, GB theory exists in four dimensions by re-scaling the GB parameter $\alpha\rightarrow\frac{\alpha}{D-4}$ under the limit $D\rightarrow4$. In this case, the integral of GB terms does give the non-trivial contribution to the dynamical equations. Finally, a spherically symmetric BH solution is obtained for the Einstein-Gauss-Bonnet (EGB) theory in four dimensions. Since a covariantly novel EGB theory in four dimensions does not exist unquestionably. So, the method of dimensional regularization considered in \cite{2020PhRvL.124h1301G} is opposed by many scientists in \cite{2020PhRvL.125n9001G,2020EPJC...80..647G,2020EPJC...80..992M,2020JCAP...07..013K,2020PhRvD.102b4029B,2021ChPhC..45a3107A,2021EPJP..136..180H}. Recently, the authors in \cite{2020JCAP...07..013K,2020PhRvD.102b4029B,2020JHEP...07..027H,2021PDU....3100770C,2020PhLB..80935717L,2020PhRvD.102b4025F} found a well-defined $4D$ EGB theory by various approaches with several degrees of freedom in general in contradiction with two degrees of freedom satisfying Lovelock's theorem. Aoki-Gorji-Mukohyama \cite{2020PhLB..81035843A,2020JCAP...09..014A} have found that we can get a well-defined EGB theory in four dimensions by breaking off the partial diffeomorphism invariance by considering Arnowitt-Deser-Misner decomposition. It means that the $3D$ spatial diffeomorphism invariance must be kept and the temporal diffeomorphism should be violated. Ultimately, Jafarzade et al. \cite{2021JCAP...04..008J} proposed that despite the minimal coupling of the matter fields with EGB theory, the limit $D\rightarrow4$ considered in \cite{2020PhRvL.124h1301G} gives a solution of the well-defined Aoki-Gorji-Mukohyama theory of four dimensional EGB gravity.

The coupling of gravity with a matter field is a common interest of scientists working in different areas of gravity theories, especially, in BH Physics where BHs surrounded by different fields are investigated. Such coupled BH solutions in $4D$ EGB gravity have also been derived in the presence of various fields \cite{2020PhLB..80835658S,2021EPJC...81..361W,2021PhLB..81836383S,2022EPJP..137..278B,2022Univ....8..232K,2022EPJP..137..969S,2022AnPhy.43768726A,2020PDU....3000730S,2021PDU....3100793G,2022IJMPD..3150058Z}. Usually, the BH solutions exhibit singularity at the origin, where all the physical laws become meaningless. To deal with such issues, the nonlinear electromagnetic field has been introduced that is coupled with gravity \cite{1998PhRvL..80.5056A,1999PhLB..464...25A,1999GReGr..31..629A,2000PhRvD..61h4003C}. In this context, BH solutions in $4D$ EGB gravity immersed in a nonlinear electrodynamics field have also been derived \cite{2020AnPhy.42168285J,2021AnPhy.42468347G,2021AnPhy.42868449K,2022Univ....8..244K}. One example is a $4D$ spherically symmetric non-rotating EGB BH immersed in the Born-Infeld field \cite{2020EPJC...80..662Y}. Furthermore, we can also find the coupling of the non-Abelian Yang-Mills (YM) field with GR and the general GB gravity in \cite{2007PhRvD..76h7501M,2008PhRvD..78f4050M,2008PhLB..659..471H,2008PhLB..665..125H}. In the coupling of nonlinear YM field in terms of the power $q$ of the YM invariant given as $\big(F_{\mu\nu}^{(a)}F^{\mu\nu(a)}\big)^q$ with GB gravity, Mazharimousavi and Halilsoy \cite{2009PhLB..681..190M} obtained a BH solution in $D$ dimensions. Recently, Biswas \cite{2022GReGr..54..161B} derived a BH solution in EGB theory with a power-Yang-Mills (PYM) field in $4D$ AdS spacetime.

Apart from the construction of BH solutions, the study of various properties of BHs is of quite interest to the scientific community. One of these properties is the shadow or the visual appearance of the BHs. Due to the strong gravity around BHs, the light particles get trapped and fall in from the unstable circular orbits giving us an image called shadow \cite{2022PhR...947....1P}. The earlier studies related to the appearance of BHs can be found in \cite{1966SvPhU...8..522Z,1966MNRAS.131..463S,1976PhRvD..14.3281Y,1979A&A....75..228L,1986SvPhU..29..215D,1993A&A...272..355V} and later, the shadows were explored for various BH solutions, see \cite{2014PhRvD..89l4004G,2016PhRvD..94b4054A,2018PhRvD..97f4021T,2019PhRvD..99d4015H,2019PhRvD.100d4055Z,2019PhRvD.100l4024K,2020PhRvD.101h4001L,2020PhRvD.102f4020J,PhysRevD.103.024013,2020PhRvD.102j4032G,2021PhRvD.103l4050C,2022PhRvD.105l4009H}. GB BHs in $D$ dimensions and EGB BHs in four dimensions have also been investigated for the shadows \cite{2017PhLB..768..373C,2020EPJC...80..180D,2020EPJC...80.1049K,2020EPJC...80..588G,2020JCAP...07..053K,2020EPJC...80..872Z}. This provides us with enough motivation to investigate the shadows and related properties for $4D$ EGB BH in the PYM field. Thus, the non-rotating BH solution for $\Lambda=0$ in Ref. \cite{2022GReGr..54..161B} will be converted into its rotating counterpart and further the shadows will be studied via the method of impact parameters. Moreover, the distortion and energy emission rate will also be studied using shadows' results.

Testing gravity theories has become a more important and interesting field of relativistic astrophysics, thanks to high-resolution observations related to BH physics. In fact, BHs are fascinating objects in the universe, first predicted mathematically as a solution of the EFE. The quest for BH candidates in observations has confirmed that supermassive BHs exist at the center of elliptical and spiral galaxies. The intermediate-mass BHs are found through high redshifts in other distant low luminous galaxies. For the first time, the Event Horizon Telescope (EHT) collaboration observed the image of the supermassive BH located at the center of the nearby galaxy Miessner M$87$ in $2019$ \cite{2019ApJ...875L...1E} and later the image of the supermassive BH Sgr. A* at the center of Milky Way in $2022$ \cite{2022ApJ...930L..12E,2022ApJ...930L..17E}. On the other hand, the observational data from the image sizes is a good tool to get constraints on the mass of the BHs and testing gravity theories that may help to understand BH physics in depth and gravity theory dominating around them. In recent years, constraints on the parameters of gravity theories have been obtained by several authors in Refs. \cite{2022IJMPD..3150058Z,2023PDU....4001178U,2023EPJC...83..250P,2023AnPhy.44869197P,2023EPJC...83..318G,2023ApJ...944..149A,2021MNRAS.504.5927A,2022ApJ...932...51A,2022ApJ...939...77W,2019PhRvD.100l4024K}. In the same way after obtaining the shadows of the BH, we will try to constrain the BH parameters for various cases of M$87$* and Sgr. A*

The outline of the paper is as follows: The section $2$ will consist of the action, the metric of the theory and the discussion of the radius of photon sphere. Further, in section $3$, we will convert the static metric into a rotating one and will study the horizon radius. In section $4$, we will develop the geodesic equations and the mathematical scheme for shadows. The effective potential, shadows and distortion will be discussed in this section. In Section $6$, we will obtain the constraints on the BH parameters by comparison with the EHT data. Finally, in section $7$, the results will be summarized, and the conclusion will be presented. Note that, we use the units $8\pi G_N=c=1$, where $G_N$ and $c$ are the gravitational constant and the cosmic speed limit, respectively.

\section{The Static Black Hole}
\begin{figure}[t]
\begin{center}
\subfigure{
\includegraphics[height=4.5cm,width=5.7cm]{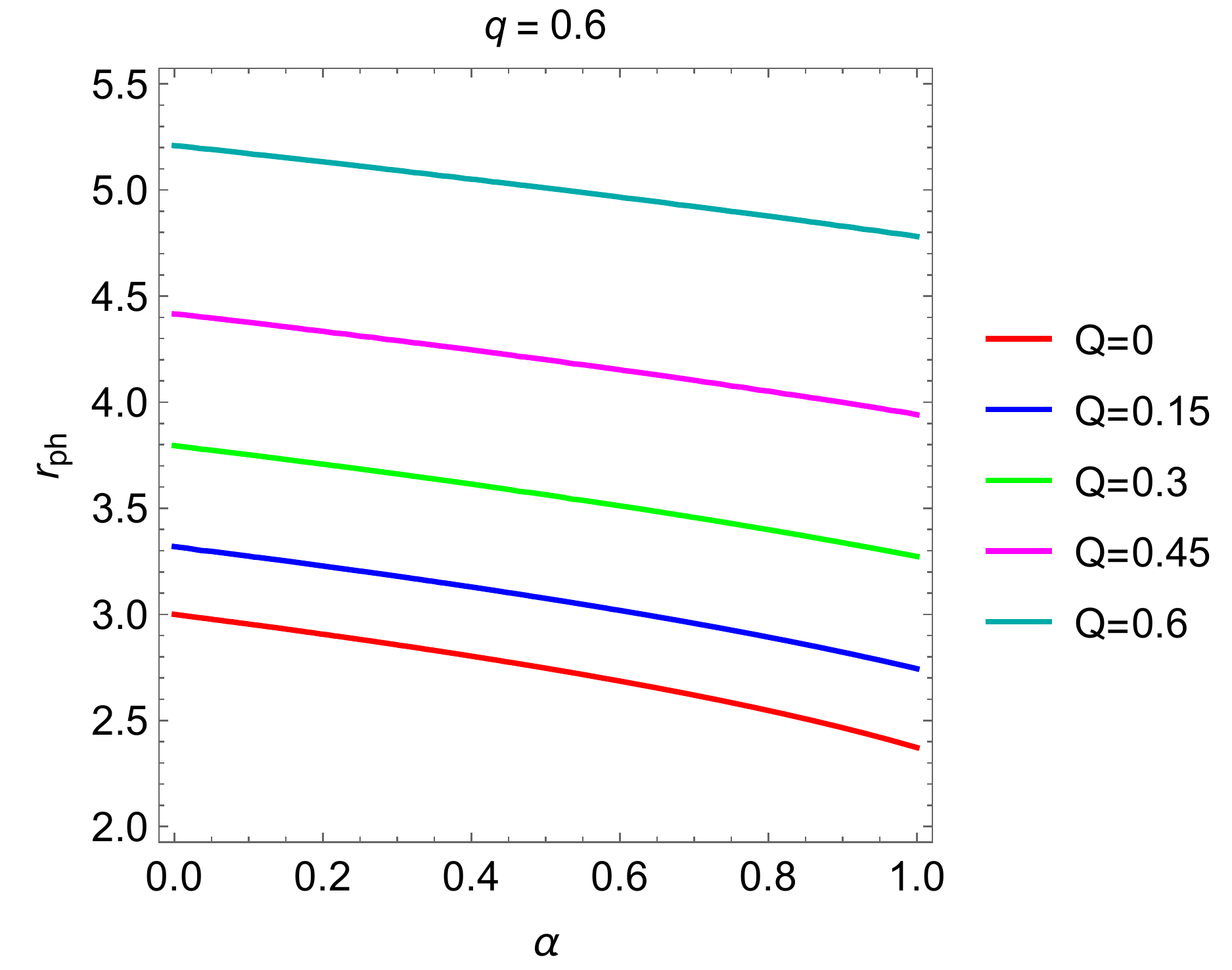}}
~
\subfigure{
\includegraphics[height=4.5cm,width=5.7cm]{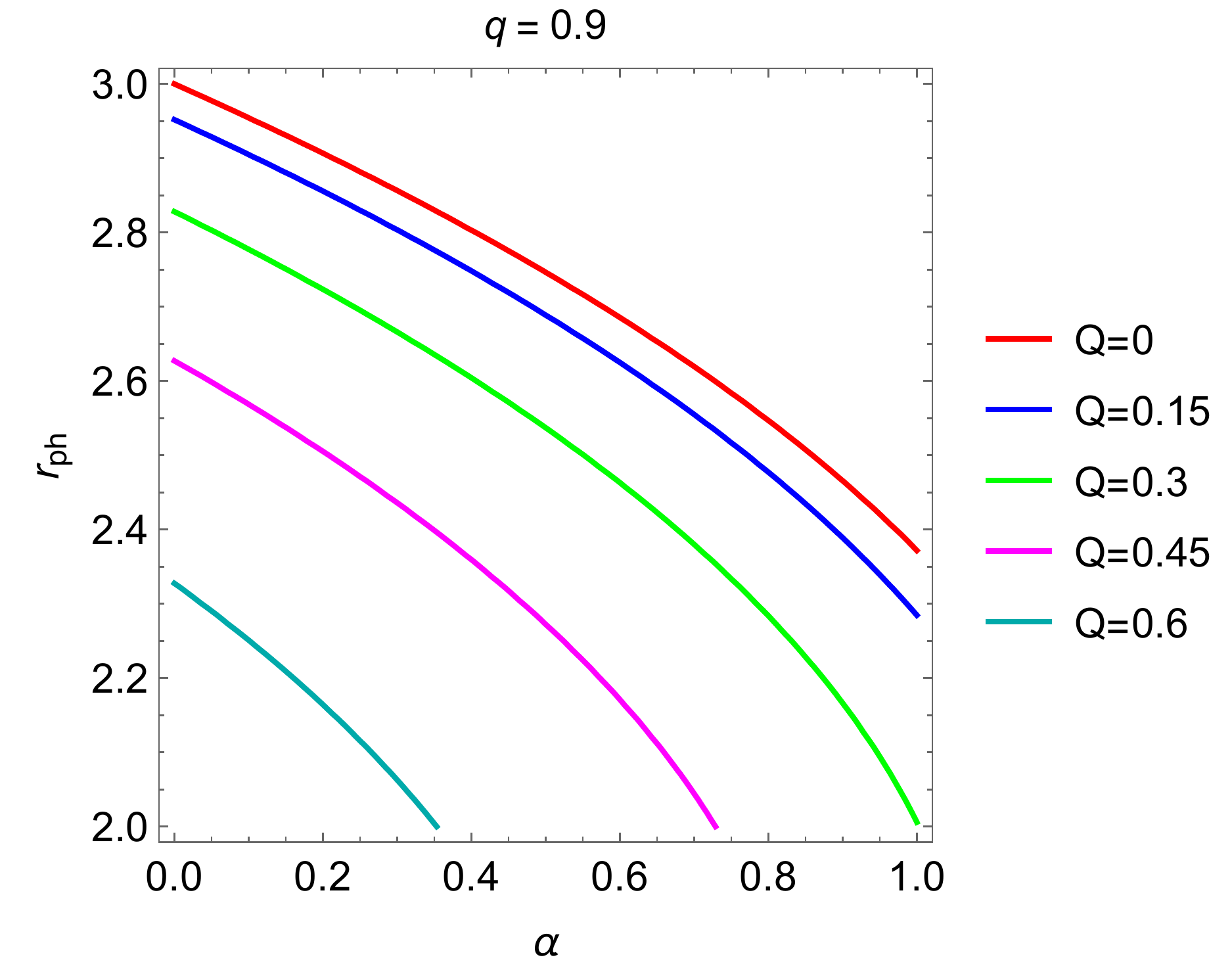}}
~
\subfigure{
\includegraphics[height=4.5cm,width=5.5cm]{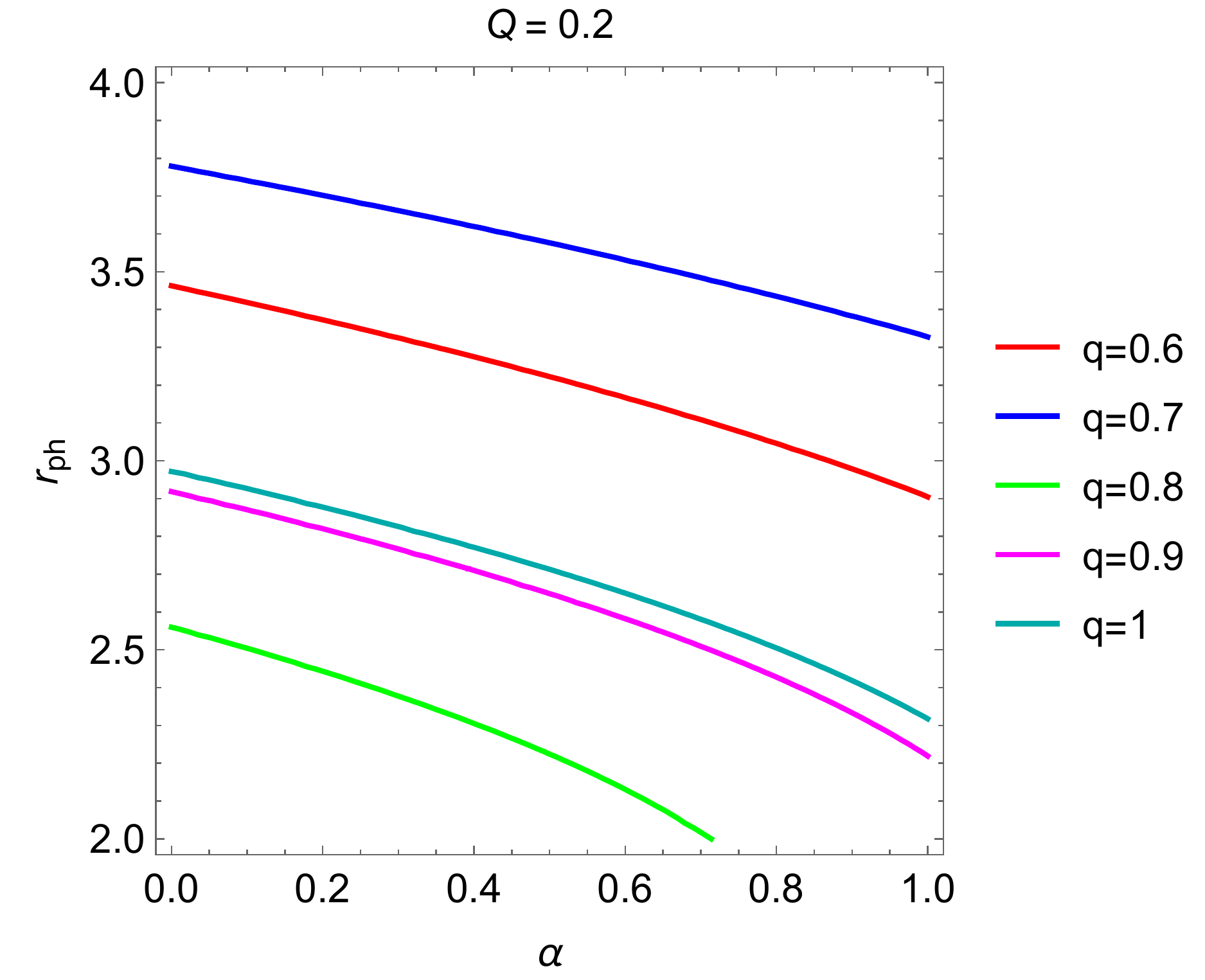}}
\subfigure{
\includegraphics[height=4.5cm,width=5.7cm]{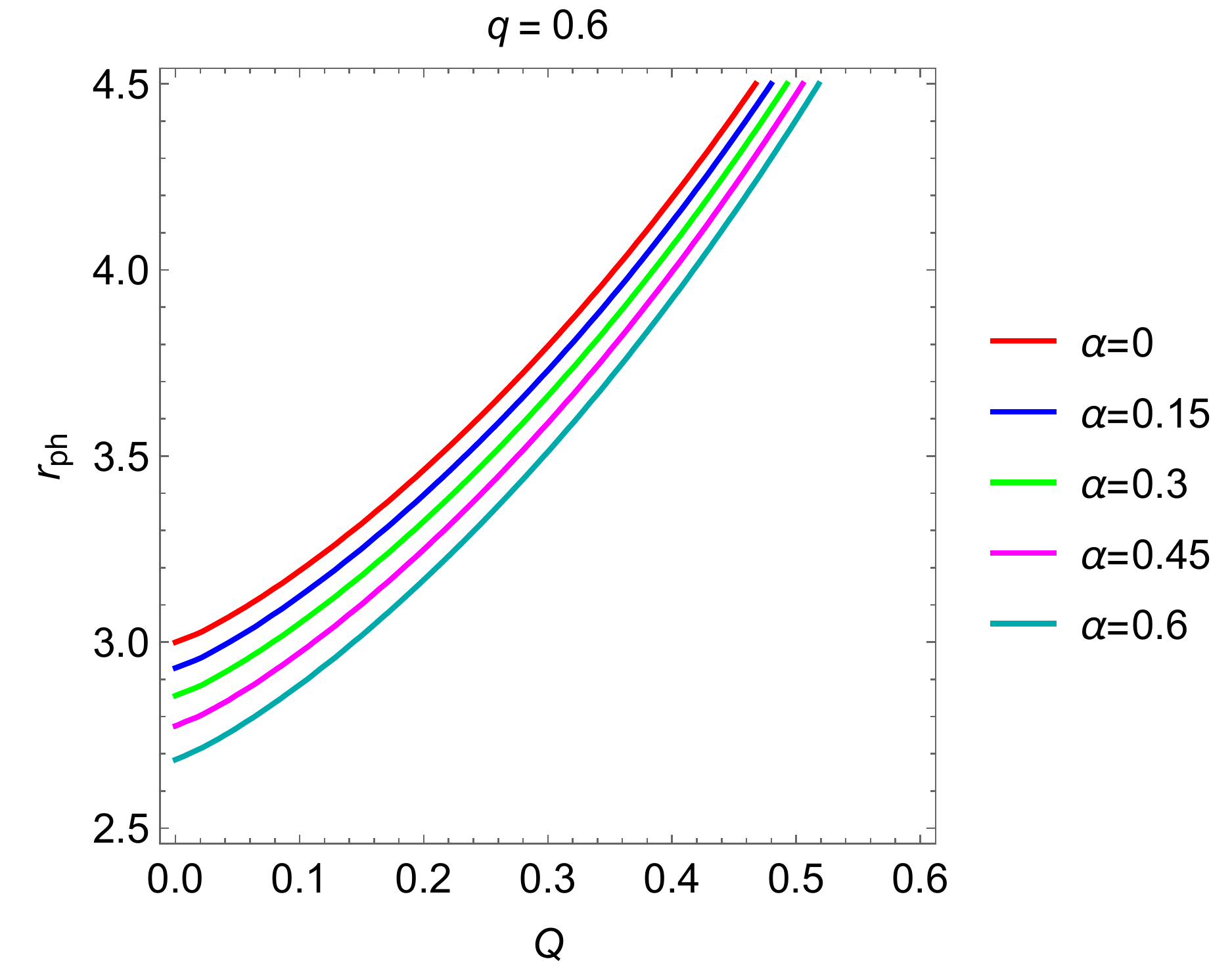}}
~
\subfigure{
\includegraphics[height=4.5cm,width=5.7cm]{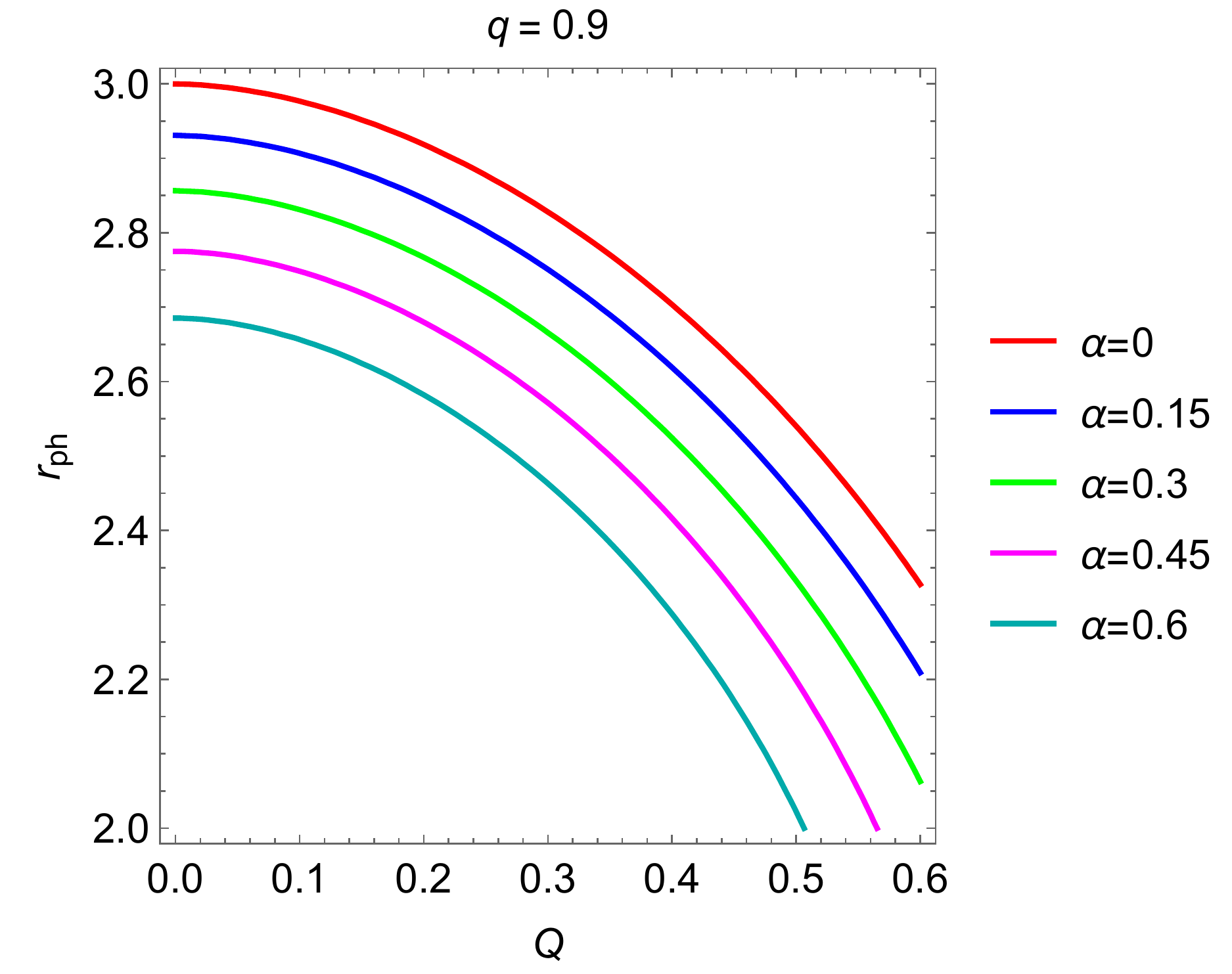}}
~
\subfigure{
\includegraphics[height=4.5cm,width=5.5cm]{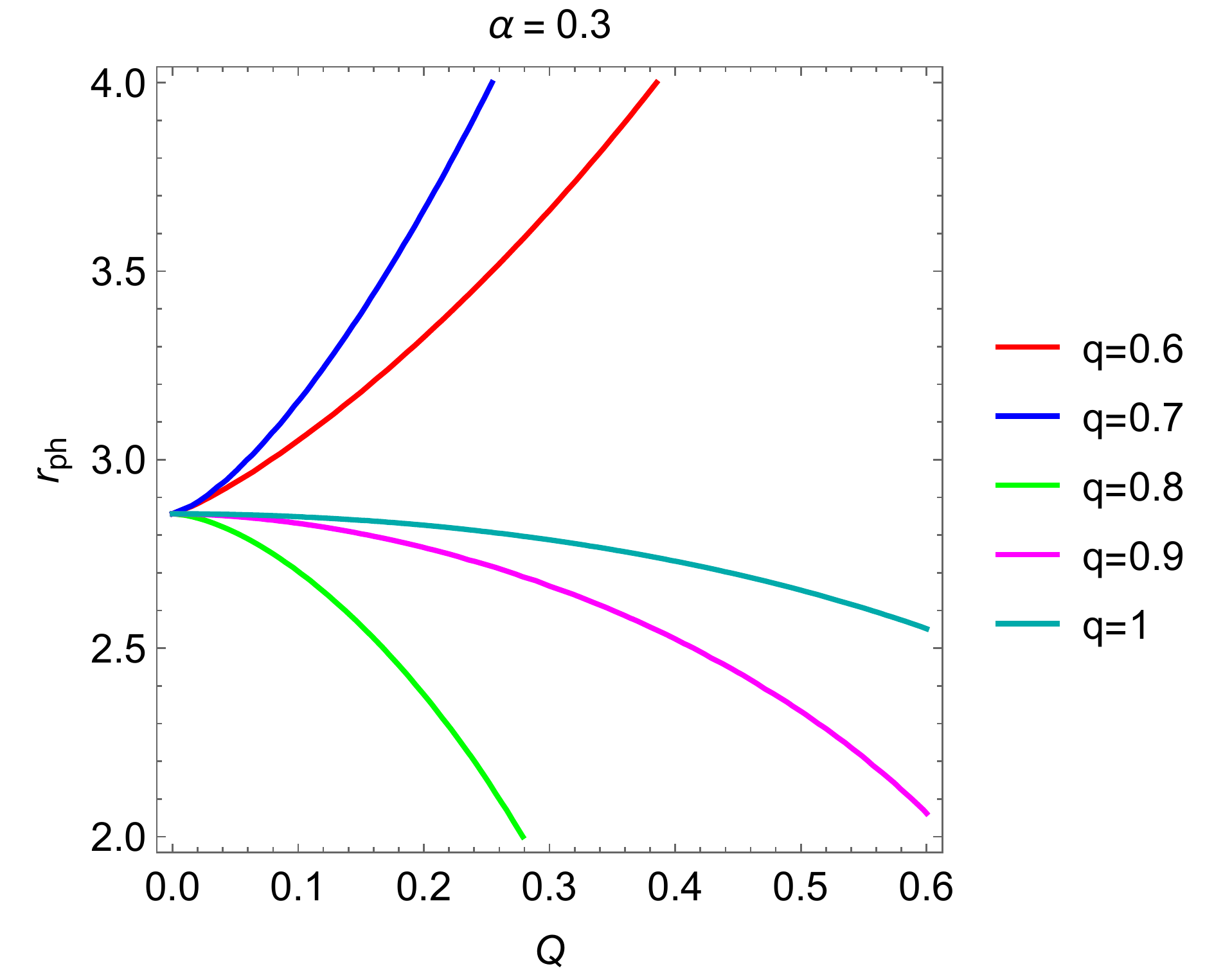}}
\subfigure{
\includegraphics[height=4.5cm,width=5.7cm]{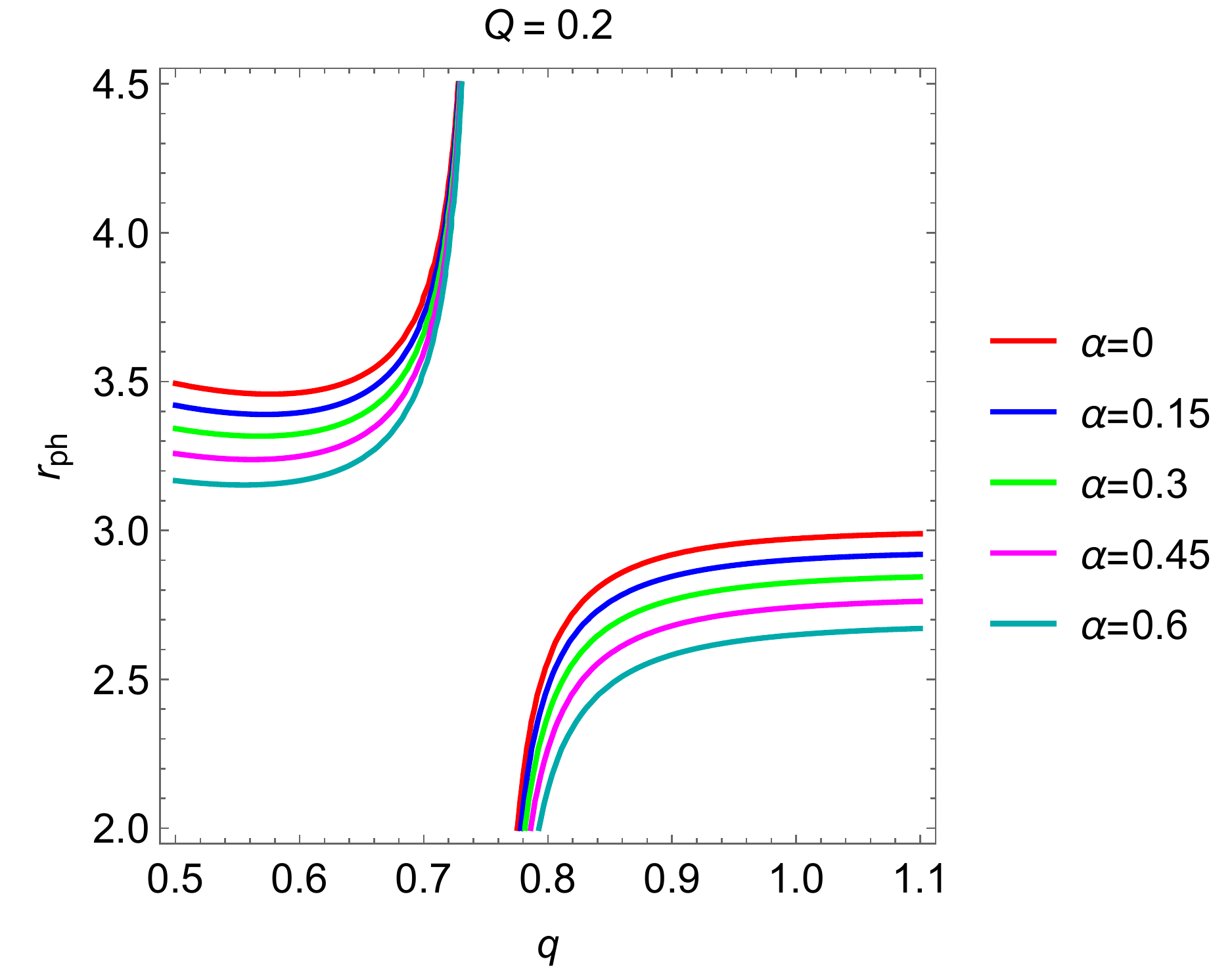}}
~
\subfigure{
\includegraphics[height=4.5cm,width=5.7cm]{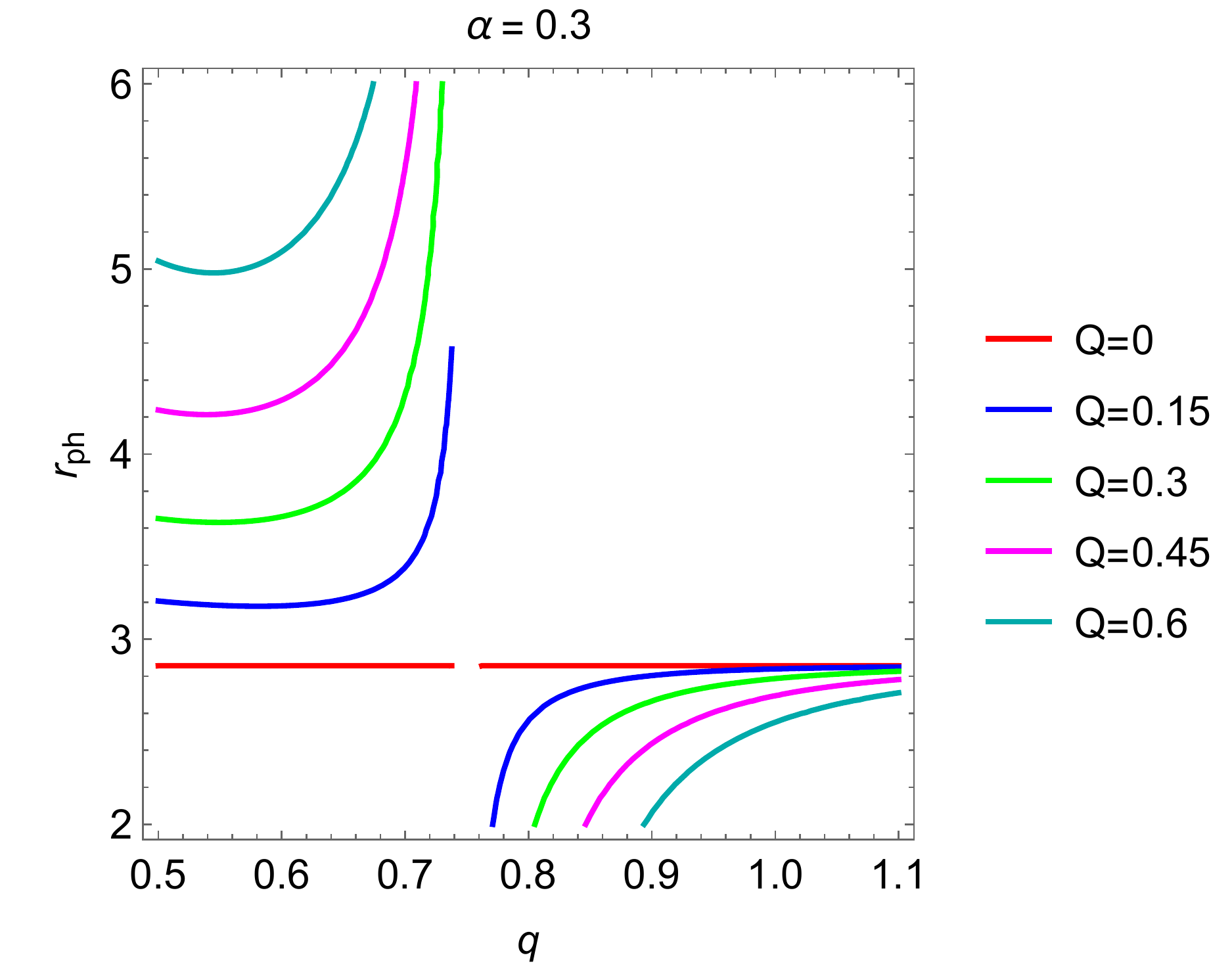}}
\end{center}
\caption{Plots for the radius of photon sphere w.r.t $q$, $Q$ and $\alpha$.} \label{phsphere}
\end{figure}

We present a brief discussion of the four-dimensional EGB BH immersed in the PYM field. The action for the GB gravity in $D$ dimensions coupled with the PYM field reads \cite{2009PhLB..681..190M,2022GReGr..54..161B}
\begin{equation}
S_G=\frac{1}{2}\int\sqrt{-det(g_{\mu\nu})} \bigg[R-\frac{\Lambda(D-2)(D-1)}{3}+\alpha\mathcal{L}_{GB}-\mathcal{F}^q\bigg]d^Dx, \label{1}
\end{equation}
where, $R$ is scalar curvature, $\Lambda$ is the cosmological constant, $\alpha$ is the GB parameter having the dimensions of $[L]^2$ such that $L$ denotes the length. In action (\ref{1})
\begin{eqnarray}
\mathcal{L}_{GB}=R^2 - 4 R_{\gamma\zeta} R^{\gamma\zeta} + R_{\gamma\zeta\pi\sigma} R^{\gamma\zeta\pi\sigma}, \label{2}
\end{eqnarray}
is the GB Lagrangian, where, $R_{\gamma\zeta\pi\sigma}$ and $R_{\gamma\zeta}$ represent the Riemann and Ricci tensors, respectively. Moreover, in the action (\ref{1}), the symbol $\mathcal{F}$ represents the YM invariant which is given as
\begin{equation}
\mathcal{F}=\sum_{a=1}^{\frac{(D-1)(D-2)}{2}}Tr\big(F_{\lambda\sigma}^{(a)}F^{\lambda\sigma(a)}\big), \label{3}
\end{equation}
with the parameter $q\in(0,\infty)$ that behaves as the power of YM charge. The expression for the YM field is given as
\begin{equation}
F_{\gamma\zeta}^{(a)}=\partial_\gamma A_\zeta^{(a)}-\partial_\zeta A_\gamma^{(a)}+\frac{1}{2\sigma}C^{(a)}_{(b)(c)}A_\gamma^{(b)}A_\zeta^{(c)}, \label{4}
\end{equation}
where $A_\zeta^{(a)}$ denote the YM potentials of the gauge group $SO(D-1)$, $\sigma$ is a constant and $C^{(a)}_{(b)(c)}$ are the structure constants of $\frac{(D-2)(D-1)}{2}$-parameter Lie group $G$. The action (\ref{1}) defines the possible coupling of modified gravity and the YM field with power, in the presence of cosmological constant. The $D$-dimensional integral is considered instead of the four dimensional integral because the required BH solution is derived from the pre-existing procedure for $D$-dimensional GB BHs by a dimensional regularization procedure. The equation of motion for potentials $A_\mu^{(a)}$ and the metric tensor $g_{\mu\nu}$ can be written as
\begin{eqnarray}
d\big(^\ast \textbf{F}^{(a)}\mathcal{F}^{q-1}\big)+\frac{1}{\sigma}C^{(a)}_{(b)(c)}\mathcal{F}^{q-1}\big(\textbf{A}^{(b)}\wedge\textbf{F}^{(c)}\big)&=&0, \label{5}\\
G_{\mu\nu}+\frac{(D-2)(D-1)}{3}\Lambda g_{\mu\nu}+\alpha H_{\mu\nu}&=&T_{\mu\nu}, \label{6}
\end{eqnarray}
where the symbols $^\ast$ and $\wedge$ represent the duality and wedge product, respectively. Moreover, $G_{\mu\nu}$, $H_{\mu\nu}$ and $T_{\mu\nu}$ are Einstein, Lanczos and energy-momentum tensors, respectively. The detailed expressions for these tensors are 
\begin{eqnarray}
G_{\mu\nu}&=&R_{\mu\nu}-\frac{1}{2}g_{\mu\nu}R, \label{7}\\
H_{\mu\nu}&=&2\big(RR_{\mu\nu}-2R_{\mu\sigma}R^\sigma_\nu-2R^{\sigma\delta}R_{\mu\sigma\nu\delta}+R^{\sigma\delta\lambda}_\mu R_{\nu\sigma\delta\lambda}\big)-\frac{1}{2}g_{\mu\nu}\mathcal{L}_{\mathcal{GB}}, \label{8}\\
T^\mu_\nu&=&2q\mathcal{F}^{q-1}Tr\big(F_{\nu\delta}^{(a)}F^{\mu\delta(a)}\big)-\frac{1}{2}\delta^\mu_\nu\mathcal{F}^q. \label{9}
\end{eqnarray}
The magnetic Wu-Yang ansatz \cite{2008PhLB..659..471H,2008PhLB..665..125H,1969pmuu.book.....M,1975PhRvD..12.2212Y} can be considered for the YM field such that the one-forms for YM magnetic potential can be written as
\begin{equation}
\textbf{A}^{(a)}=\frac{Q}{r^2}\big(x_idx_j-x_jdx_i\big), \label{10}
\end{equation}
where $r^2=\sum_{p=1}^{D-1}x^2_p$. The symbols $a$, $i$ and $j$ have the ranges $2\leq 2a\leq (D-1)(D-2)$ and $1\leq j\leq i-1\leq D-2$. Moreover, the YM magnetic charge is represented by $Q$. The general metric in $D$ dimensions is
\begin{equation}
g_{\gamma\zeta}dx^{\gamma}dx^{\zeta}=-f(r)dt^2+f(r)^{-1}dr^2+r^2\big(h_{rs}dx^rdx^s\big), \label{11}
\end{equation}
where $h_{rs}dx^rdx^s$ describes the geodesic distance between two points on a $(D-2)$-hypersphere. Then by re-scaling $\alpha$ to $\frac{\alpha}{D-4}$ and by utilizing the limit $D\rightarrow4$, the BH solution is obtained with the metric function as
\begin{equation}
f(r)=1+\frac{r^2\bigg[1\pm\sqrt{1+2\alpha\bigg(\frac{4M}{r^3}-\frac{(2Q^2)^q}{(4q-3)r^{4q}}\bigg)}\bigg]}{2\alpha}. \label{12}
\end{equation}
Note that, in the BH solution (\ref{12}), we have considered $\Lambda=0$. The metric function (\ref{12}) corresponds to two branches of solution. However, we proceed with only the $-$ branch of the solution because the solution corresponding to the $+$ sign does not have any physical description. Furthermore, when $q=1$, the BH solution becomes
\begin{equation}
f(r)=1+\frac{r^2\bigg[1\pm\sqrt{1+\frac{4\alpha}{r^3}\bigg(2M-\frac{Q^2}{r}\bigg)}\bigg]}{2\alpha}. \label{13}
\end{equation}
The solution (\ref{13}) is identical to the one derived by Fernandes \cite{2020PhLB..80535468F} with the only difference that $Q$ is electrical charge instead of YM magnetic charge. Without magnetic charge $Q$, the solution reduces to the novel BH solution derived by Glavan and Lin \cite{2020PhRvL.124h1301G}
\begin{equation}
f(r)=1+\frac{r^2}{2\alpha}\bigg[1\pm\sqrt{1+\frac{8\alpha M}{r^3}}\bigg]. \label{14}
\end{equation}
For $\alpha\rightarrow0$, we get the standard BH solution \cite{2009PhLB..681..190M}
\begin{equation}
f(r)=1-\frac{2M}{r}+\frac{2^{q-1}Q^{2q}}{(4q-3)r^{4q-2}}, \label{15}
\end{equation}
We can verify that the metric (\ref{12}) is asymptotically flat for $q>\frac{1}{2}$ and $q\neq\frac{3}{4}$. Hence, we proceed with those values of $q$ for which the BH solution under consideration is asymptotically flat. Next, we present an analysis of the radius of photon sphere. For this, we need the condition
\begin{equation}
\bigg[\frac{d}{dr}\big(\varrho(r)\big)\bigg]_{r=r_{ph}}=0, \label{16}
\end{equation}
where
\begin{equation}
\varrho(r)=\frac{g_{22}}{g_{00}}=\frac{r^2}{f(r)}. \label{17}
\end{equation}
The behaviour of the size of photon sphere in terms of its radius $r_{ph}$ has been plotted w.r.t GB parameter $\alpha$, power $q$ and YM charge $Q$ in Fig. \ref{phsphere}. In the upper panel, we can find the plots for $r_{ph}$ vs $\alpha$ in which $Q$ varies for each curve in the left and middle plots for fixed $q$. Whereas $Q$ has been fixed for the right plot corresponding to the different values of $q$ for each curve. The plots show that the photon sphere becomes smaller for all curves as $\alpha$ increases in the entire panel. However, in the middle plot, the photon sphere becomes smaller relatively faster as compared to the other two plots. Moreover, as $Q$ increases, the curves are shifted upwards for the left plot and downwards for the middle plot. Whereas, in the right plot, the curves are also shifted upwards for both intervals of $q$ i.e. for $q<\frac{3}{4}$ and $q>\frac{3}{4}$. However, it is observed that the set of curves for both intervals is separated by an arbitrary curve corresponding to uncharged EGB BH. In the middle panel, we find the plots for $r_{ph}$ vs $Q$ whereas $\alpha$ varies for different curves in the left and middle plots for fixed values of $q$. In the right plot, the curves correspond to $q$ and $\alpha$ has been fixed. It can be observed that the photon sphere becomes enlarged for all curves as $Q$ increases in the left plot. Whereas, the size of photon sphere decreases for all curves as $Q$ increases in the middle plot. The curves expand in the left plot and shrink in the right plot as $\alpha$ grows larger. The right plot is relatively different as compared to the other plots observed so far. For $q<\frac{3}{4}$, the size of photon sphere increases and for $q>\frac{3}{4}$, the size of photon sphere decreases w.r.t increase in $Q$. Moreover, for $Q=0$, the curves converge at a fixed value of $r_{ph}$ corresponding to the uncharged EGB BH. For $q<\frac{3}{4}$, the size of photon sphere increases faster w.r.t increasing $q$ and for fixed $Q$. Whereas, for $q>\frac{3}{4}$, the size of photon sphere decreases faster w.r.t increasing $q$ and for fixed $Q$. In the lower panel, we plotted for $r_{ph}$ vs $q$ for the curves corresponding to $\alpha$ for fixed values of $Q$ for the left plot and for the curves corresponding to $Q$ for fixed values of $\alpha$ for the right plot. A discontinuity has been observed in curves for $r_{ph}$ when $q\rightarrow\frac{3}{4}$ in the entire panel. However, as we move away from $q=\frac{3}{4}$, the radius becomes stable and is approximately constant. In the left plot, $r_{ph}$ grows rapidly as $q\rightarrow\frac{3}{4}^-$ whereas $r_{ph}\rightarrow0$ as $q\rightarrow\frac{3}{4}^+$ for all values of $\alpha$. When $q\lessapprox0.7$ and $q\gtrapprox0.8$, the size of photon sphere becomes small as $\alpha$ increases for each curve for fixed $q$. In the right plot, a similar kind of discontinuity is observed at $q=\frac{3}{4}$. However, for $Q=0$, a constant value of $r_{ph}$ is observed. For $q<\frac{3}{4}$, the size of photon sphere increases with increase in $Q$ whereas for $q>\frac{3}{4}$, the size of photon sphere decreases with increase in $Q$.

\section{The Rotating Black Hole}
\begin{figure}[t]
\begin{center}
\subfigure{
\includegraphics[height=5.5cm,width=6.6cm]{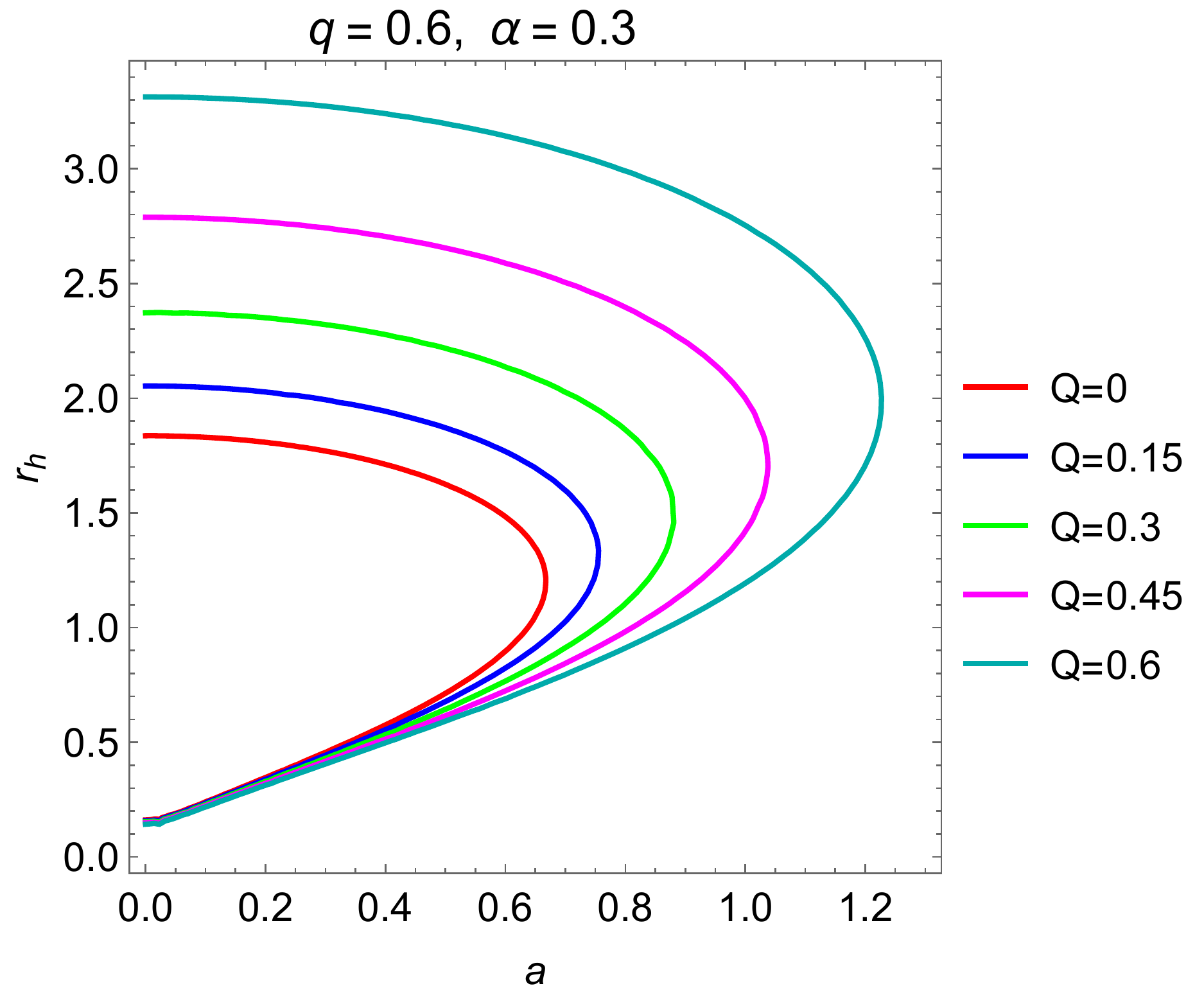}}
~
\subfigure{
\includegraphics[height=5.5cm,width=6.6cm]{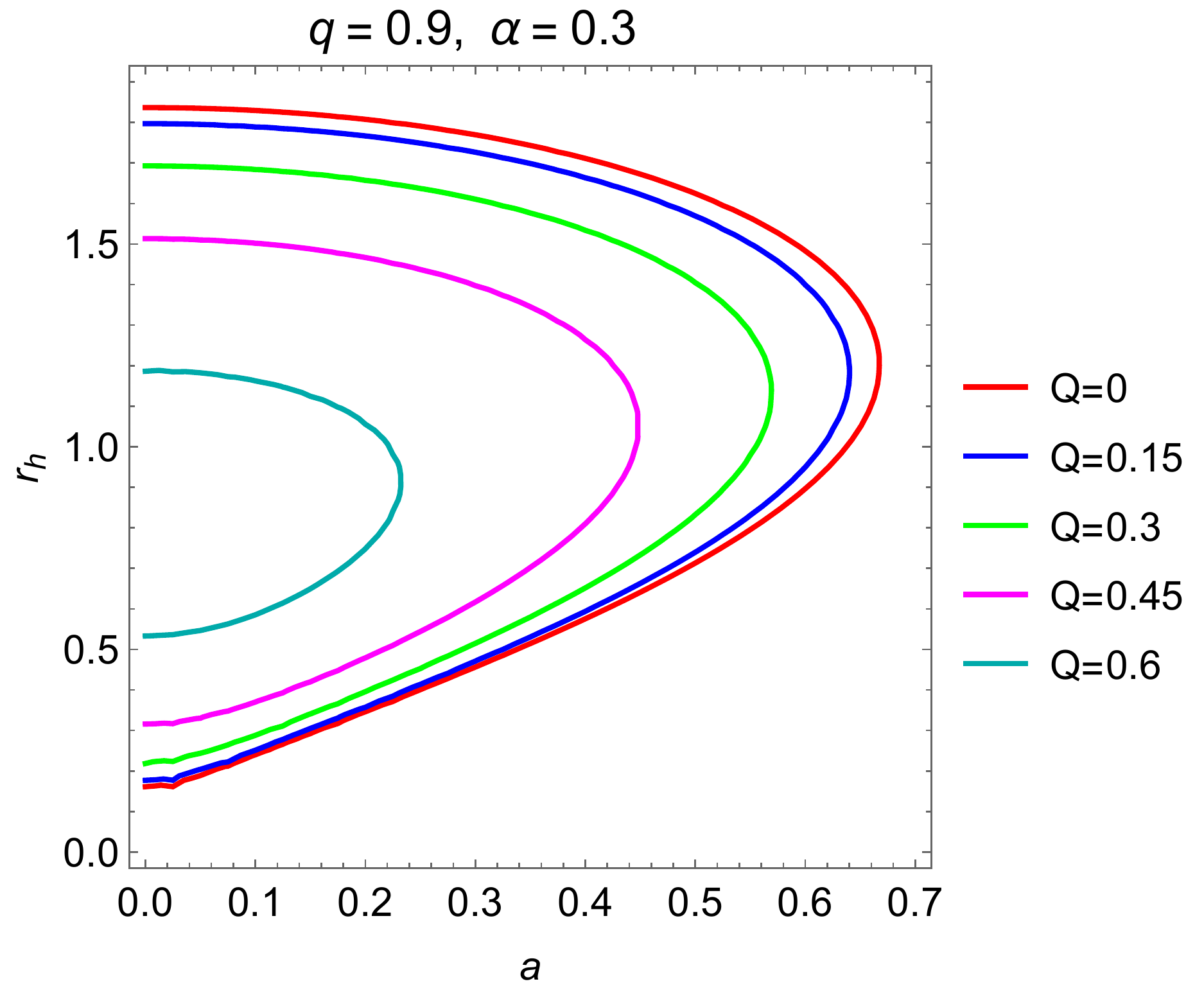}}
~
\subfigure{
\includegraphics[height=5.5cm,width=6.6cm]{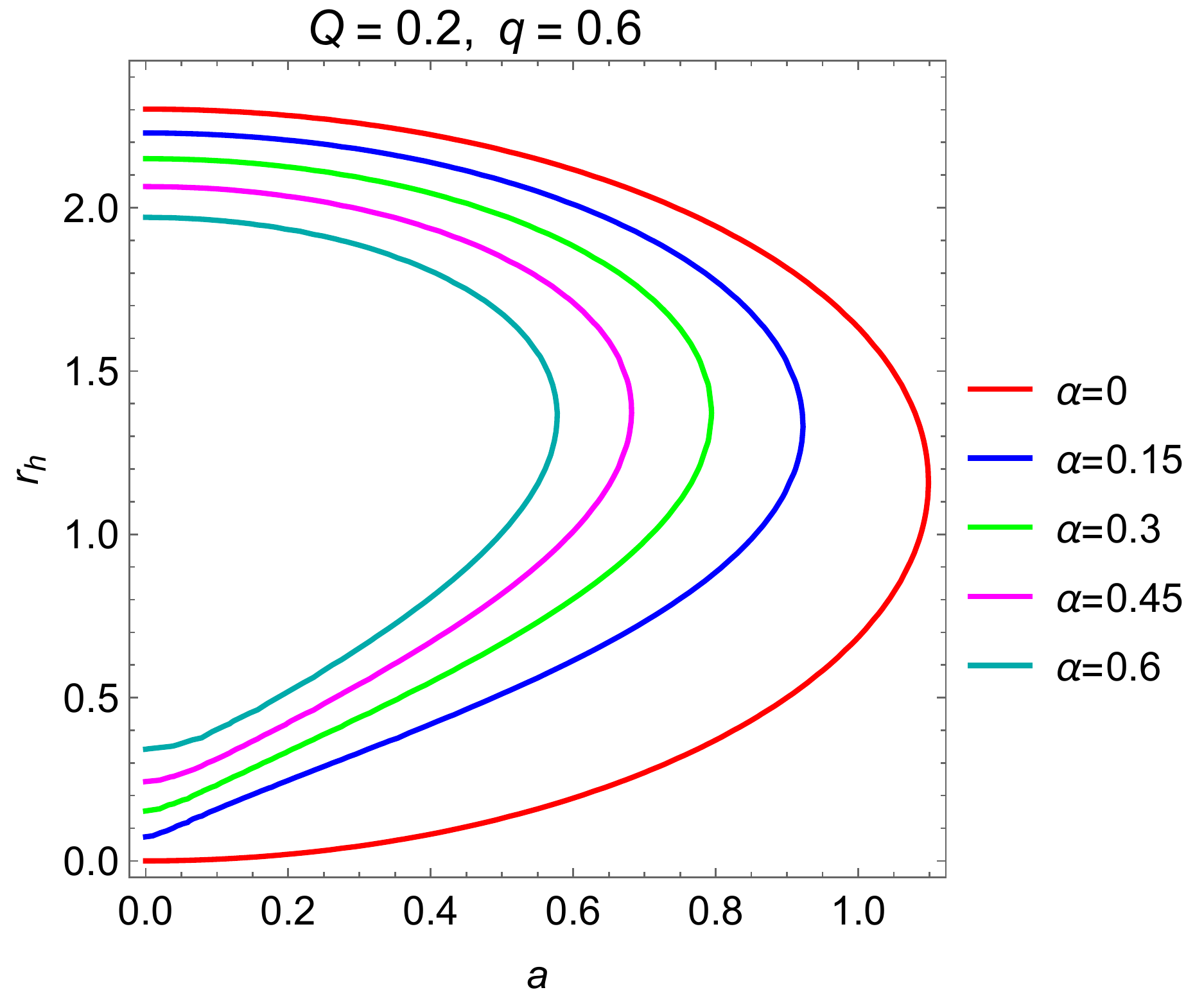}}
\subfigure{
\includegraphics[height=5.5cm,width=6.6cm]{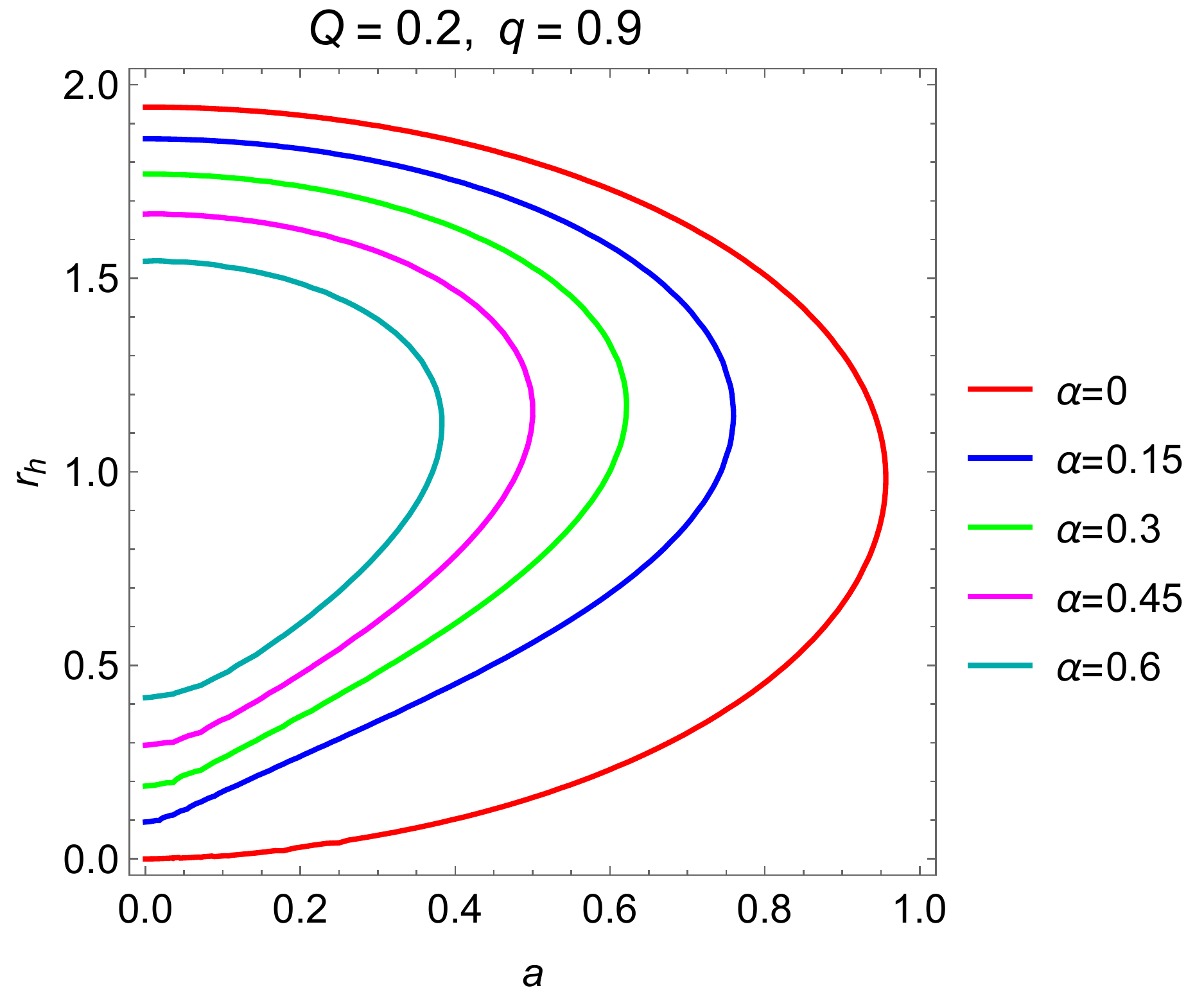}}
~
\subfigure{
\includegraphics[height=5.5cm,width=6.4cm]{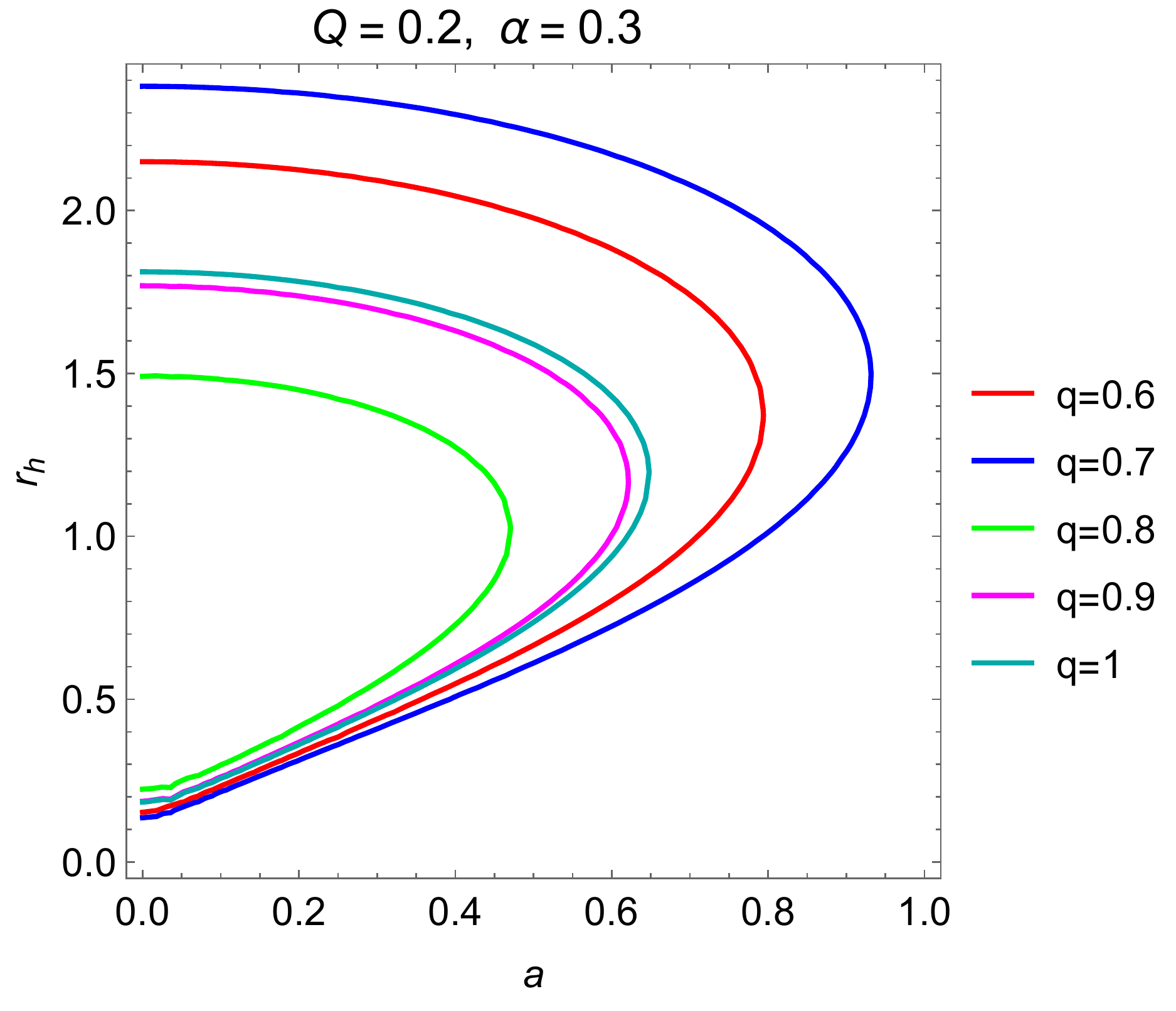}}
\end{center}
\caption{Plots for the horizon radius w.r.t $a$ for various values of $q$, $Q$ and $\alpha$.} \label{horizonrad}
\end{figure}

The Newman-Janis algorithm (NJA) \cite{1965JMP.....6..915N,1965JMP.....6..918N} has been developed for deriving the rotating BH solutions from their static counterparts within GR and has been successfully used over the years. The rotating metrics were obtained by applying NJA to Schwarzschild and Reissner-Nordstr\"om metrics, respectively. We know that the source for Schwarzschild and Kerr BHs is a vacuum and for Reissner-Nordstr{\"o}m and Kerr-Newman BHs is the electric charge. It can be deduced that the algorithm applied within GR does not affect the source. However, it is not true for the BHs in modified gravity theories. Hansen and Yunes \cite{2013PhRvD..88j4020H} proposed that some additional sources appeared through the application of NJA in non-GR theories. In non-GR theories, to overcome this issue, the rotating metrics can be derived by the application of a modified NJA \cite{2014PhRvD..90f4041A,2014EPJC...74.2865A} to their static counterparts. This algorithm does not comprise any radial complexification procedure. Instead, it consists of functions of the spin parameter whose values can be determined within the algorithm. For this reason, the modified algorithm becomes straightforward that can be applied to numerous BH solutions in GR and modified theories. It has been successfully applied to obtain the rotating BH solutions for imperfect fluids and some generic rotating regular BH metrics \cite{2021Galax...9...43F,2022EPJC...82..547W}. Moreover, the rotating metrics in $4D$ EGB theory have also been obtained by this method \cite{2020JCAP...07..053K,2022PDU....3500916P}. We also apply the modified algorithm to the static metric (\ref{11}) by considering the Eddington-Finkelstein (EF) coordinates $(u,r,\theta,\phi)$ such that
\begin{eqnarray}
dt-du=\frac{dr}{f(r)}. \label{18}
\end{eqnarray}
By using the transformation (\ref{18}) in the static BH solution, we get
\begin{equation}
ds^{2}=-f(r)du^{2}-2dudr+r^{2}d\Omega_2^2, \label{19}
\end{equation}
where $d\Omega_2^2$ is the geodesic distance between two points on the ordinary Euclidean sphere. The null tetrad representation for the metric is given by
\begin{eqnarray}
l^a&=&\delta^a_r, \label{20}\\
n^a&=&\delta^a_u-\frac{f(r)}{2}\delta^a_r, \label{21}\\
m^a&=&\frac{1}{\sqrt{2}r}\bigg(\delta^a_\theta+\frac{i}{\sin{\theta}}\delta^a_\phi\bigg). \label{22}
\end{eqnarray}
The conjugate metric tensor can also be written as
\begin{equation}
g^{ab}=-l^an^b-n^al^b+m^a\bar{m}^b+\bar{m}^am^b, \label{23}
\end{equation}
where, the representation $\bar{m}^a$ is the complex conjugate of $m^a$. Moreover, the above null tetrad obeys the conditions:
\begin{eqnarray}
l_al^a=n_an^a=m_am^a=\bar{m}_a\bar{m}^a&=&0, \label{24}\\
l_am^a=l_a\bar{m}^a=n_am^a=n_a\bar{m}^a&=&0, \label{25}\\
-l_an^a=-l^an_a=m_a\bar{m}^a=m^a\bar{m}_a&=&1. \label{26}
\end{eqnarray}
The rotation in $(u,r)$-plane is governed by the following transformations with $a$ as spin parameter:
\begin{eqnarray}
u'-u&\rightarrow&-ia\cos{\theta}, \label{27}\\
r'-r&\rightarrow&ia\cos{\theta}. \label{28}
\end{eqnarray}
We introduce the undetermined metric functions as
\begin{eqnarray}
f(r)&\rightarrow& \mathcal{A}(r,\theta,a), \label{29}\\
r^2&\rightarrow& \mathcal{B}(r,\theta,a). \label{30}
\end{eqnarray}
Using the complex transformations without the radial complexification and the undetermined metric functions, the null tetrad transforms as
\begin{eqnarray}
l'^{a}&=&\delta^a_r, \label{31}\\
n'^{a}&=&\delta^a_u - \frac{\mathcal{A}}{2}\delta^a_r, \label{32}\\
m'^{a}&=&\frac{1}{\sqrt{2\mathcal{B}}}\bigg(ia\sin{\theta}\big(\delta^a_u-\delta^a_r\big)+\delta^a_\theta + i\csc{\theta}\delta^a_\phi\bigg), \label{33}
\end{eqnarray}
Now, using the transformed null tetrad and the metric tensor (\ref{23}), the metric in EF coordinates reads
\begin{eqnarray}
ds^{2}&=&-\mathcal{A}du^{2}-2dudr+2a(\mathcal{A}-1)\sin^{2}\theta dud\phi+\mathcal{B}d\theta^{2}+2a\sin^{2}\theta drd\phi \nonumber \\
&&+(\mathcal{B}\sin^{2}\theta-(\mathcal{A}-2)a^2\sin^{4}\theta)d\phi^{2}. \label{34}
\end{eqnarray}
We omit the primes for avoiding confusion in Eq. (\ref{34}). The metric (\ref{34}) is to be converted into the Boyer-Lindquist coordinates with the transformations
\begin{eqnarray}
dt-du&=&\frac{a^2+r^2}{a^2+r^2f(r)}dr, \label{35}\\
d\phi'-d\phi&=&\frac{a}{a^2+r^2f(r)}dr. \label{36}
\end{eqnarray}
Then by assuming
\begin{eqnarray}
\mathcal{A}=\frac{r^2f(r)+a^2\cos^2{\theta}}{\mathcal{B}}, ~~~ \mathcal{B}=a^2\cos^2{\theta}+r^2, \label{37}
\end{eqnarray}
we get
\begin{eqnarray}
ds^{2}&=&\frac{a^2\sin^2{\theta}-\Delta(r)}{\rho^2}dt^{2}+\frac{\rho^2}{\Delta(r)}dr^2+\rho^2d\theta^{2}-\frac{\Delta(r)a^2\sin^{4}\theta-\big(r^2+a^2\big)^2\sin^{2}\theta}{\rho^2}d\phi^{2} \nonumber \\
&&-\frac{2a\sin^{2}\theta\big(a^2+r^2-\Delta(r)\big)}{\rho^2}dtd\phi, \label{38}
\end{eqnarray}
where
\begin{eqnarray}
\Delta(r)&=&a^2+r^2f(r)=a^2+r^2+\frac{r^4\bigg[1\pm\sqrt{1+2\alpha\bigg(\frac{4M}{r^3}-\frac{(2Q^2)^q}{(4q-3)r^{4q}}}\bigg]}{2\alpha}, \label{39}\\
\rho^2&=&r^2+a^2\cos^2{\theta}. \label{40}
\end{eqnarray}
Since it must be kept in mind that the rotating BH solution (\ref{38}) may not exactly satisfy the initial EFE due to the possible generation of the extra sources. However, the BH solution (\ref{38}) may be regarded as a regularized $4D$ EGB BH solution with a PYM field corresponding to an unknown EFE that is distinct from the EFE (\ref{6}). Moreover, when the spin vanishes, the metric (\ref{38}) reduces to its static counterpart (\ref{12}). If $\alpha\rightarrow0$, we get the Kerr BH solution in the same way as we get the Schwarzschild BH solution for non-rotating cases. Due to these reasons, we can regard the solution (\ref{38}) as a possible rotating BH solution. Next, we study the nature of horizon radii of the rotating BH metric (\ref{38}) by solving the equation $\Delta(r)=0$ for its roots. For a rigorous analysis, we solve $\Delta(r)=0$ numerically to obtain the radius of horizon $r_h$ w.r.t spin $a$ depicted in Fig. \ref{horizonrad}. It is found that as $a$ increases, the event horizon (EH) becomes smaller, whereas the Cauchy horizon (CH) grows larger for all curves in all plots. The curves in the upper panel correspond to different values of $Q$ for fixed values of $q$ and $\alpha$. The curves in the left plot are shifted outwards with an increasing value of $Q$. Whereas in the right plot, the curves are shifted inwards with an increasing value of $Q$. Therefore, the extremal values of $a$ increase in the left plot and decrease in the right plot w.r.t $Q$. Moreover, for the slowly rotating case, the CH does not vary significantly w.r.t $Q$ in the left plot. The curves in the middle panel correspond to $\alpha$ for fixed $Q$ and $q$. It is found that the size of EH becomes smaller and CH becomes larger with increasing $\alpha$ for both plots. Therefore, the extremal values of spin decrease w.r.t $\alpha$ for both plots. However, in the left plot, when $a=0$, $r_h\rightarrow r_h^0$ as $\alpha$ increases, where $r_h^0$ is slightly less than $2$. Whereas, in the right plot, when $a=0$, $r_h\rightarrow r_h^0$ as $\alpha\rightarrow0$. This is how the curves for $q<\frac{3}{4}$ are separated by the curves for $q>\frac{3}{4}$. The plot in the lower panel corresponds to the curves for various values of $q$ and a fixed value of $Q$ and $\alpha$. We can see that the curves are expanded as $q\rightarrow\frac{3}{4}$ and for $q>\frac{3}{4}$. In the limit $q\rightarrow\frac{3}{4}$, the EH remains above the curve corresponding to the Schwarzschild BH and increases w.r.t $q$. Whereas, for $q>\frac{3}{4}$, the curves approach the curve corresponding to the Schwarzschild BH.

\section{Geodesics and Shadows}
We study the effective potential, shadow and distortion for rotating and non-rotating BH in $4D$ EGB gravity with PYM field. For this, we require the equations of geodesics that can be derived using the Hamilton-Jacobi (HJ) approach \cite{1968PhRv..174.1559C}. The Lagrangian in tensorial form can be written as
\begin{equation}
\mathcal{L}=\frac{1}{2}g_{\mu\nu}\dot{x}^{\mu}\dot{x}^{\nu}. \label{41}
\end{equation}
The generalized momenta are given by
\begin{equation}
p_\mu=g_{\mu\nu}\dot{x}^{\nu}. \label{42}
\end{equation}
The two constants of motion are energy $E$ and angular momentum $l$ (z-component) are obtained by solving the Lagrangian, generalized momenta and metric coefficient from Eqs. (\ref{41}), (\ref{42}) and (\ref{39}), respectively and are given as
\begin{eqnarray}
E&:=&-\frac{\partial\mathcal{L}}{\partial\dot{t}}=-g_{tt}\dot{t}-g_{\phi t}\dot{\phi}=p_t, \label{43} \\
l&:=&\frac{\partial\mathcal{L}}{\partial\dot{\phi}}=g_{\phi t}\dot{t}+g_{\phi \phi}\dot{\phi}=p_\phi, \label{44}
\end{eqnarray}
where, $\dot x=\frac{\partial x}{\partial\tau}$ and $\tau$ is the affine parameter. The HJ equation is given by
\begin{equation}
2\frac{\partial \mathcal{S_J}}{\partial\tau}+g^{\mu\nu}\frac{\partial \mathcal{S_J}}{\partial x^\mu}\frac{\partial \mathcal{S_J}}{\partial x^\nu}=0, \label{45}
\end{equation}
such that the Jacobi action $\mathcal{S_J}$ can be taken as
\begin{equation}
\mathcal{S_J}=\frac{1}{2}m_p\tau-Et+l\phi+\mathcal{A}_r(r)+\mathcal{A}_\theta(\theta), \label{46}
\end{equation}
where $m_p$ denotes the moving particle's mass, $\mathcal{A}_r(r)$ and $\mathcal{A}_\theta(\theta)$ are arbitrary functions. Now, inserting Eq. (\ref{46}) in (\ref{45}) and then simplifying further by implementing the separation of variables procedure \cite{1968PhRv..174.1559C} leads to the equations for null geodesics $(m_p=0)$ as
\begin{eqnarray}
\rho^4\dot{t}^2&=&\bigg[\frac{E\big(r^2+a^2\big)^2-al\big(r^2+a^2\big)}{\Delta}+a\sin^2{\theta}\big(l\csc^2{\theta}-aE\big)\bigg]^2, \label{47} \\
\rho^4\dot{r}^2&=&\mathcal{R}, \label{48} \\
\rho^4\dot{\theta}^2&=&\Theta, \label{49} \\
\rho^4\dot{\phi}^2&=&\bigg[\frac{aE\big(r^2+a^2\big)-la^2}{\Delta}+\big(l\csc^2{\theta}-aE\big)\bigg]^2, \label{50}
\end{eqnarray}
where
\begin{eqnarray}
\mathcal{R}(r)&=&\big(al-(a^2+r^2)E\big)^2-\Delta\mathcal{Z}-\Delta(aE-l)^2, \label{51} \\
\Theta(\theta)&=&\cot^2{\theta}\big(a^2E^2\sin^2{\theta}-l^2\big)+\mathcal{Z}, \label{52}
\end{eqnarray}
such that $\mathcal{Z}$ is the Carter constant and $\mathcal{Z}+(aE-l)^2$ is the modified Carter constant. The effective potential for an observer residing in the equatorial plane at $\big(\theta=\frac{\pi}{2}\big)$ is related to the radial geodesic equation that is given by
\begin{eqnarray}
V_{eff}=-\frac{\mathcal{R}(r)}{2r^4}. \label{53}
\end{eqnarray}
\begin{figure}[ht!]
\begin{center}
\subfigure{
\includegraphics[height=5cm,width=7.14cm]{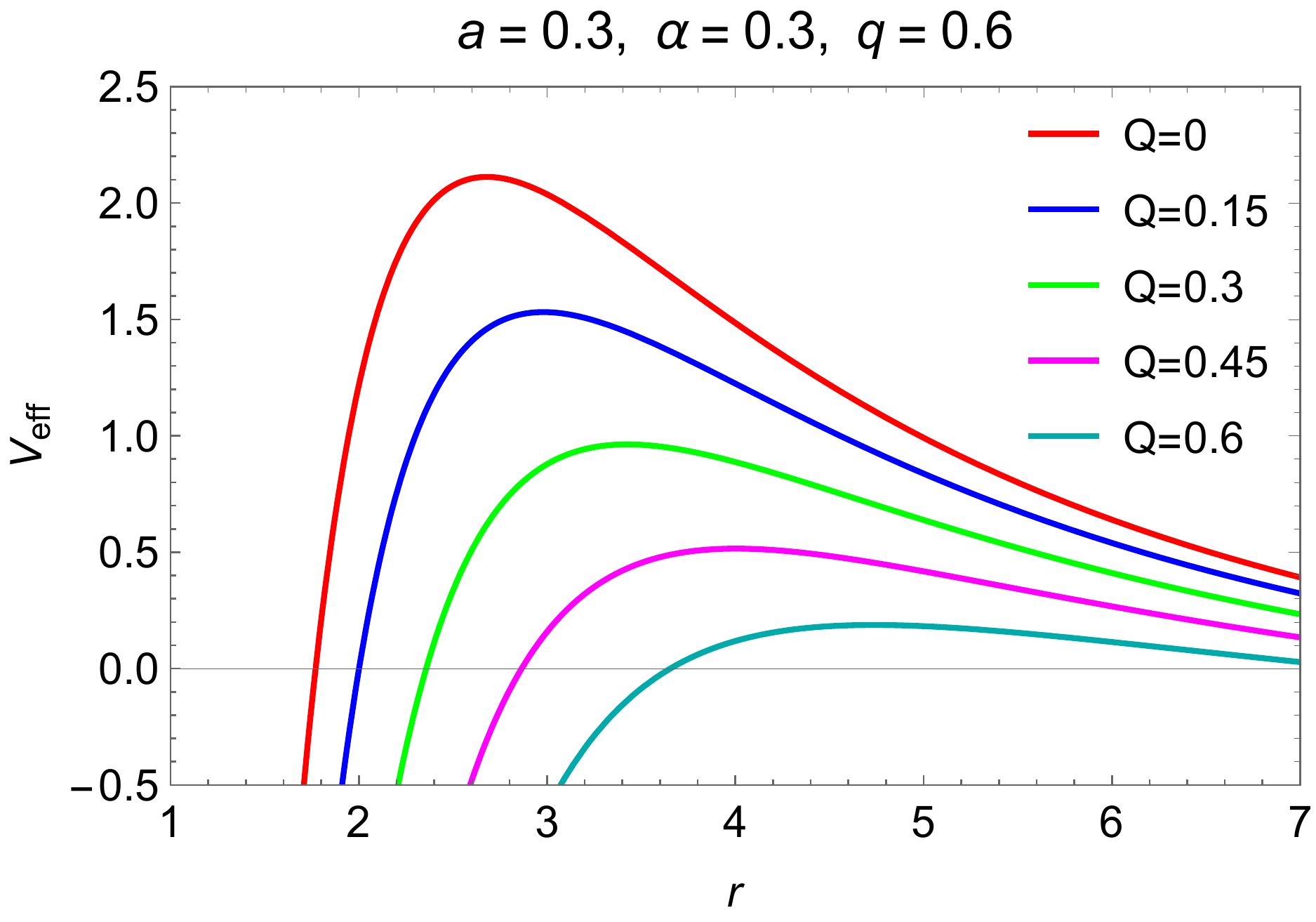}}
~~~
\subfigure{
\includegraphics[height=5cm,width=6.83cm]{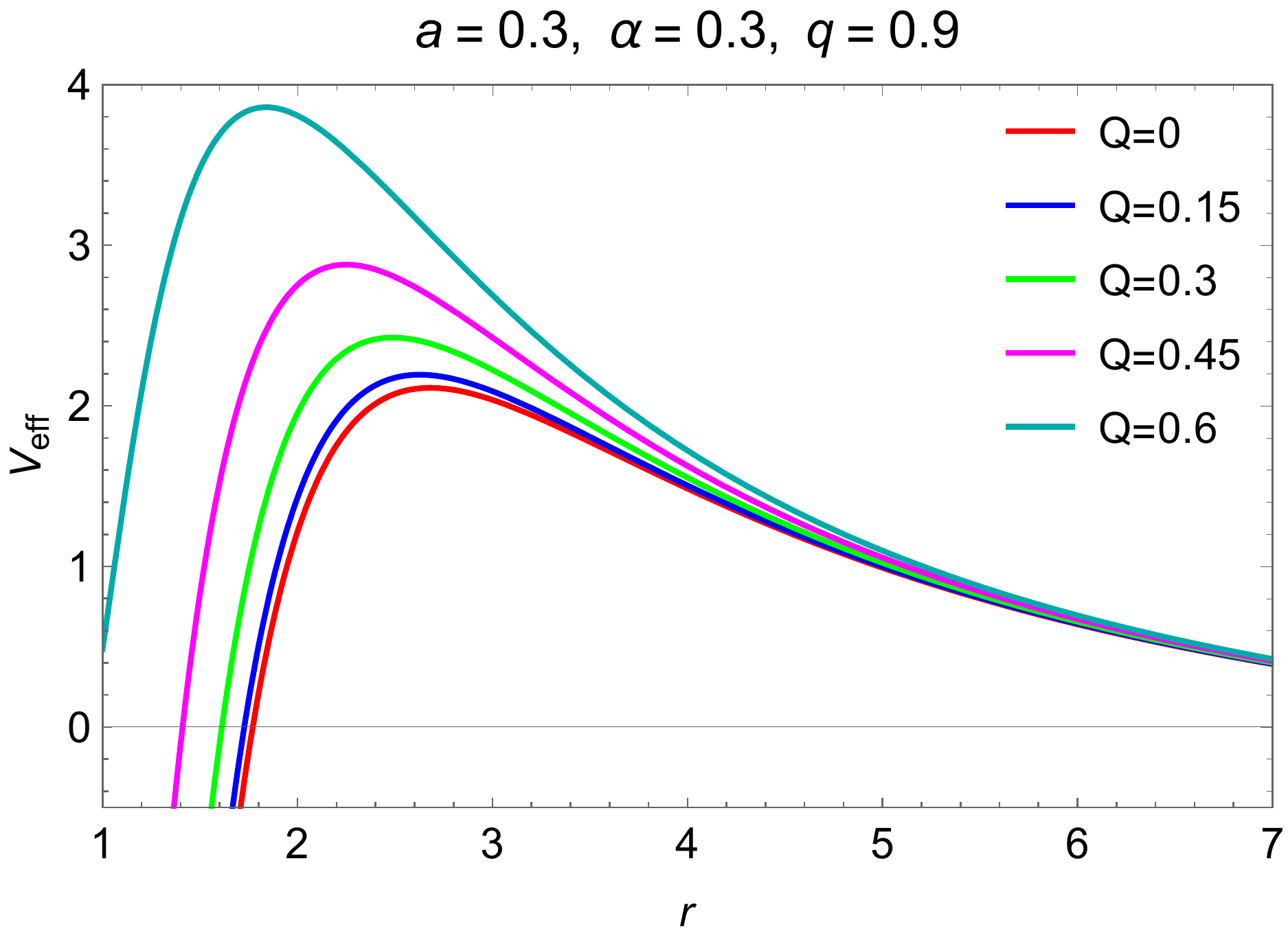}}
\subfigure{
\includegraphics[height=5cm,width=7.14cm]{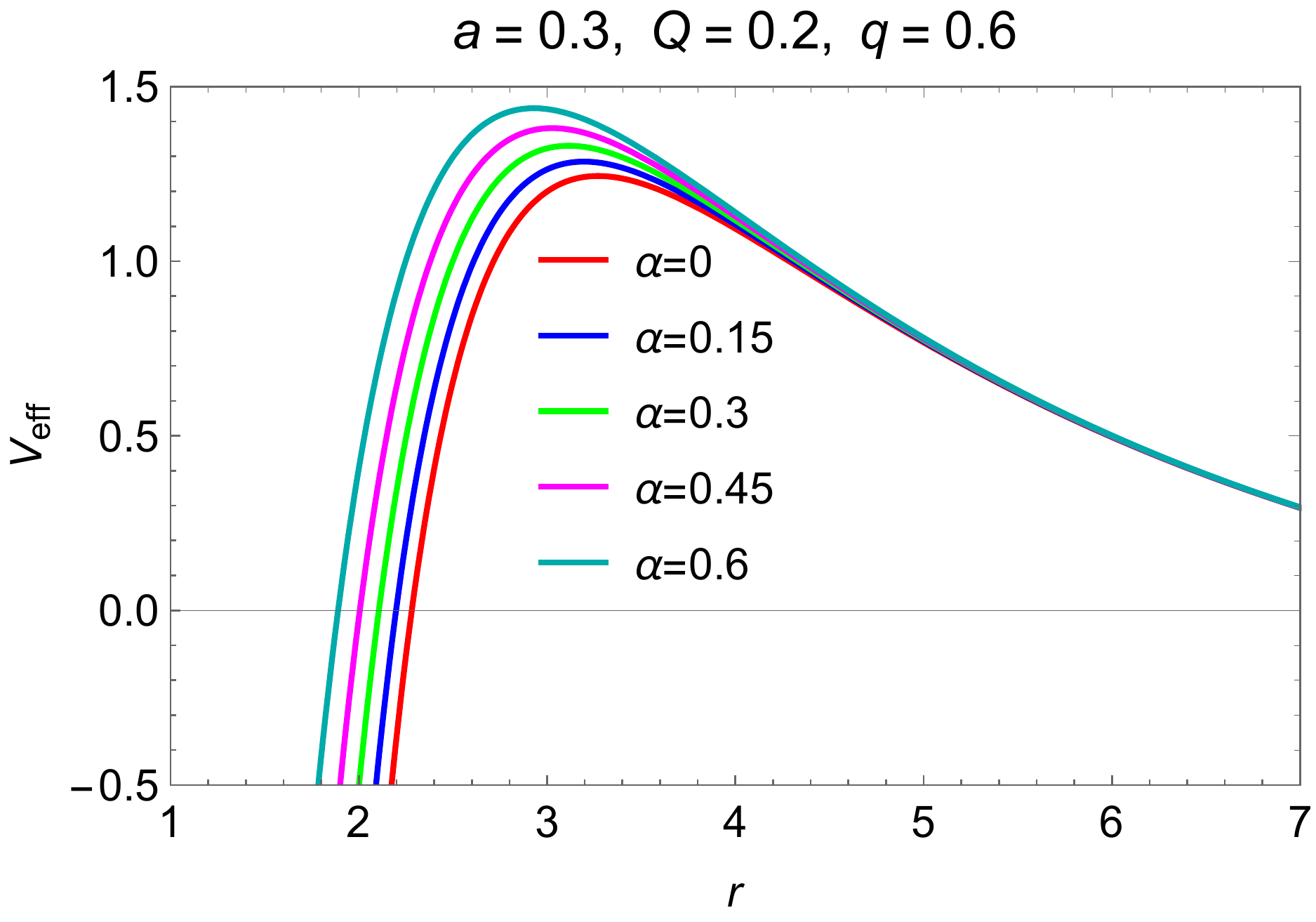}}
~~~
\subfigure{
\includegraphics[height=5cm,width=7.14cm]{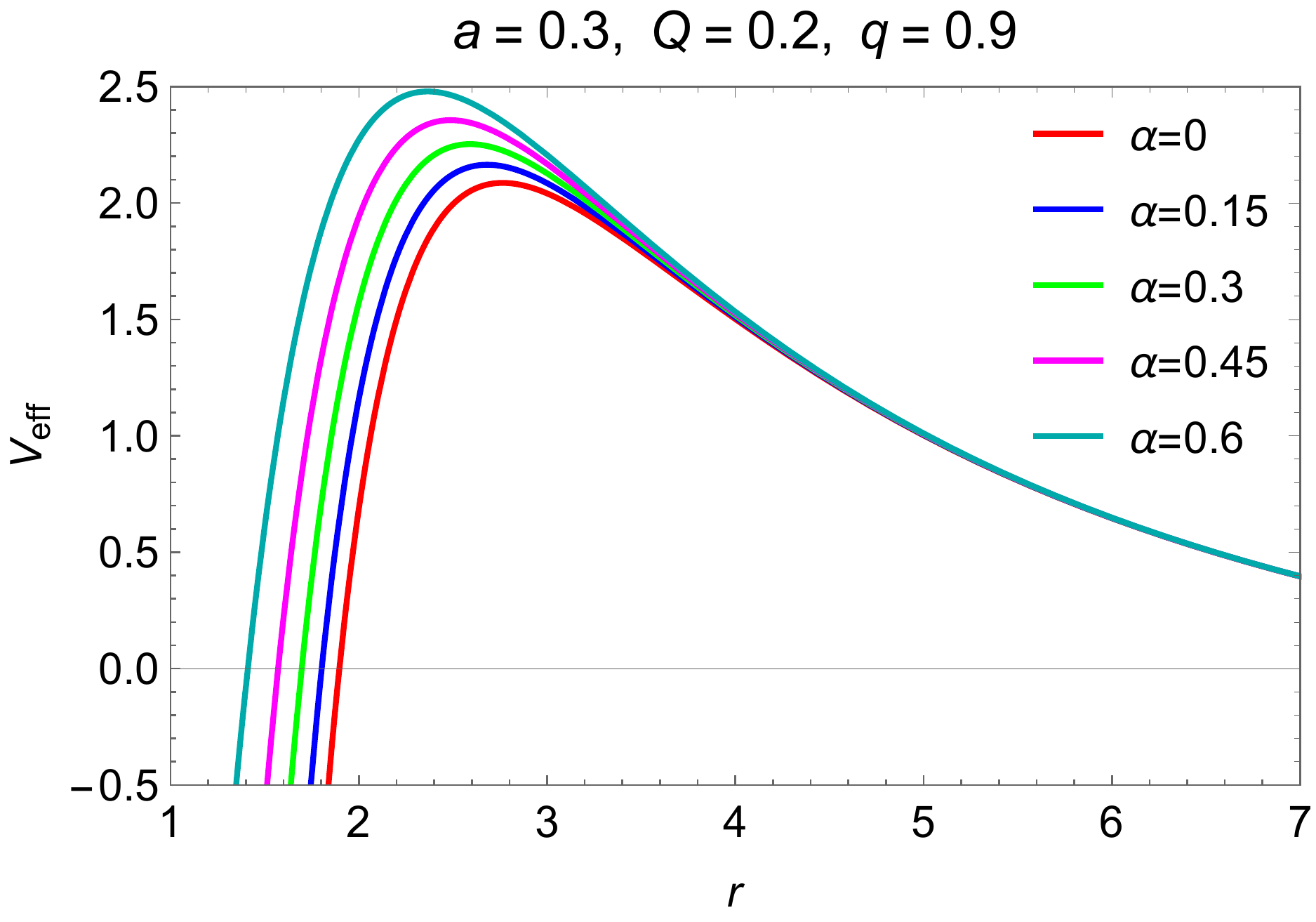}}
\subfigure{
\includegraphics[height=5cm,width=7.14cm]{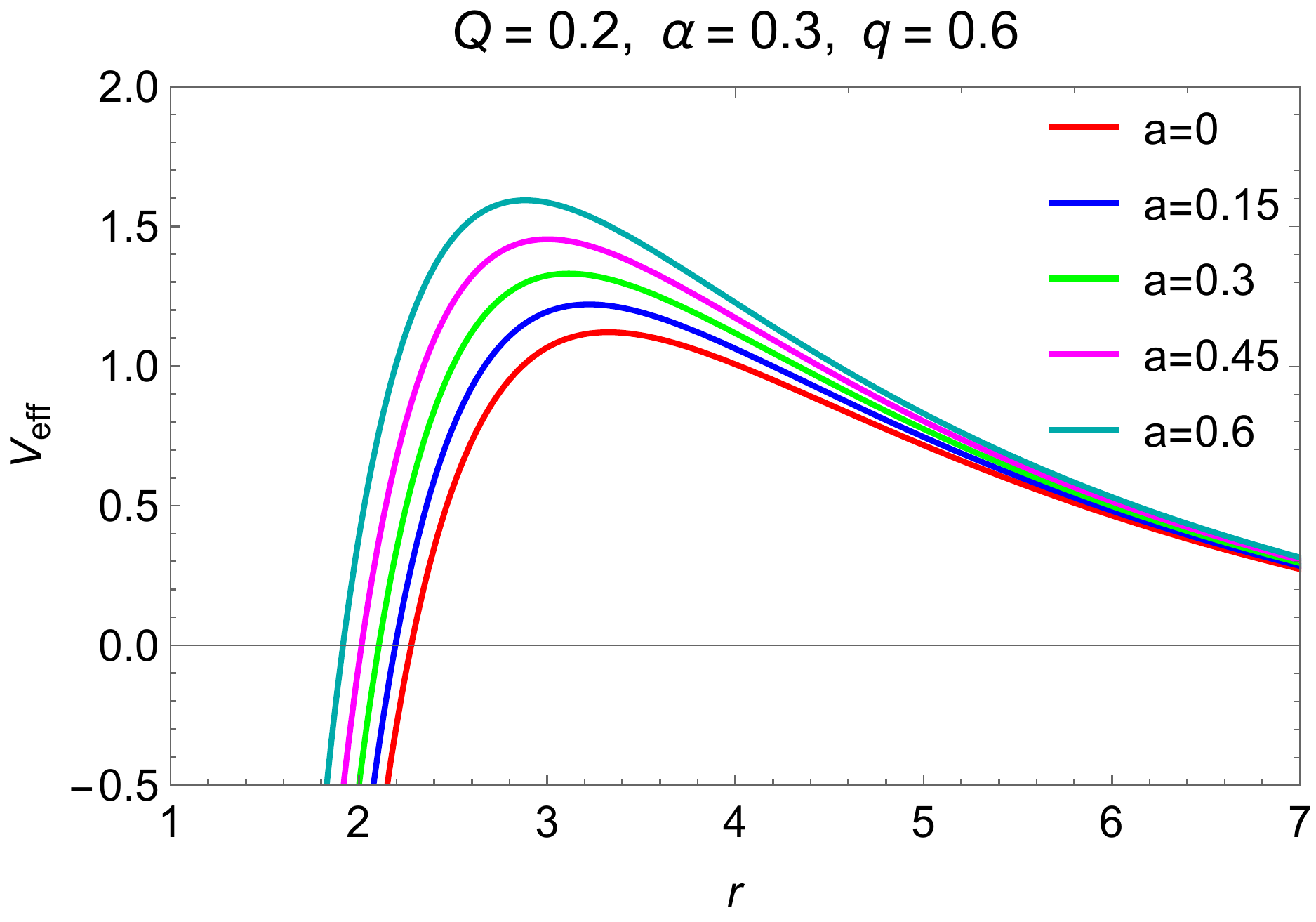}}
~~~
\subfigure{
\includegraphics[height=5cm,width=7.14cm]{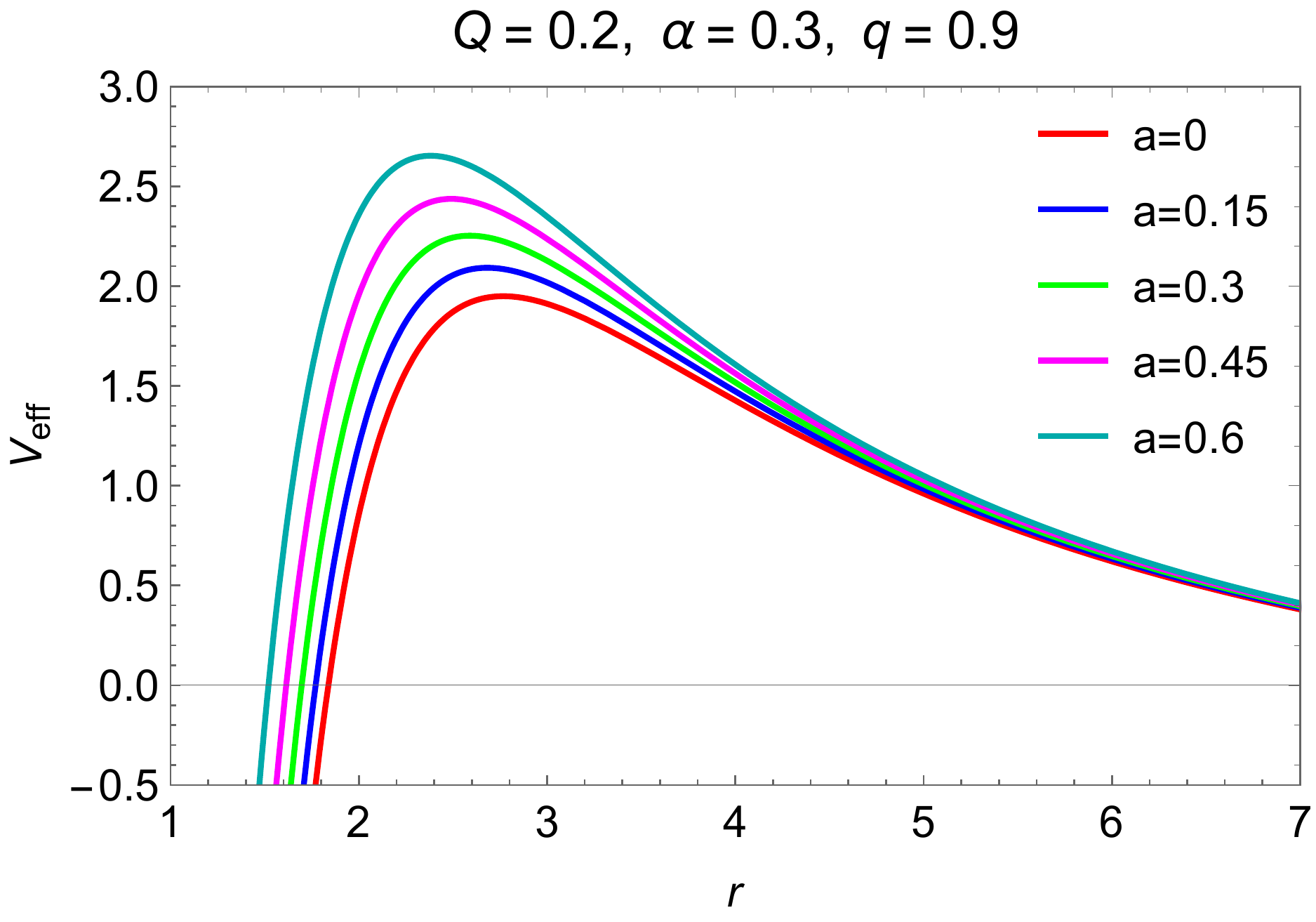}}
\subfigure{
\includegraphics[height=5cm,width=6.92cm]{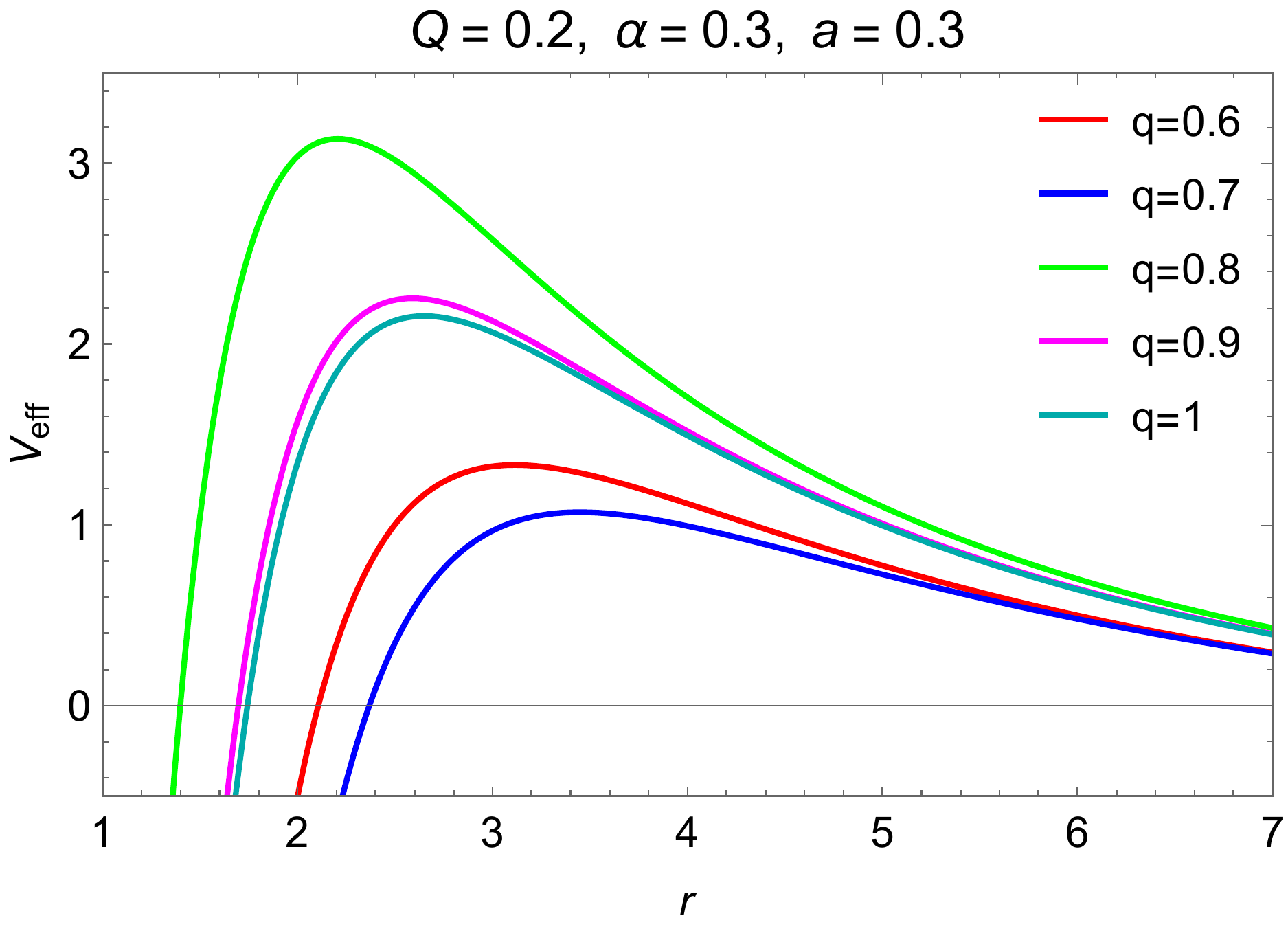}}
\end{center}
\caption{Plots for the effective potential w.r.t $r$ for different values of $q$, $Q$ and $\alpha$.} \label{effpotfig}
\end{figure}
The effective potential has been plotted w.r.t $r$ in Fig. \ref{effpotfig} for various parametric values of $\alpha$, $q$ and $Q$. The curves in the top panel correspond to the different values of $Q$ and fixed $a$, $q$ and $\alpha$. In the left plot, it can be seen that the peak of the curves drops and shifts away from the horizon with increasing $Q$. This suggests that as we increase the value of $Q$, the unstable circular orbits get enlarged and are attained at lower values of the potential. Whereas an exactly converse behaviour is seen in the right plot. In this plot, the unstable circular orbits get shifted towards the BH and are attained at higher potentials with increasing values of $Q$. The curves in the second panel correspond to different values of $\alpha$ with fixed $a$, $q$ and $Q$. For both plots, as we increase $\alpha$, the unstable circular orbits are attained at higher values and are reduced in size. However, the minimum peak in the right plot is above $V_{eff}=2$, whereas the peaks in the left plot approach $V_{eff}=2$. Thus, separating the curves for $q<\frac{3}{4}$ and $q>\frac{3}{4}$. The behaviour of effective potential in the third panel is very much similar to that in the second panel. The unstable circular orbits in both plots get closer to the horizon w.r.t increase in $a$ and are attained at higher values of the spin. Again the separation of curves is evident for $q<\frac{3}{4}$ and $q>\frac{3}{4}$. The plot in the bottom panel is for the different values of $q$ and fixed $a$, $Q$ and $\alpha$. By increasing $q$ in the interval $(0.5,0.75)$ and $(0.75,\infty)$, the peaks in the effective potential curves drop and hence causing the unstable circular orbits to become smaller. However, due to the singular nature of the metric function at $q=0.75$, the curves in both intervals are separated.

\begin{figure}[ht!]
\begin{center}
\subfigure{
\includegraphics[height=5.2cm,width=6.24cm]{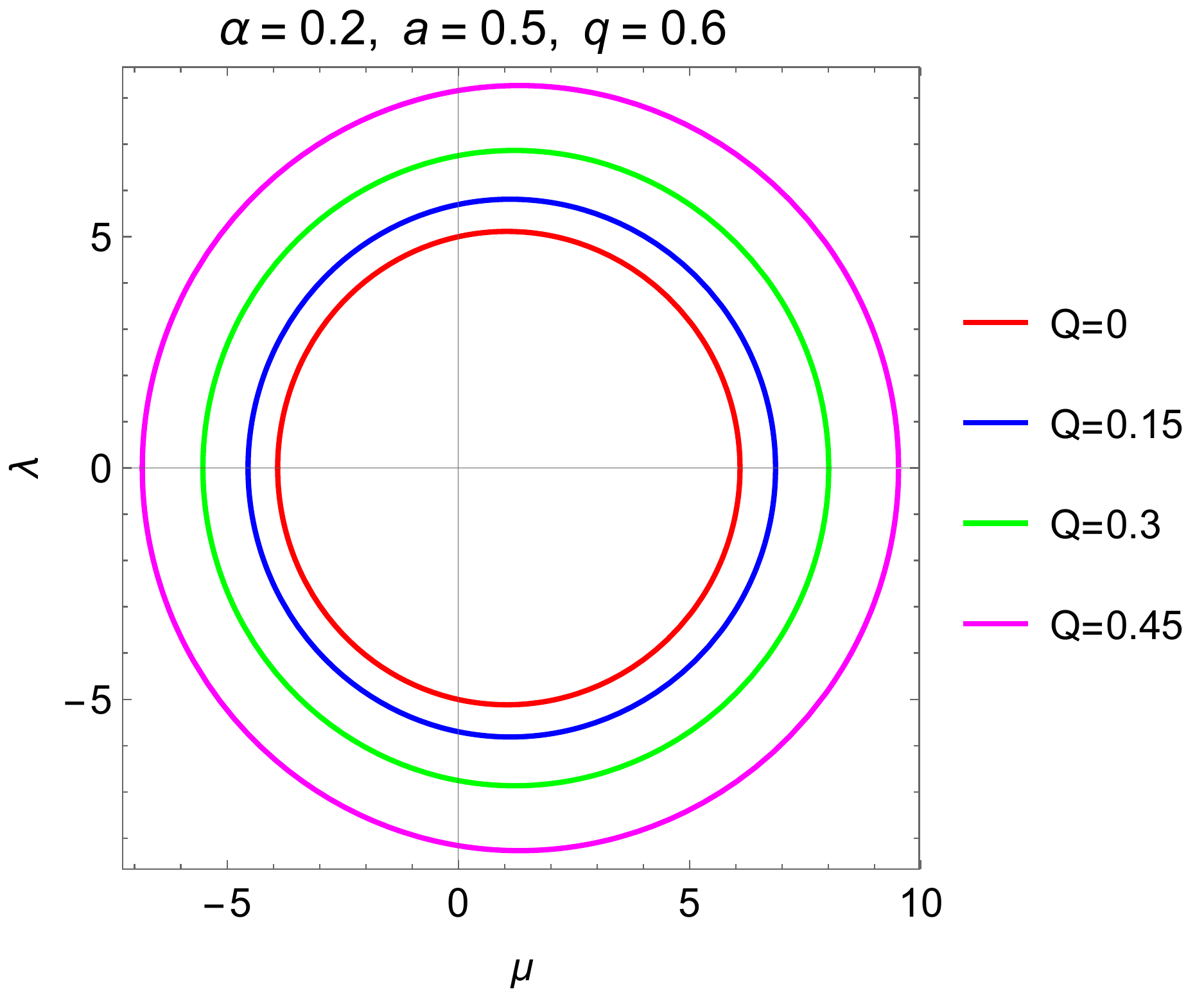}}
~~~~
\subfigure{
\includegraphics[height=5.2cm,width=6.24cm]{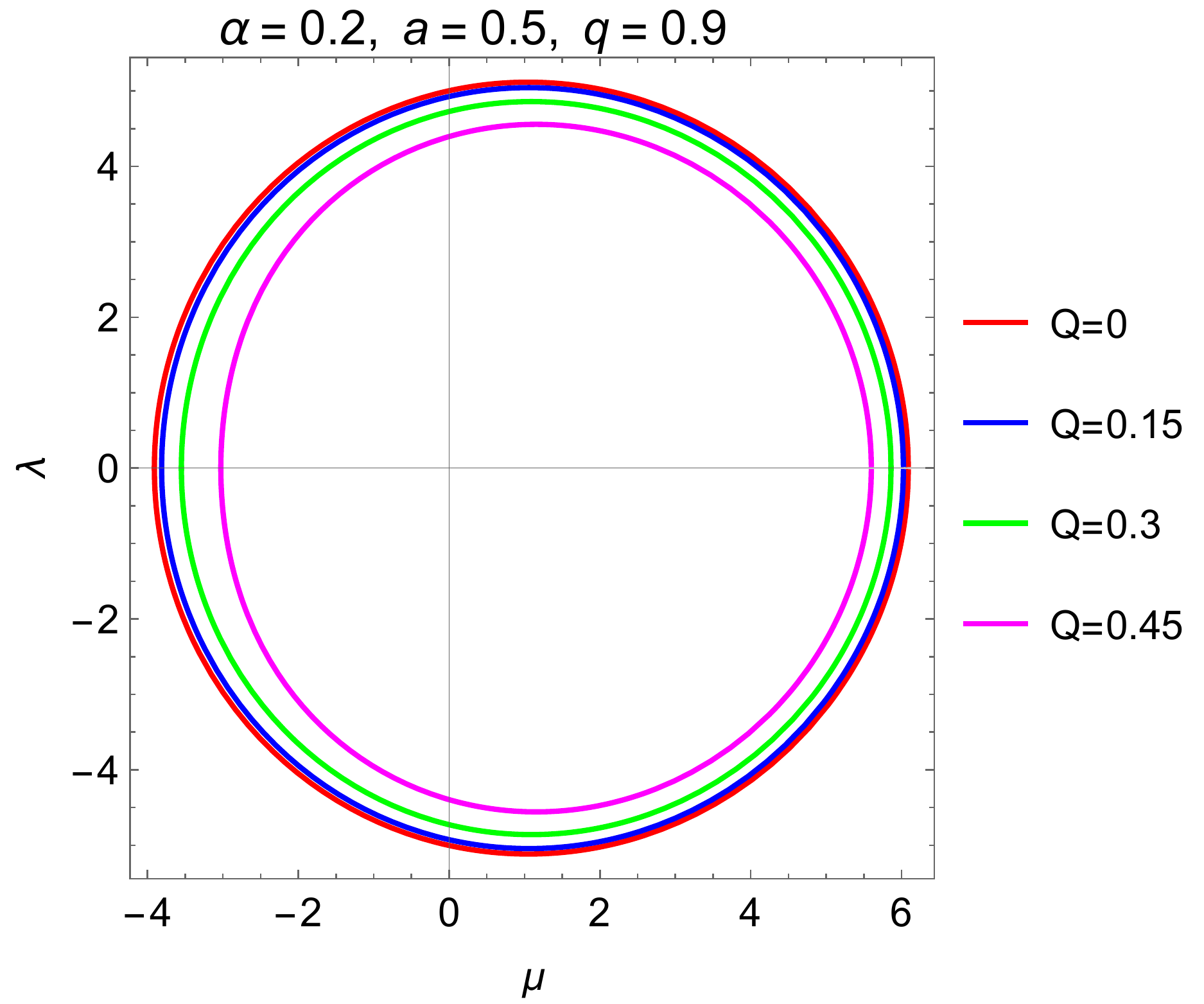}}
\subfigure{
\includegraphics[height=5.2cm,width=6.24cm]{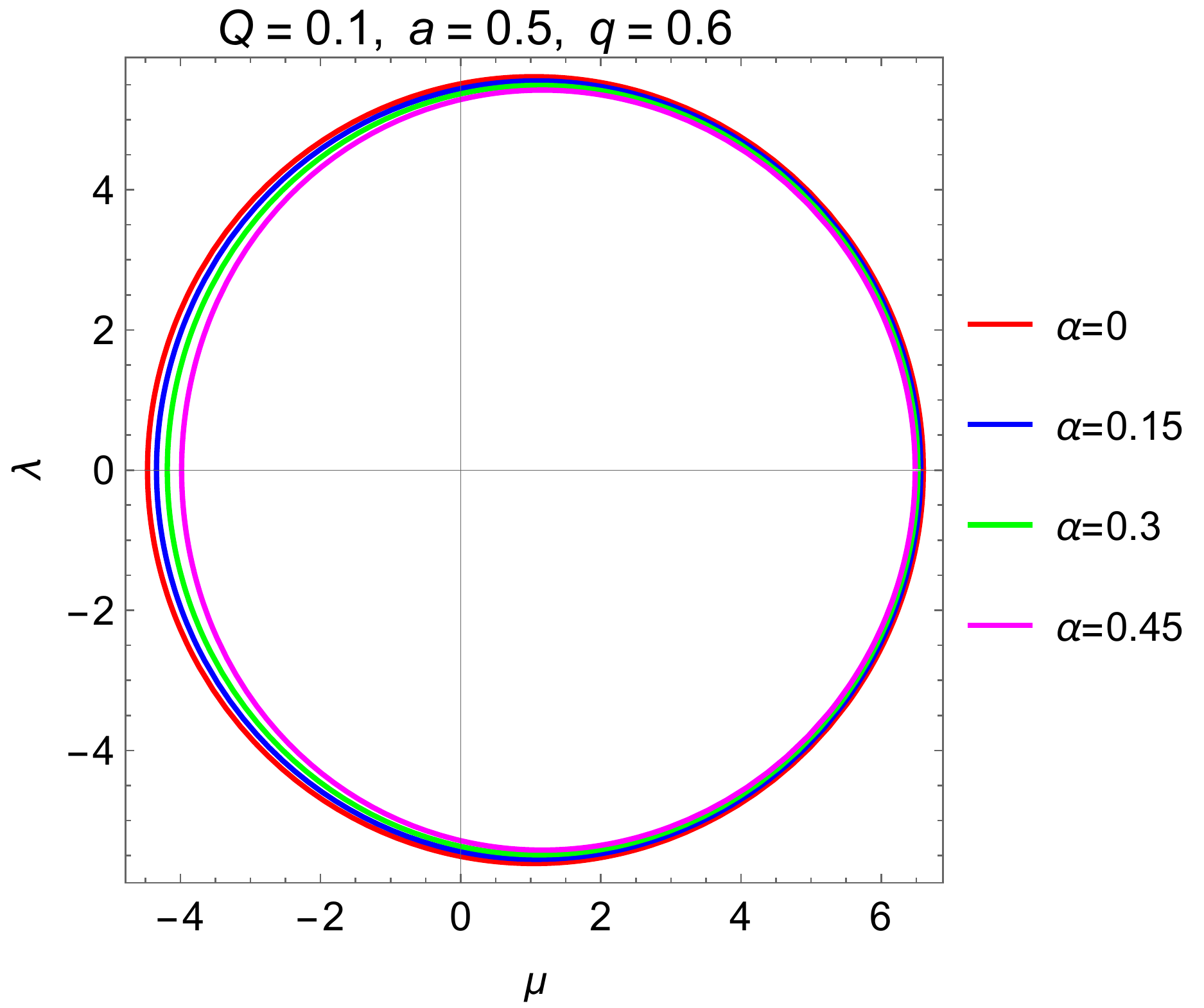}}
~~~~
\subfigure{
\includegraphics[height=5.2cm,width=6.24cm]{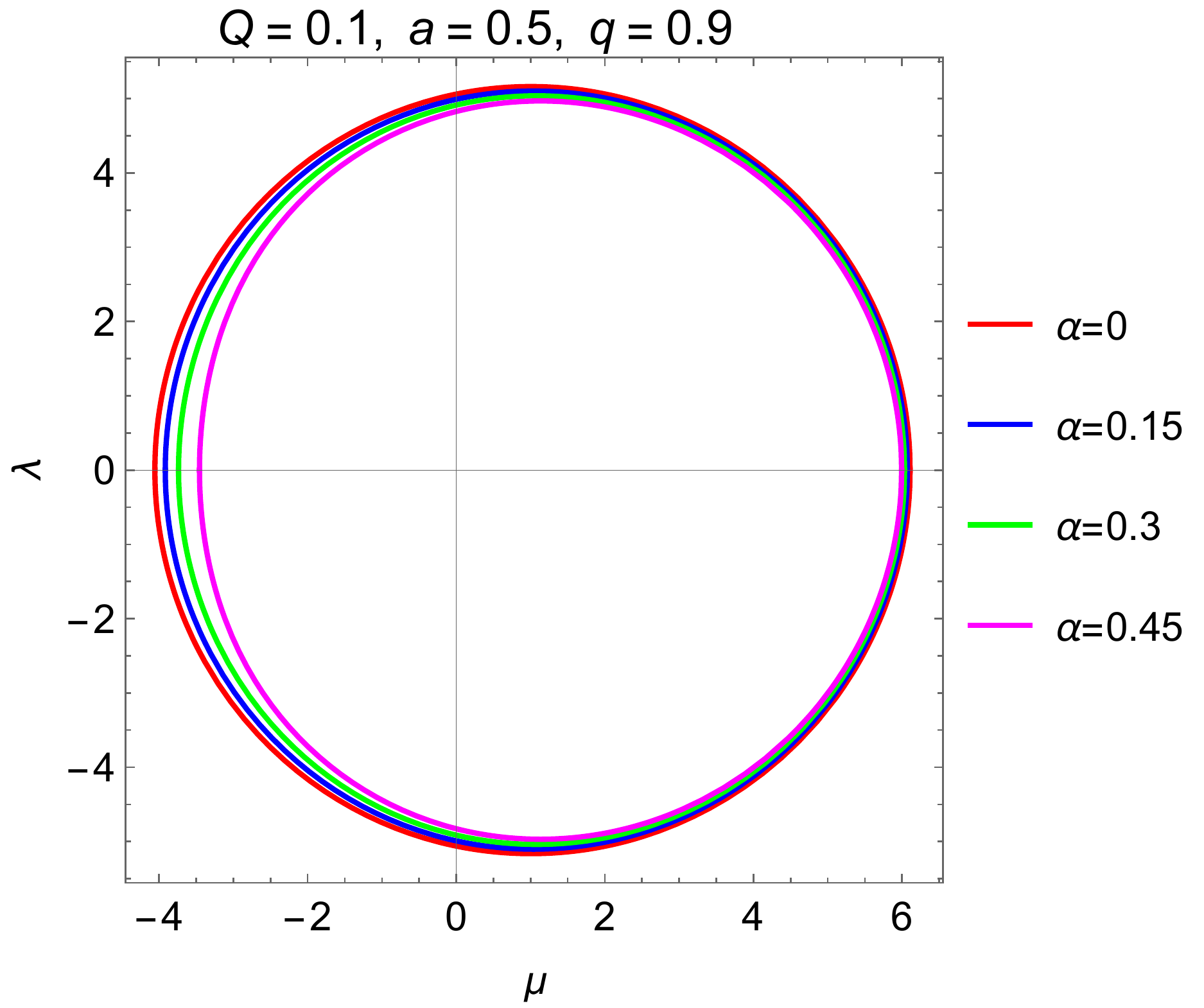}}
\subfigure{
\includegraphics[height=5.2cm,width=6.47cm]{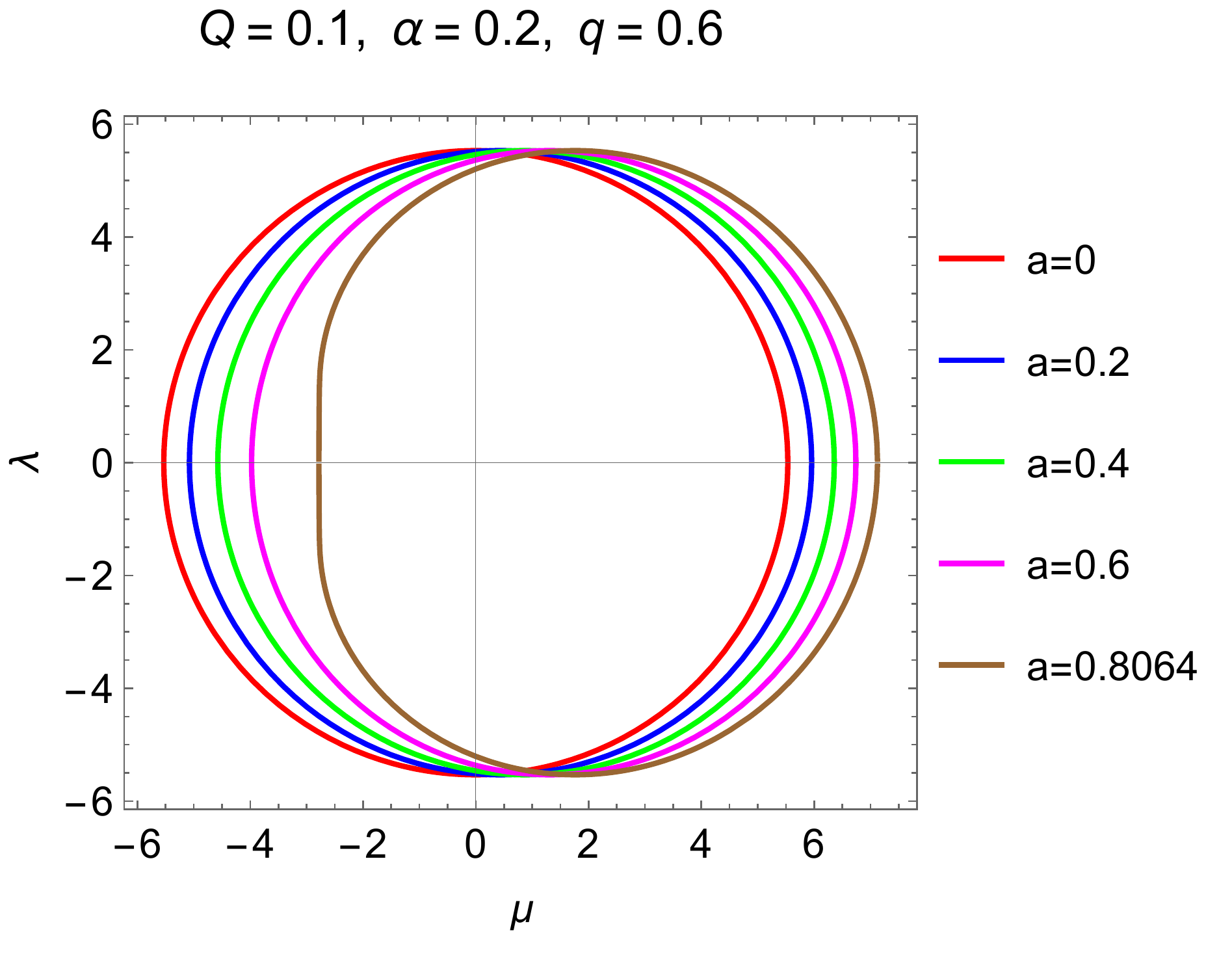}}
~~~~
\subfigure{
\includegraphics[height=5.2cm,width=6.47cm]{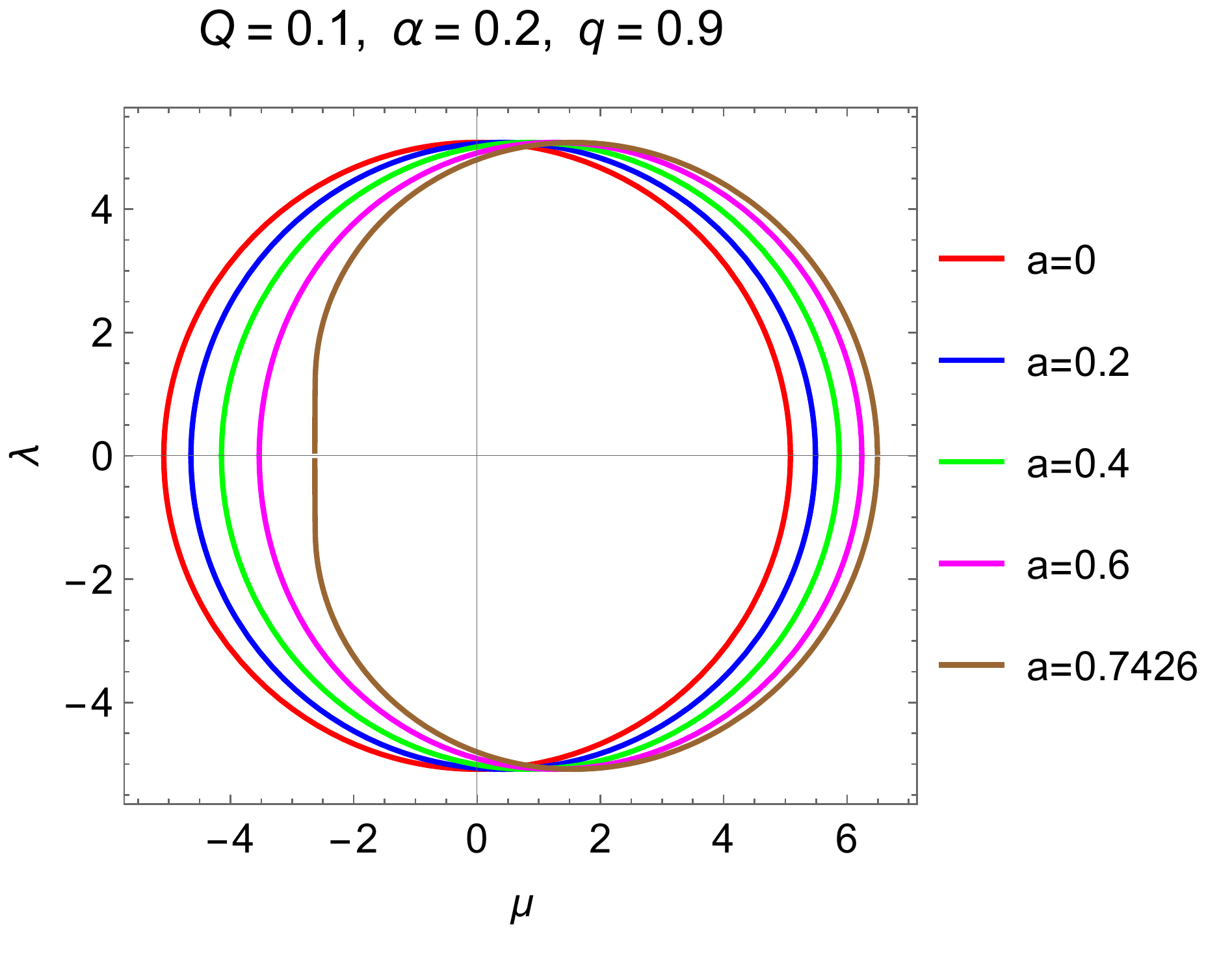}}
\subfigure{
\includegraphics[height=5.2cm,width=6.05cm]{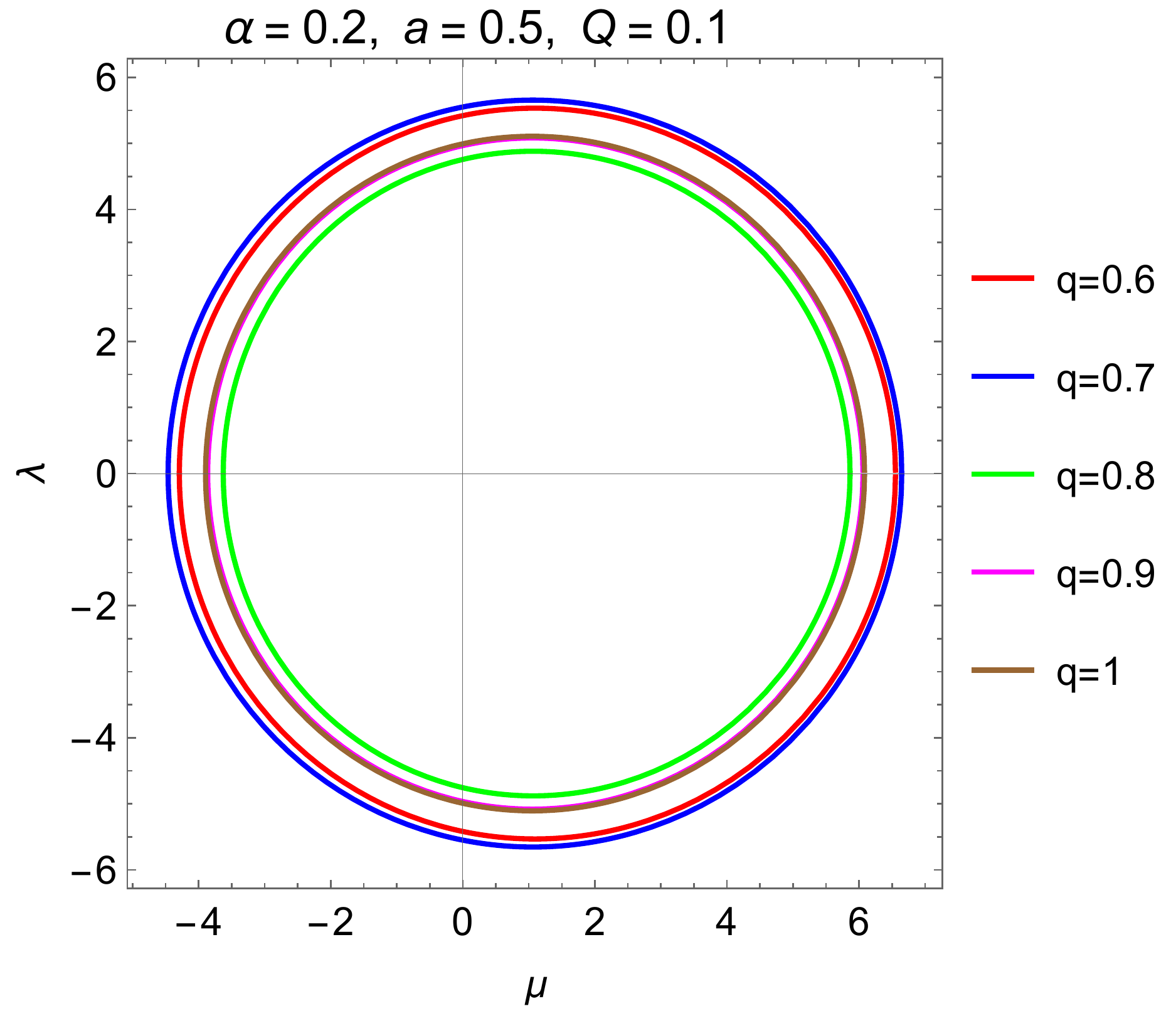}}
\end{center}
\caption{Plots for the shadows w.r.t different values of $a$, $q$, $Q$ and $\alpha$.} \label{shadowfig}
\end{figure}

The photons move around the BH in circular orbits, out of which the unstable circular orbits are of prime importance. Thus, the condition $r=constant$ must be satisfied by photons, i.e. they must reside on the surface of a sphere. The conditions for such orbits are the equations $\dot{r}=0=\ddot{r}$ or equivalently $\mathcal{R}(r)=0=\mathcal{R}'(r)$, where prime represents the derivative w.r.t $r$. The conditions for unstable circular orbits can also be written as $V_{eff}(r)=0$ and $V_{eff}'(r)=0$. Then by using these conditions, we get
\begin{eqnarray}
K_E=\frac{16r^2\Delta(r)}{(\Delta'(r))^2}, \qquad L_E=\frac{a^2+r^2}{a}-\frac{4r\Delta(r)}{a\Delta'(r)}, \label{55}
\end{eqnarray}
where the symbols $L_E=l/E$ and $K_E=\big(\mathcal{Z}+(l-aE)^2\big)/E^2$ are useful in calculating the shadows for the static BHs. In terms of the modified Carter constant, the above symbols reduce to $\xi=l/E$ and $\eta=\mathcal{Z}/E^2$, known as Bardeen's impact parameters. Then, for unstable orbits, the impact parameters can be written as
\begin{eqnarray}
\xi&=&\frac{r^2+a^2}{a}-\frac{4r\Delta(r)}{a\Delta'(r)}, \label{56} \\
\eta&=&\frac{r^2\big(8r\Delta(r)\Delta'(r)-16\Delta(r)\big(\Delta(r)-a^2\big)-r^2(\Delta'(r))^2\big)}{a^2(\Delta'(r))^2}. \label{57}
\end{eqnarray}
For a rotating BH, the shell of the photon sphere does not have a unit width, instead, it forms a thick shell with a variable radius $r_p$. The null rays spiral asymptotically towards this photon region. In such a case, the shadows are governed by the parametric equations with parameter $r_p\in\big[r_{p,min},r_{p,max}\big]$. The boundary points of this interval are the extremal values of photon region's radius. These extreme values of $r_p$ are the real positive roots of $\eta=0$. For a non-rotating BH, the photon sphere is a shell with unit width and hence $r_p$ does not behave as a parameter. We obtain a unique value of $K_E(r_p)$ by using the constant $r_p$. However, the value of $L_E(r_p)$ cannot be determined because taking $a=0$ in Eq. (\ref{55}) gives an indeterminate value. Hence, $L_E(r_p)$ remains undetermined and can be used as the parameter whose values exist in the interval with endpoints obtained by solving $\Theta(\theta_0)=0$. Since the shadow is a $2$D image, therefore, it is plotted by projecting it on a celestial plane with the help of two coordinates $\mu$ and $\lambda$ defined as
\begin{eqnarray}
\mu&=&-\lim\limits_{r\rightarrow\infty}\bigg[\frac{d\phi}{dr}r^2\sin\theta\bigg]_{\theta\rightarrow\theta_0}, \label{58} \\
\lambda&=&\lim\limits_{r\rightarrow\infty}\bigg[\frac{d\theta}{dr}r^2\bigg]_{\theta\rightarrow\theta_0}, \label{59}
\end{eqnarray}
where $\theta_0$ denotes the observer's angle of inclination. Using $\theta$ and $\phi$ equations of motion and further solving the limits, the celestial parameters become
\begin{eqnarray}
\mu&=&-\xi\csc\theta_0, \label{60} \\
\lambda&=&\pm\sqrt{\eta+a^2\cos^2\theta_0-\xi^2\cot^2\theta_0}. \label{61}
\end{eqnarray}
Since we are working for an equatorial observer ($\theta_0=\frac{\pi}{2}$). Thus, the celestial parameters reduce to
\begin{eqnarray}
\mu&=&-\xi, \label{62} \\
\lambda&=&\pm\sqrt{\eta}. \label{63}
\end{eqnarray}
The shadows observed at the equatorial region by an observer at infinity are shown in Fig. \ref{shadowfig}. The plots in the top panel correspond to the fixed values of $q$, $a$ and $\alpha$ with the curves corresponding to the different values of $Q$. With an increase in $Q$, the shadow size increases in the left plot and decreases in the right plot. This converse behaviour of the shadow in the panel is due to the value of $q$ lying in the intervals $(0.5,0.75)$ and $(0.75,\infty)$. It is also obvious from the effective potential plots because the unstable circular orbits define the photon sphere and thus the boundary of the shadow loop. In the second panel, the shadow curves correspond to the different values of $\alpha$ with fixed $Q$, $q$ and $a$. The effect of $\alpha$ and $q$ on the shadow size and variation is almost the same in both plots. The shadow size becomes smaller w.r.t $\alpha$ but no significant variation is seen w.r.t $q$ in both plots apart from a slight reduction in shadow size in the right plot. The flatness on either side is also increased w.r.t $\alpha$. The shadow curves in the third panel refer to the static and rotating metrics for $a=0$ and $a\neq0$, respectively. The shadow curves for the static case in both plots are pure circles whereas with the increase in $a$, the curves are shifted towards the right. For $a=0.8064$ and $a=0.7426$ in the left and right plots, respectively, the shadow loops are corresponding to the extremal BHs with clear flatness similar to that for the extremal Kerr BH shadow. The size of the shadow in the right plot is slightly smaller than that in the left plot due to the effect of $q$. The plot in the bottom panel corresponds to the fixed values of $a$, $\alpha$ and $Q$ with curves corresponding to distinct values of $q$. We can see that with an increase in $q$ in the intervals $(0.5,0.75)$ and $(0.75,\infty)$, the size of the shadows increases. However, the curves corresponding to both intervals are separated significantly as seen in the plot. 

\subsection{Distortion}
The distortion is a measure of the quantity of the difference in size and flatness of the shadow as compared to that of a non-rotating BH. However, the distortion is measured in terms of an observable known as the linear radius of the shadow \cite{2016PhRvD..94b4054A,2009PhRvD..80b4042H}, defined as
\begin{equation}
R_{sh}=\frac{(\mu_t-\mu_r)^2+\lambda_t^2}{2|\mu_t-\mu_r|}. \label{64}
\end{equation}
The linear radius is the radius of an assumptive circle that coincides with the shadow loop at the coordinates $(\mu_t,\lambda_t)$, $(\mu_b,\lambda_b)$ and $(\mu_r,0)$ corresponding to the top, the bottom and the right most point on the shadow curve, respectively. Any arbitrary point on the shadow curve has the coordinates ($\mu,\lambda$) where the subscripts $t$, $b$ and $r$ correspond to the top, the bottom and the rightmost point, respectively. For the visual arrangement of the points on the shadow curve, refer to Fig. $\textbf{9}$ in \cite{2016PhRvD..94b4054A}. The relation (\ref{64}) corresponds to only rotating BHs because, for the static BHs, the hypothetical circle will also coincide with the fourth point with coordinates $(-\mu_r,0)$. Hence, the obtained curve is a circle at the origin. In this way, the shadow and the imaginary circle will have the same radius. We use the value of $R_{sh}$ in Eq. (\ref{64}) in order to calculate the distortion. The distortion can be obtained by the relation
\begin{equation}
\delta_s=\frac{|\bar{\mu}_l-\mu_l|}{R_{sh}}, \label{65}
\end{equation}
where $(\mu_l,0)$ and $(\bar{\mu}_l,0)$ are the points on the shadow curve and imaginary circle, respectively, intersecting the $-\mu$-axis. The shadow curve on the negative axis is represented by the subscript $l$ in equations, while, $~\bar{}~$ denotes the points on the hypothetical circle.
\begin{figure}[ht!]
\begin{center}
\subfigure{
\includegraphics[height=5cm,width=7.2cm]{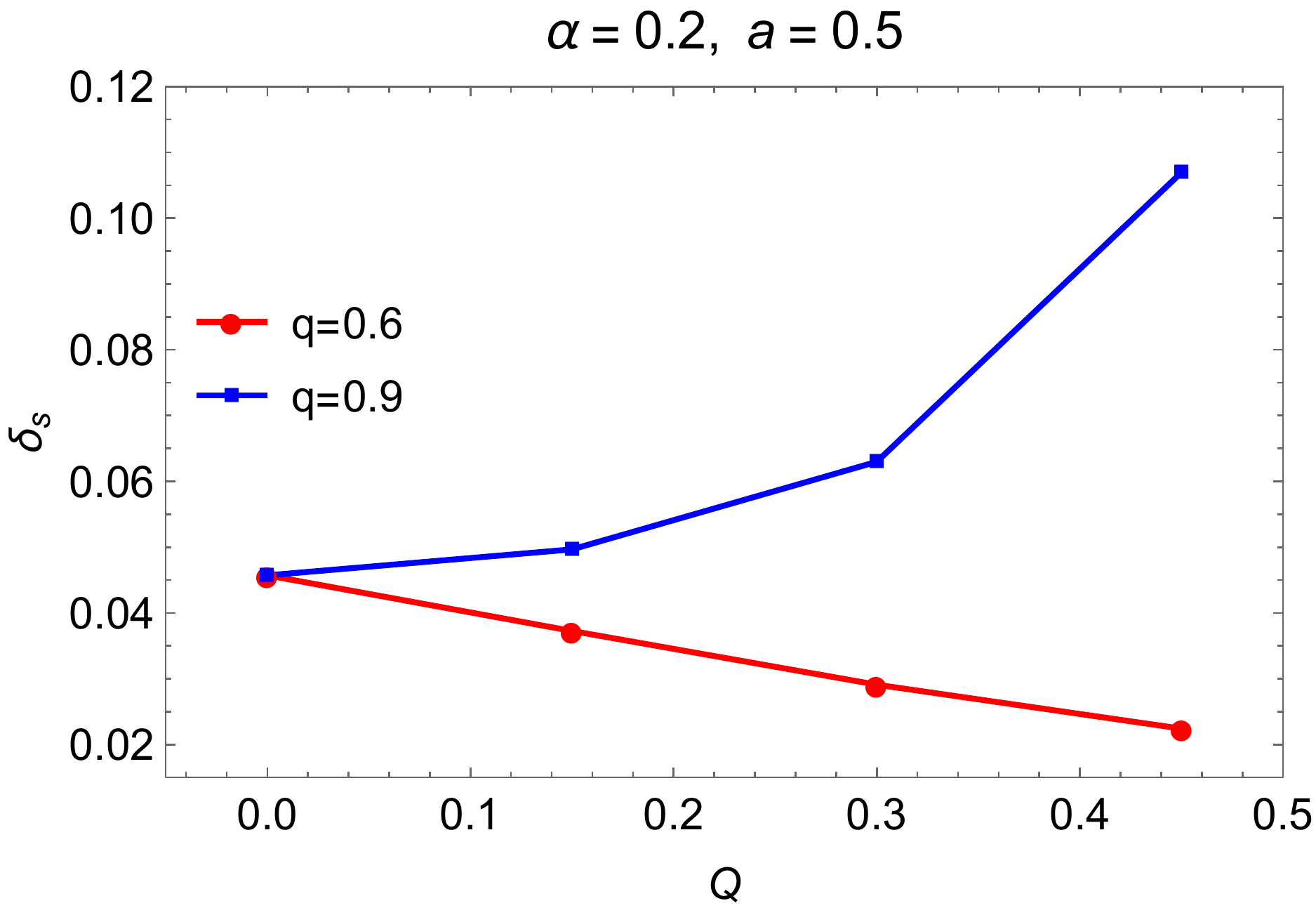}}
~~~~
\subfigure{
\includegraphics[height=5cm,width=7.4cm]{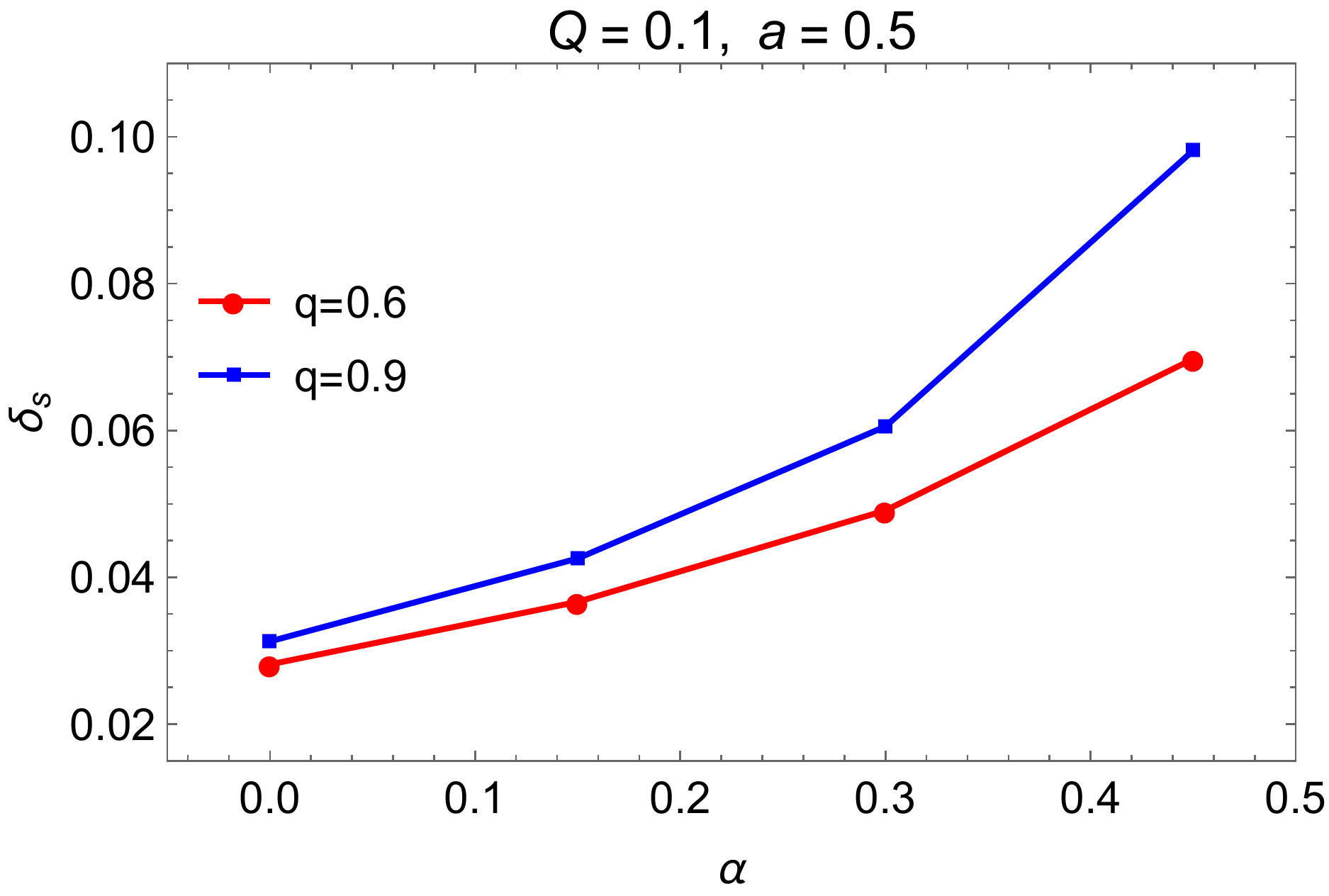}}
\subfigure{
\includegraphics[height=5cm,width=7.1cm]{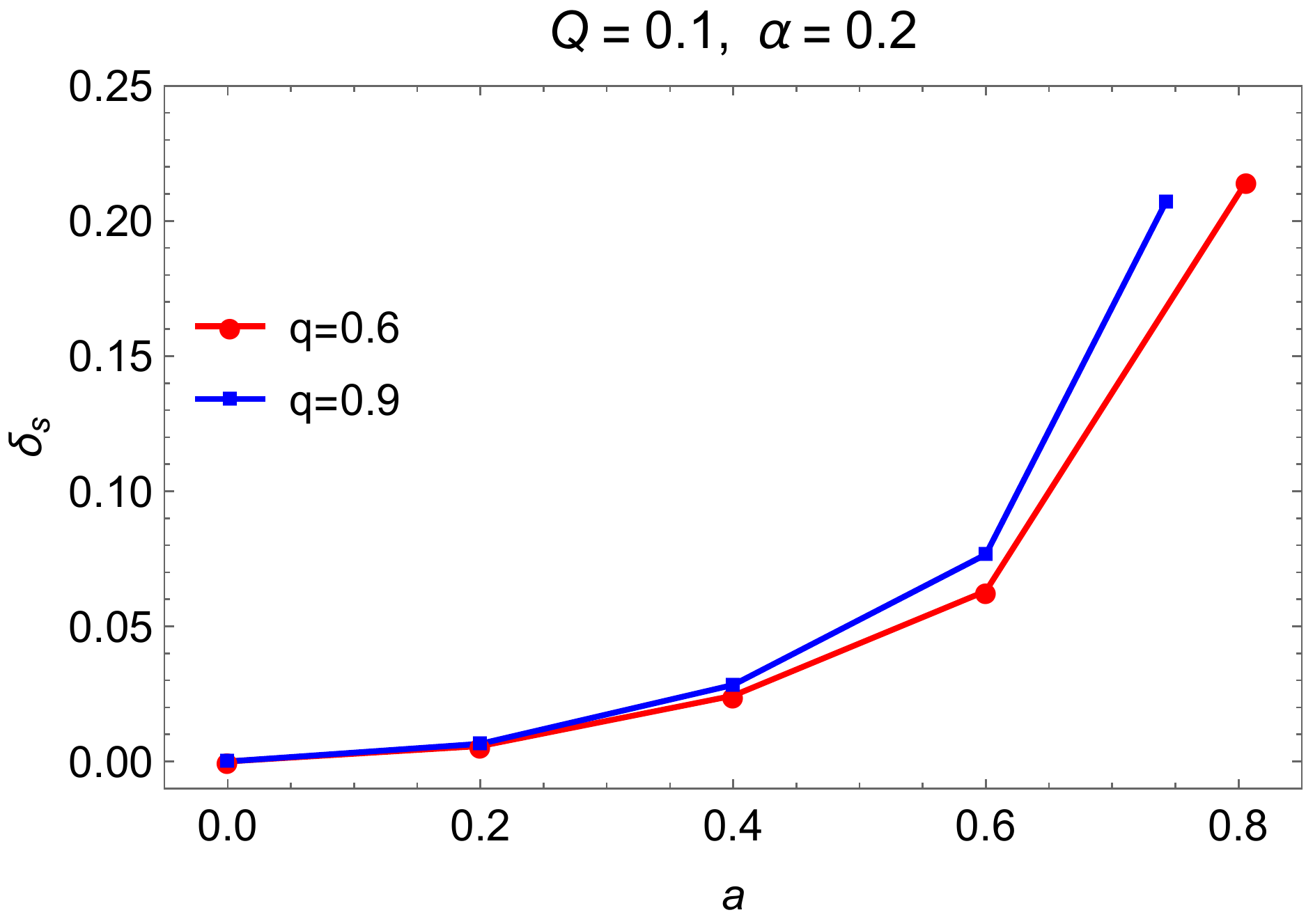}}
~~~~
\subfigure{
\includegraphics[height=5cm,width=7.2cm]{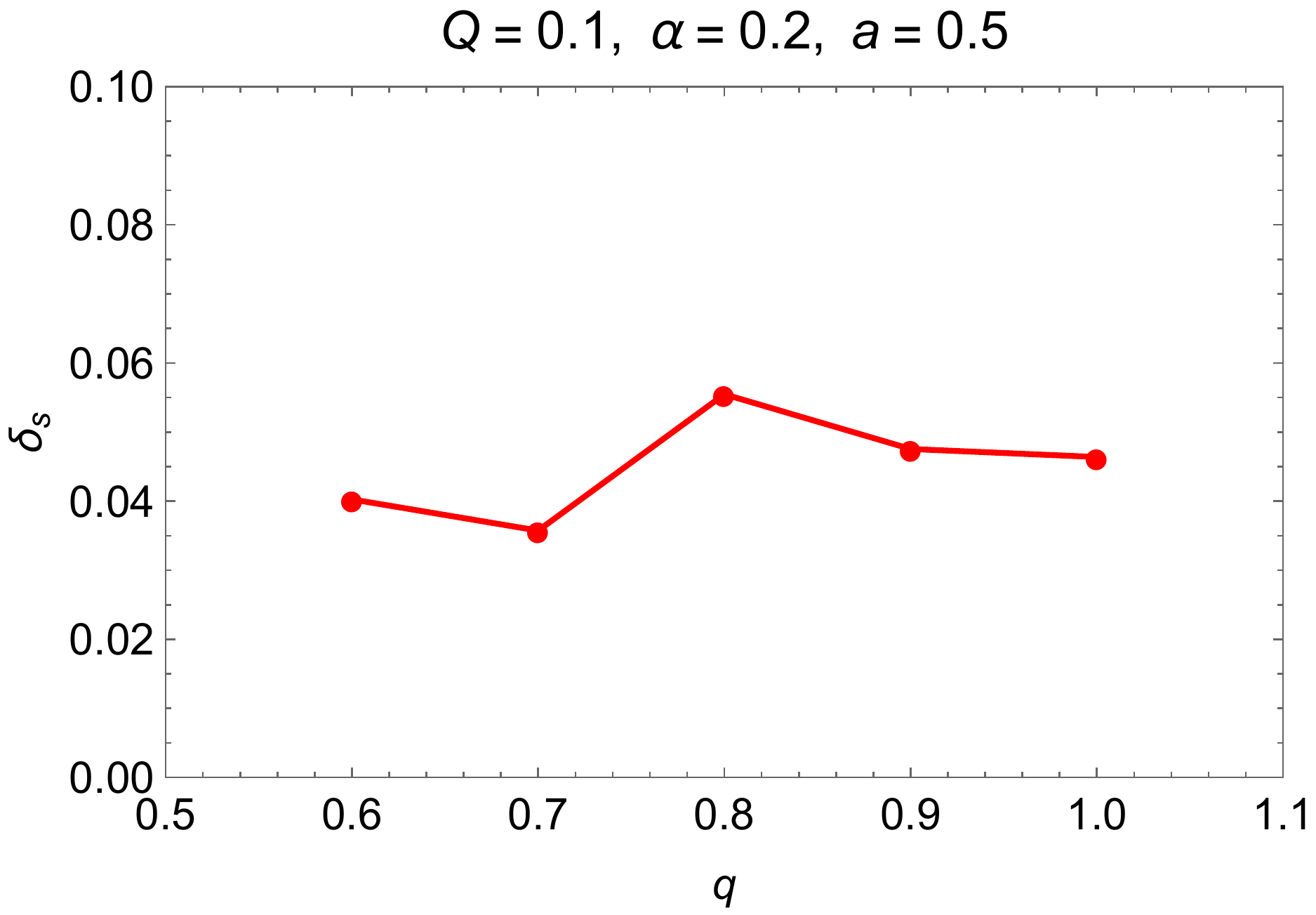}}
\end{center}
\caption{Plots for the distortion in shadows w.r.t different values of $a$, $q$, $Q$ and $\alpha$.} \label{distrfig}
\end{figure}
The distortion is plotted in Fig. \ref{distrfig}. The left plot in the upper panel corresponds to the distortion $\delta_s$ vs $Q$ with curves for different $q$ whereas $\alpha$ and $a$ have been kept fixed. Note that the red and blue curves correspond to the left and right shadow plots in the upper panel, respectively. It shows that the distortion drops almost linearly when $q=0.6$ and for $q=0.9$, the distortion increases rapidly w.r.t increase in $Q$. The right plot in the upper panel shows the behaviour of distortion w.r.t $\alpha$ for fixed values of $Q$ and $a$. The red and blue curves correspond to the left and right shadow plots, respectively, in the second panel. The distortion for the case of $q=0.6$ increases almost linearly w.r.t $\alpha$. Whereas for $q=0.9$, the distortion is higher than the case of $q=0.6$ and increases more rapidly as $\alpha$ grows larger. The left plot in the lower panel shows the behaviour of distortion w.r.t $a$ with fixed $Q$ and $\alpha$. The red and blue curves correspond to the left and right shadow plots, respectively, in the third panel. The distortion for the case of $q=0.6$ and $q=0.9$ behave very much similar to each other w.r.t $a$. However, for $q=0.9$, the distortion is slightly higher than that for the case of $q=0.6$. These plots are up to the extremal values of $a$, therefore, drastic increases are observed in both curves due to the flatness in the shadow curves. The right plot in the lower panel shows the distortion w.r.t $q$ with fixed $a$, $\alpha$ and $Q$. It corresponds to the shadow curve in the bottom panel. We can see that for $q<\frac{3}{4}$ and $q>\frac{3}{4}$, the distortion decreases with increase in $q$. However, the distortion is seen to be higher for $q>\frac{3}{4}$ as compared to that for $q<\frac{3}{4}$.

\section{Comparison with EHT Data}
In this section, we focus on obtaining the constraints of the supermassive BHs Sgr. A* and M$87$* based on recent observational data reported by the EHT collaboration. First of all, we define the observable parameters that we use to describe the size and shape of the BH shadow. It is well-known that the shape of the shadow of a spherically symmetric BH is a perfect circle, but for axially symmetric BHs there are some distortions in the BH image. As said before, for the first time, Hioki $\&$ Maedia have introduced two observational parameters of the BH shadow: the shadow radius that describes its size, and the deviation parameter to describe the distortion from the circle \cite{2009PhRvD..80b4042H}. However, there have been other approaches, and we can use a simple method for estimations of the BH parameters using the shadow observables: the shadow area $A$, and the oblateness $D$, based on the coordinate-independent formalism \cite{2015MNRAS.454.2423A,2020ApJ...892...78K}. One may introduce the area enclosed within the shadow silhouette and the shadow oblateness observable using the following integral \cite{2020ApJ...892...78K,2022EPJC...82..771S}
\begin{eqnarray}
A&=&2\int_{r_-}^{r_+}\bigg(\beta(r)\frac{d\alpha(r)}{dr}\bigg)dr, \label{eq:Area} \\
D&=&\frac{\Delta\alpha}{\Delta\beta}, \label{eq:oblat}
\end{eqnarray}  
where $r_\pm$ refers to the radii of the prograde and retrograde stable circular orbits which are obtained as the non-imaginary (real) roots of the equation $\eta_c=0$, located outside the EH \cite{2021GReGr..53...10T}. Using the Eqs. (\ref{eq:Area}) and (\ref{eq:oblat}), one can express the observational parameter of the shadow as functions of the BH parameters by numerical calculations. The point is that these functions can also be used to have information about values of the BH parameters in the framework of various gravitational theories by comparing them with EHT observational data from their shadow.

\subsection{Constraints on parameters from M$87$*}
Now we can determine the constraints for the parameters of the supermassive BH M$87$* considering it as a $4D$ EGB BH in the PYM field, where the angular diameter of its shadow image that is measured by an observer located at a distance $d$ from the BH is as follows \cite{2020JCAP...07..053K,2023ApJ...944..149A}
\begin{eqnarray}
\theta_d =\frac{2R_a}{d}, \qquad R_a^2 = {\frac{A}{\pi}}, \label{angdia}
\end{eqnarray}
where $R_a$ is the areal shadow radius. Taking into account the Eq. (\ref{eq:Area}), one may define the angular diameter of the shadow depending on the BH parameters and the inclination angle relative to the observer. Besides, it implicitly depends on the BH mass. The mass of M$87$* and the distance from the solar system can be considered as $M=6.5\times10^9M_\odot$ and $d=16.8Mpc$, respectively \cite{2019ApJ...875L...6E,2019ApJ...875L...5E}. It is worth to note that, for simplicity, we do not take into account the uncertainties of the mass and distance measurements of the BHs. The angular diameter of the image of the SMBH M$87$* is $\theta_d=42\pm3\mu as$ \cite{2019ApJ...875L...1E} in $1$-$\sigma$ confidential level. One can calculate the angular diameter using Eq. (\ref{angdia}) for the $4D$ EGB BH in the PYM field and estimate the BH parameters by comparing with the angular diameter of the image size of the SMBH M$87$*. 
\begin{figure*}[ht!]\centering
\includegraphics[width=0.35\textwidth]{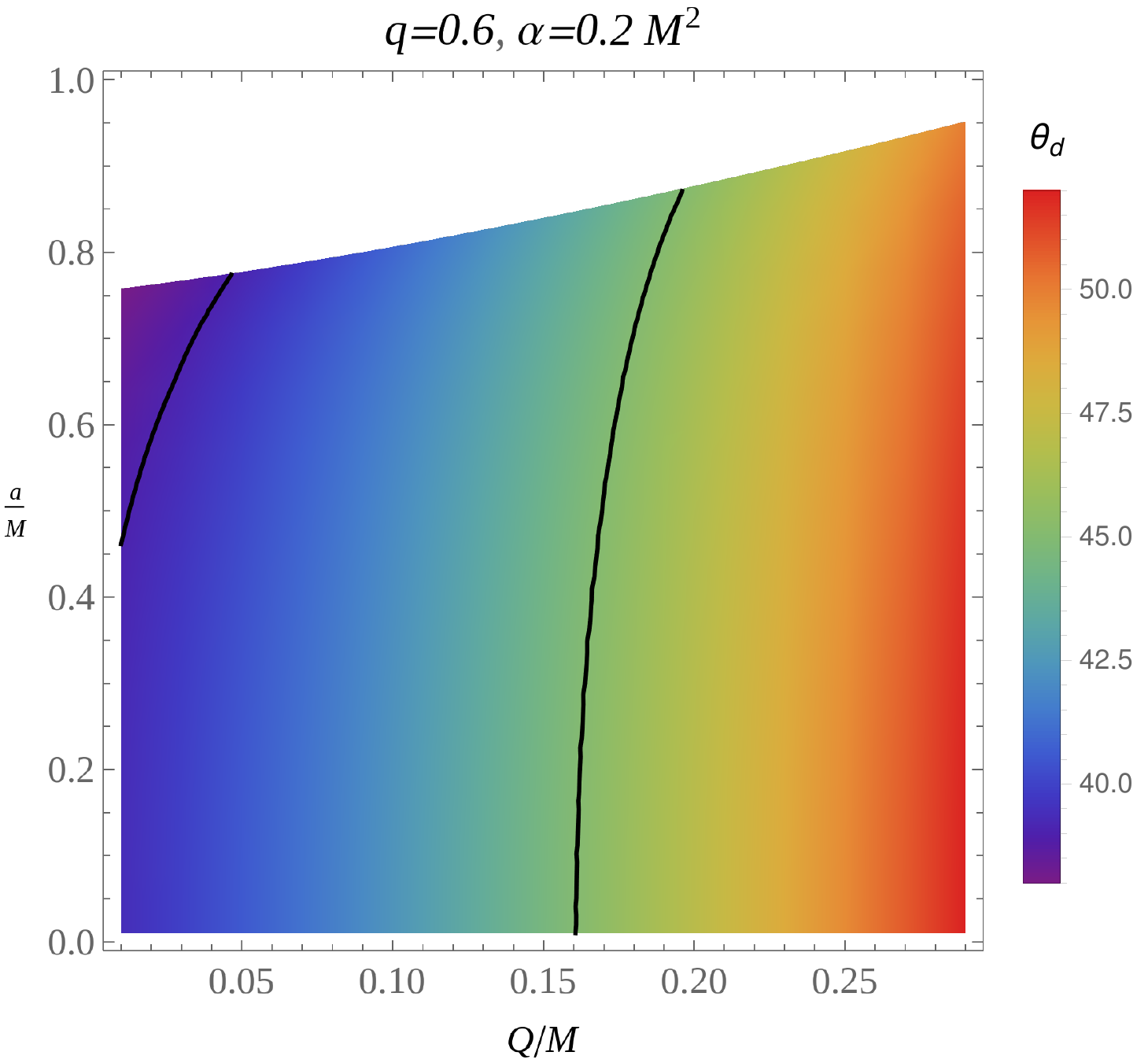}
\includegraphics[width=0.35\textwidth]{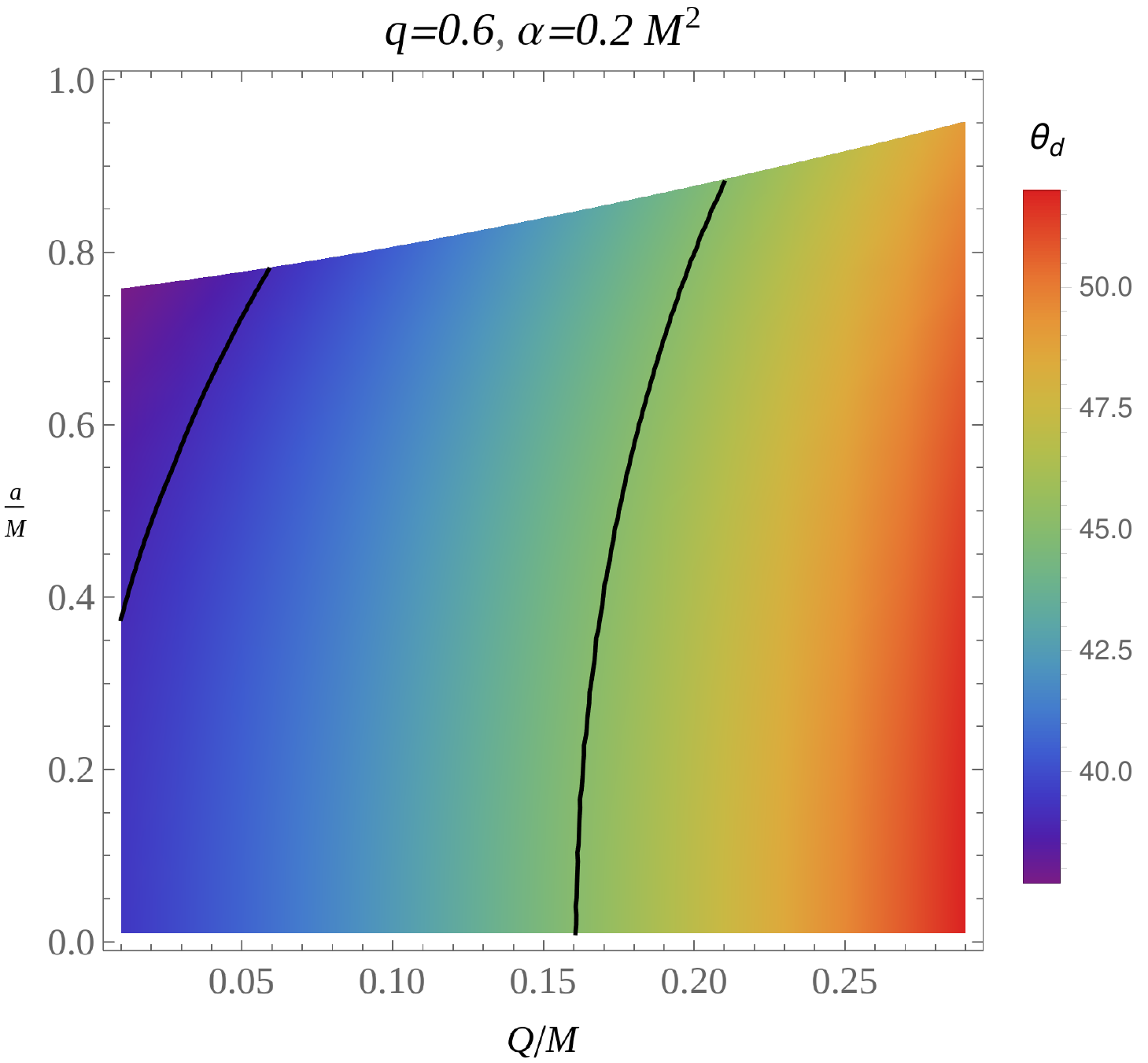}\\
\includegraphics[width=0.35\textwidth]{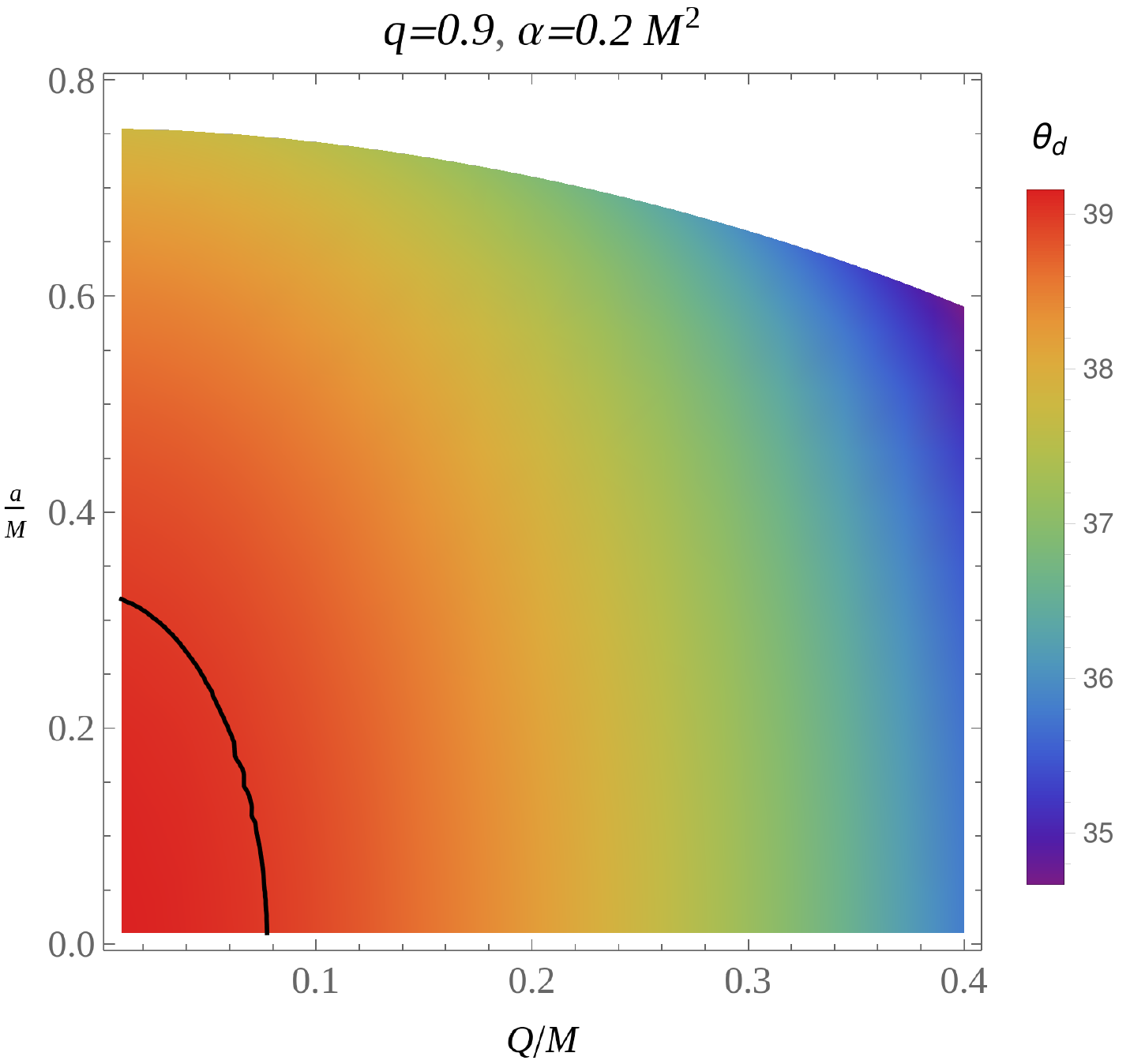}
\includegraphics[width=0.35\textwidth]{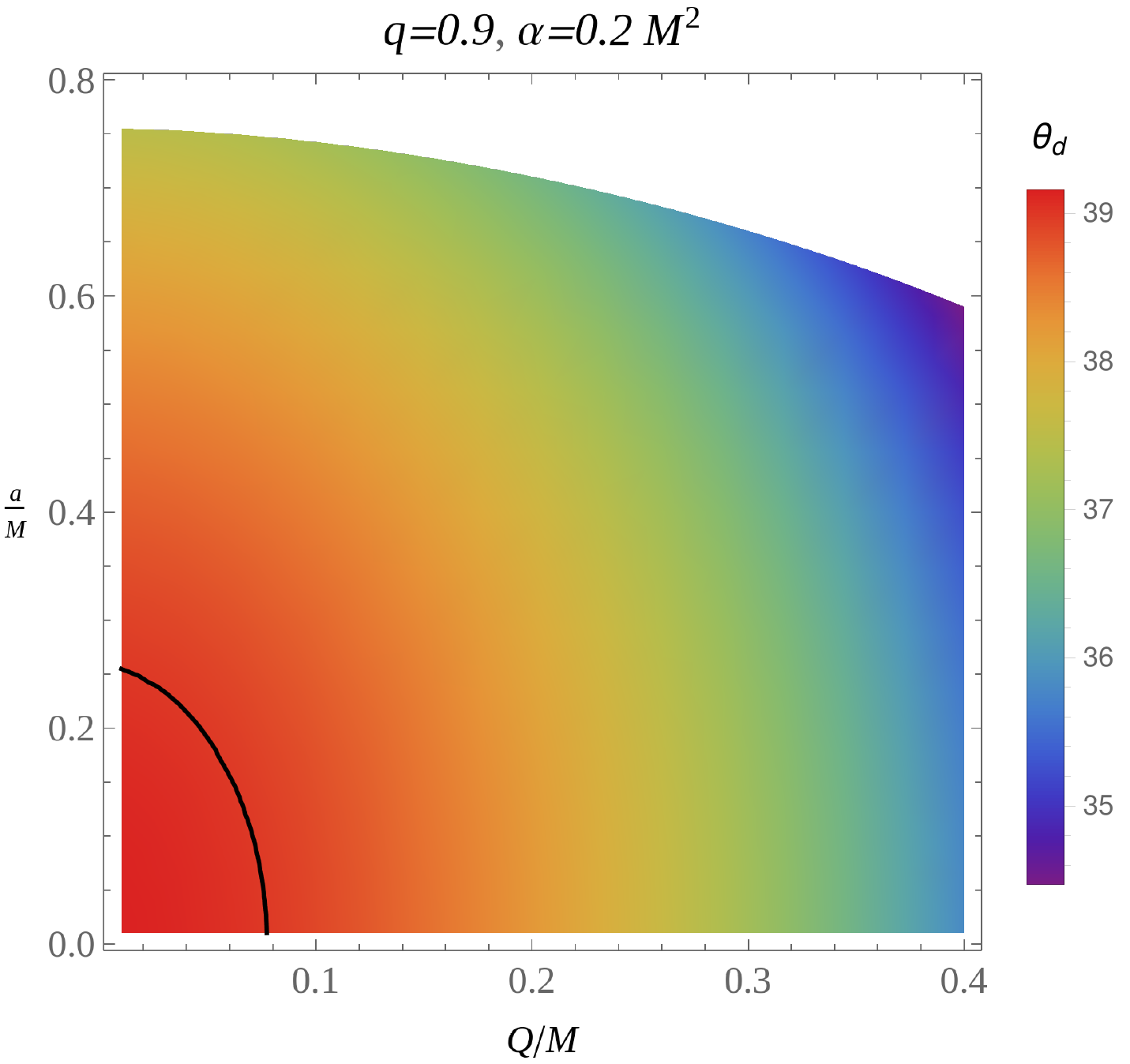}
\caption{Angular diameter observable $\theta_d$ for the BH shadows as a function of parameters $a/M$ and $Q/M$ at inclinations $90^\circ$ (left panels) and $17^\circ$ (right panels). The black curves describe the borders of the image size $\theta_d=42\pm3\mu as$ measured angular diameter of the SMBH M$87$* reported by the EHT.} \label{constraintsM1}
\end{figure*}
Figure \ref{constraintsM1} shows the density plots of the angular diameter $\theta_d$ in the $a$-$Q$ space at $90^\circ$ (left panel), and $17^\circ$ (right panel) inclination angles for the fixed values of $q$, and $\alpha$ parameters. The black curves here correspond to the limits of the measured angular diameter of the image size of the SMBH M$87$*. Thus, they define the regions of the plots fitted with the EHT results in observations of the shadow of the SMBH M$87$*. An interesting feature of these plots is that the constraints in the cases of $q=0.6$ and $q=0.9$ are entirely distinct from each other due to the converse behaviour of the shadow on the dependency of the $q$ parameter. We can have upper and lower limits for the spin and charge values of the BH at $q=0.6$ \& $\alpha=0.2M^2$, however, when $q=0.9$ \& $\alpha=0.2M^2$ we can only have lower limits for the BH spin and charge parameters. 

{Moreover, it is also observed from the figure that the black curves for limits of observable parameters shifts slightly right and the decrease in spin parameter when we change the observational angle from 90$^\circ$ to 17$^\circ$.}
\begin{figure*}[ht!]\centering
\includegraphics[width=0.35\textwidth]{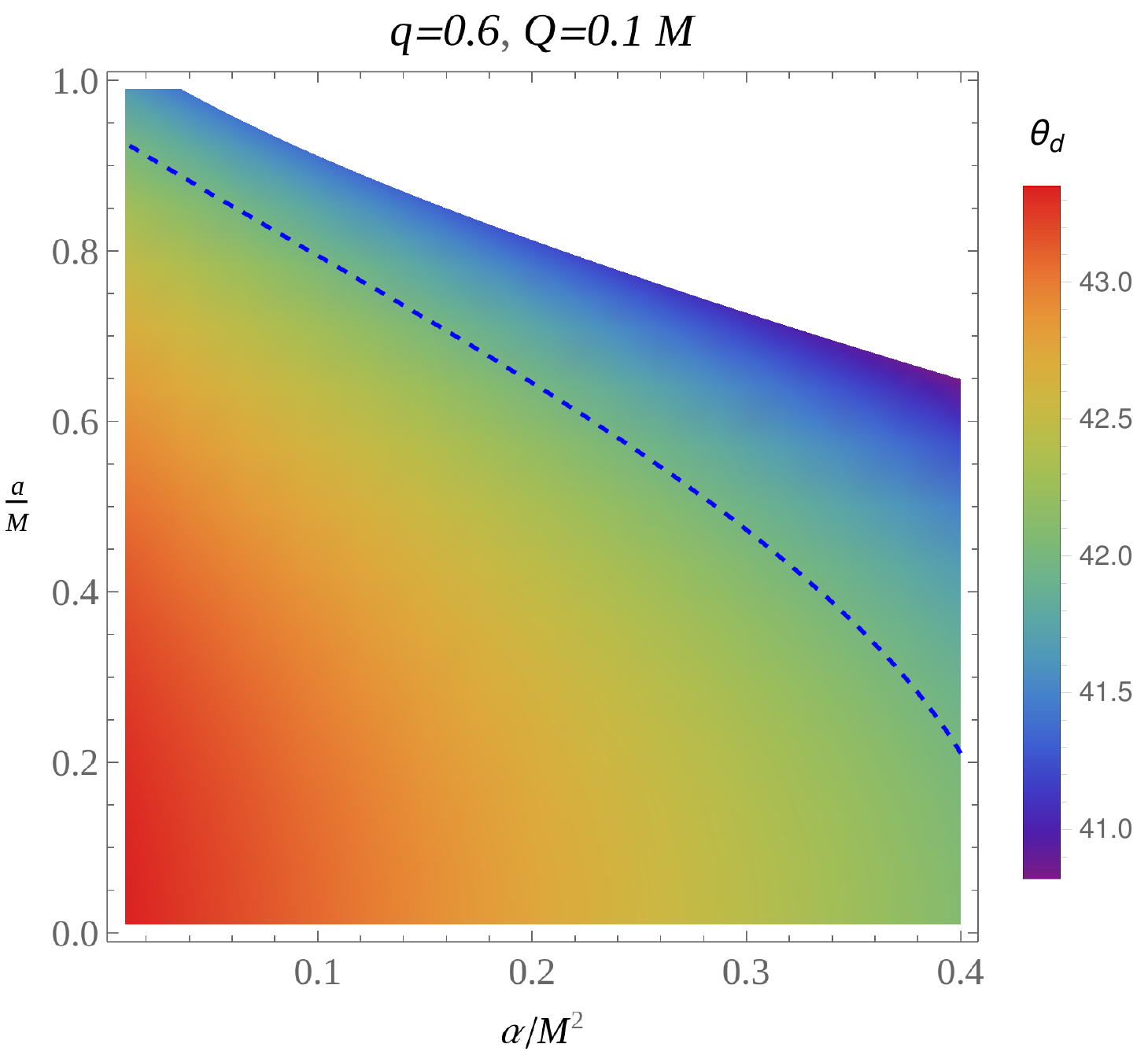}
\includegraphics[width=0.35\textwidth]{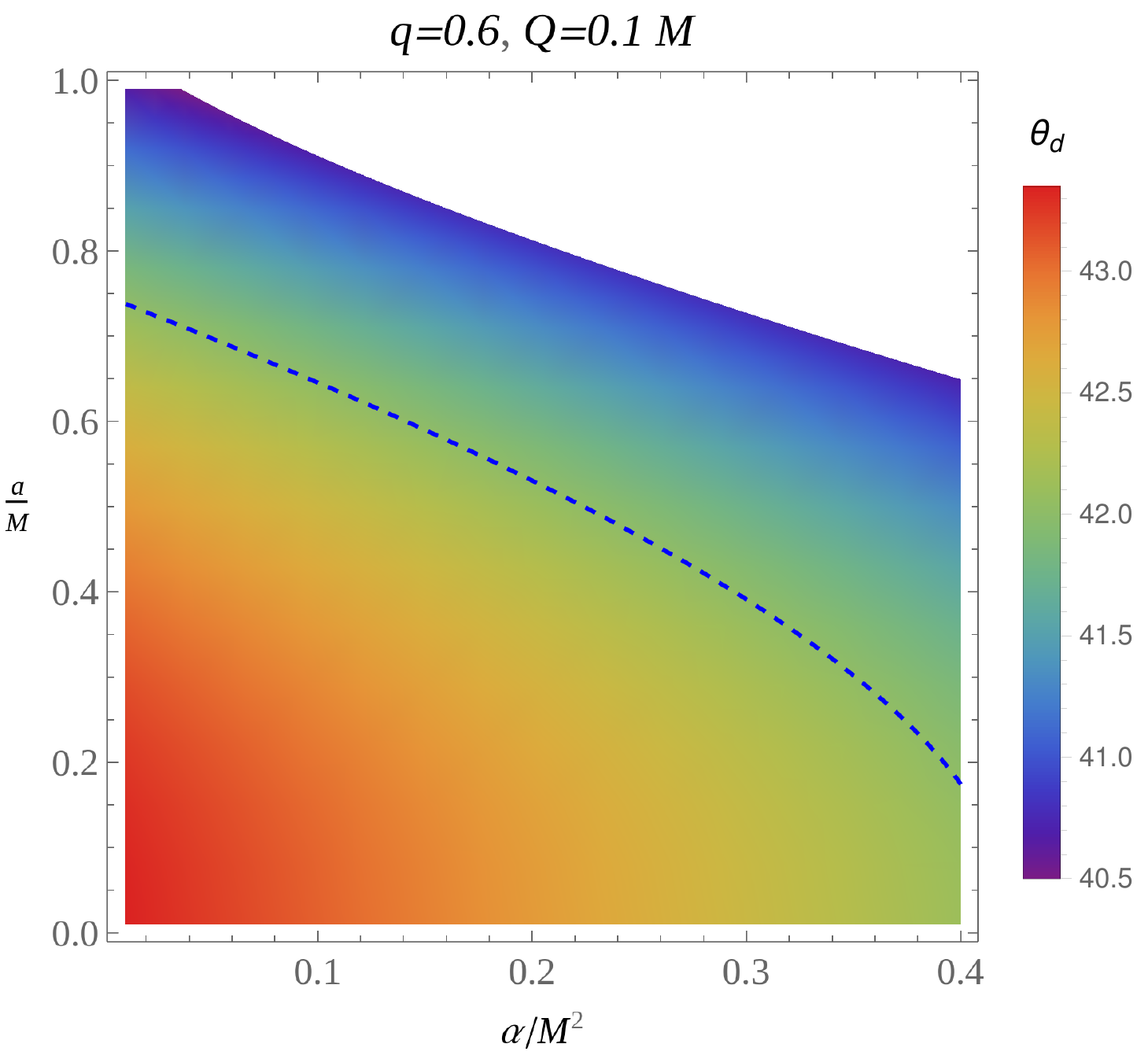}\\
\includegraphics[width=0.35\textwidth]{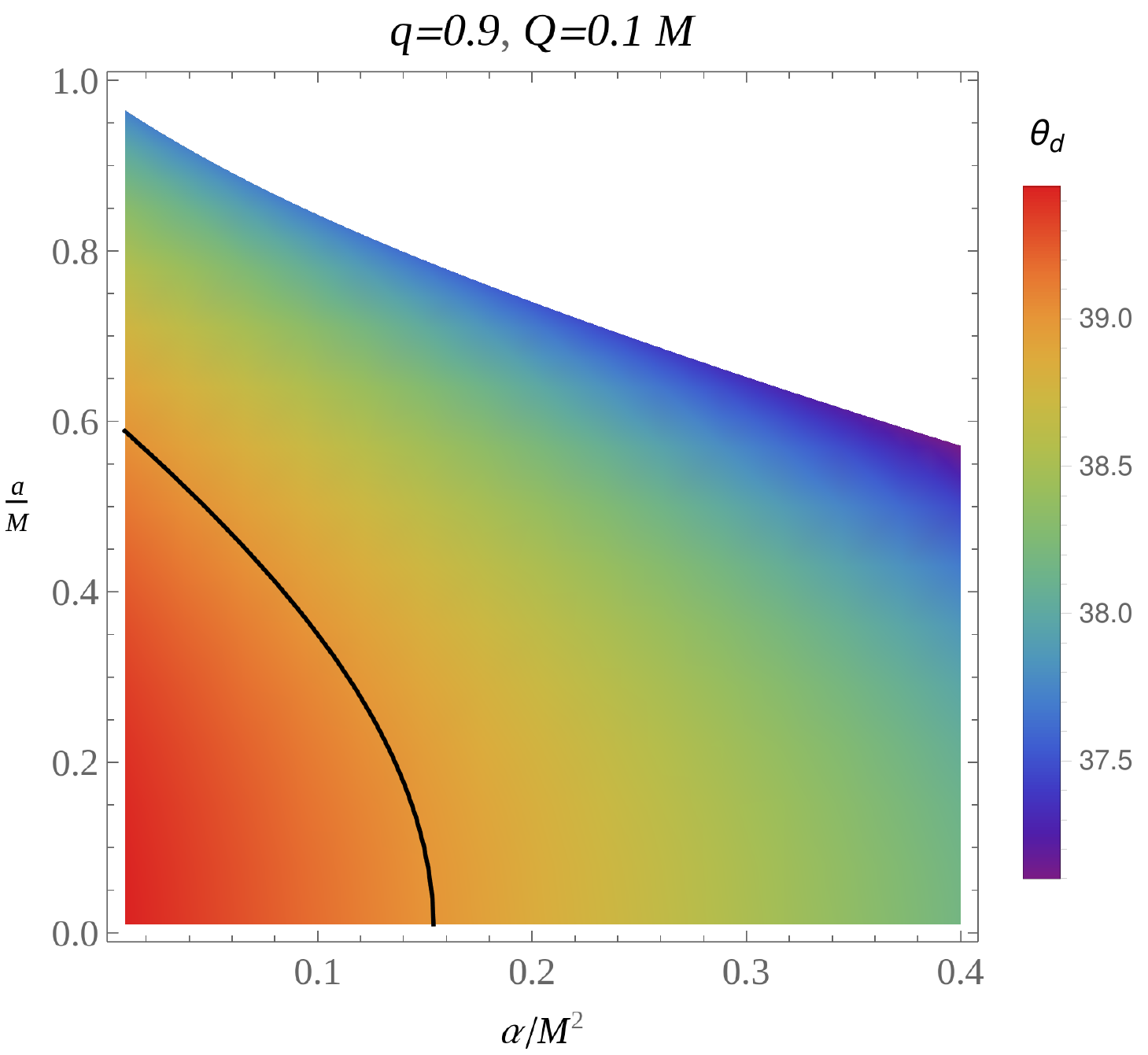}
\includegraphics[width=0.35\textwidth]{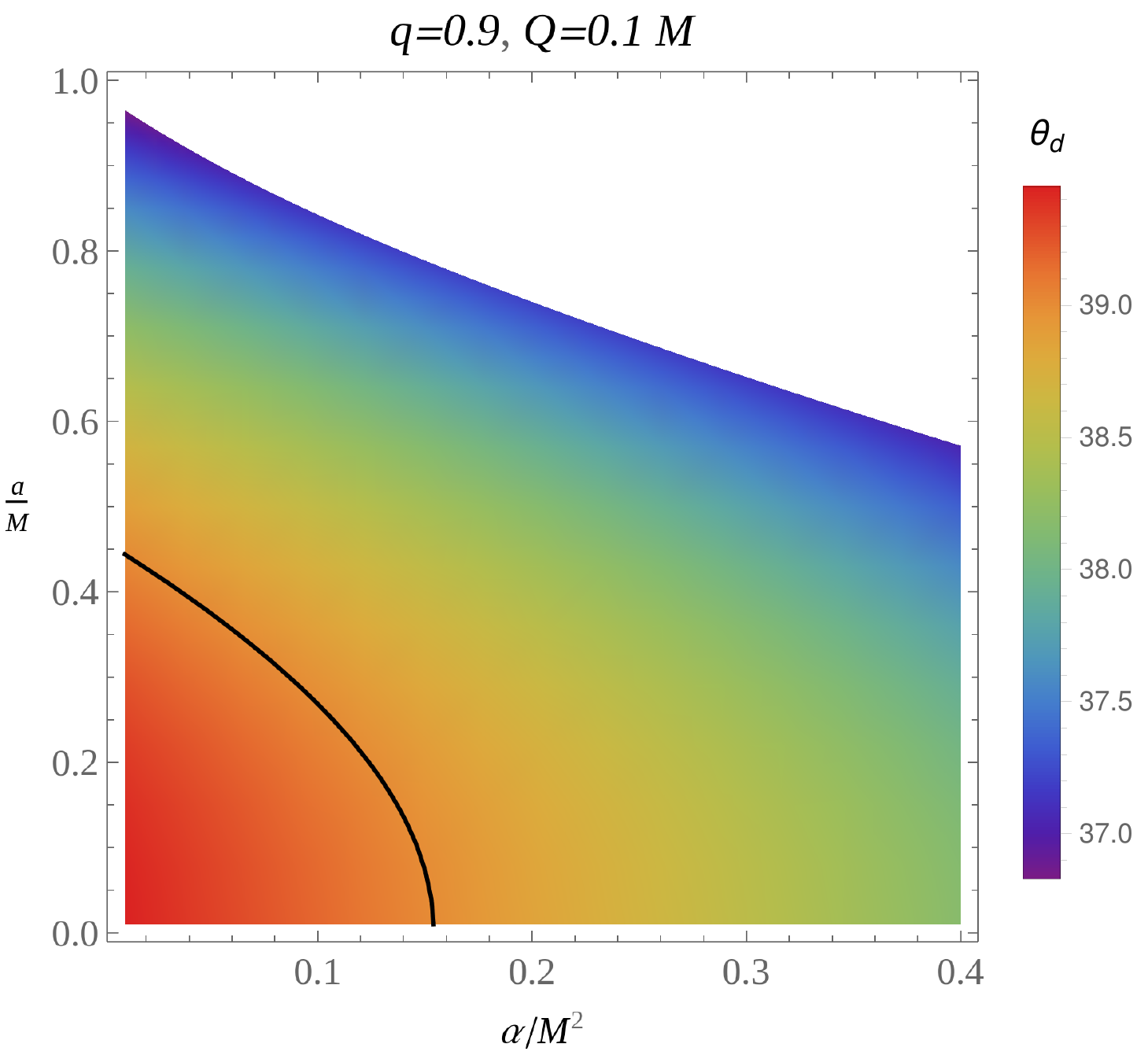}
\caption{The same figure as Fig. \ref{constraintsM1}, but for the constraints for BH spin and the GB coupling parameters and fixed values of $q=0.6$ (top panel) and $q=0.9$ (bottom panel) at $Q=0.1 M$.} \label{constraintsM2}
\end{figure*}
In Fig. \ref{constraintsM2}, we present another constraint from EHT observational results on the image size of the SMBH M$87$*. In this case, we illustrate the angular diameter of the BH shadow as a function of spin and GB parameters. The top panel for $q=0.6$, corresponds to the BH M$87$*, so we provide the relationships between the BH spin and EGB coupling parameter for the mean value of the measured diameter of the SMBH M$87$* ($42\mu as$) with blue dashed curves to describe the distribution of the observable parameter effortlessly. While in the case of $q=0.9$ (the bottom panel), we can get lower limits for the spin and GB coupling parameters. From all these plots, one can see minor differences within the change of inclination angle. One can easily see from the figure that an increase of $q$, causes a decrease in the constrained value of the spin.

\subsection{Constraints on Parameters from Sgr. A*}
The same procedure can be incorporated for the observational data of the SMBH Sgr. A* is located at the centre of the Milky Way galaxy. The angular diameter of the shadow of the BH Sgr. A* is $\theta_d=48.7\pm7\mu as$ \cite{2022ApJ...930L..12E}. The mass of Sgr. A* and the distance from Earth be considered as $M\simeq4\times10^6M_\odot$ and $d\simeq 8$ kpc, respectively \cite{2022ApJ...930L..12E,2022ApJ...930L..17E}.
\begin{figure*}\centering
\includegraphics[width=0.35\textwidth]{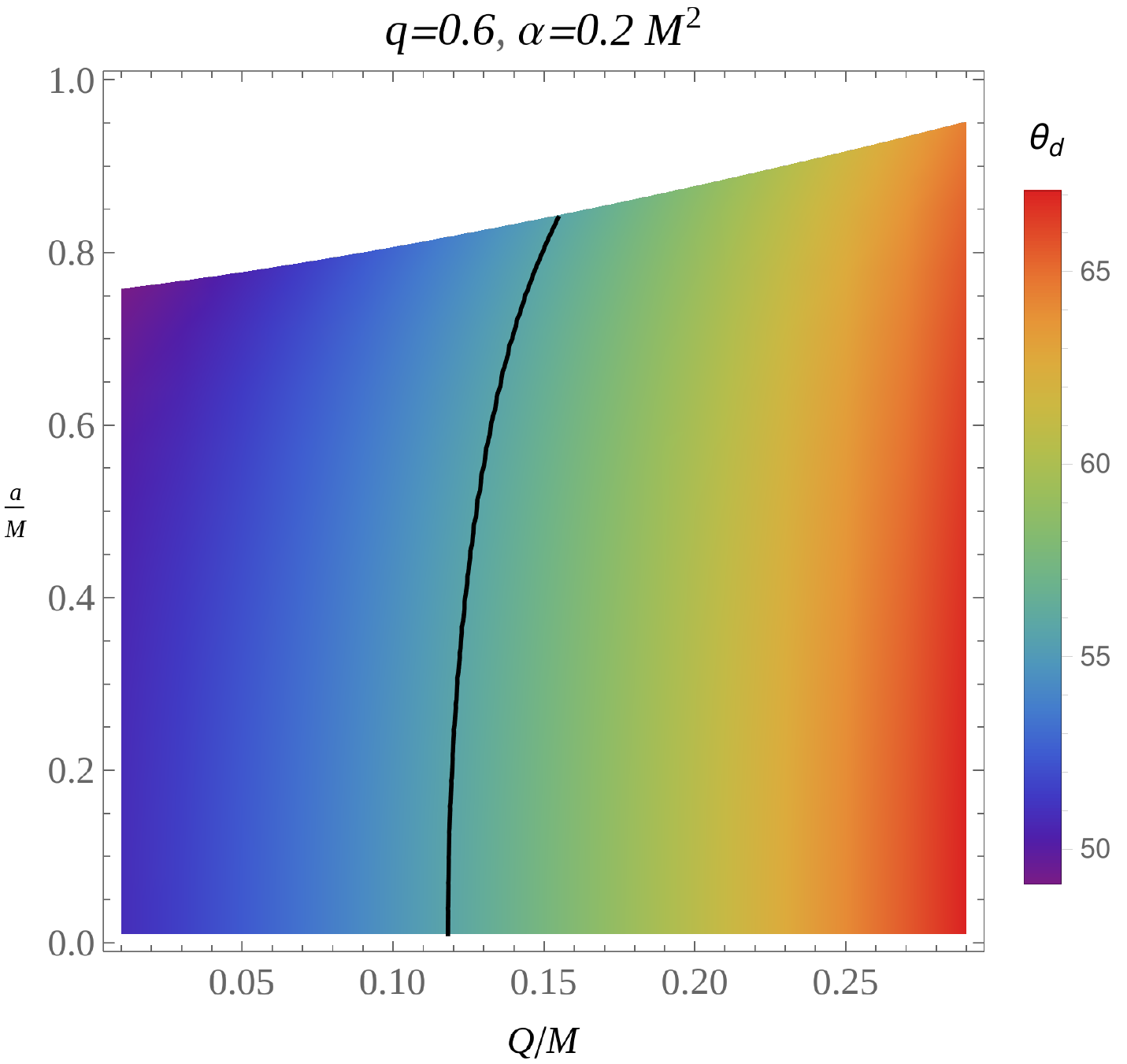}
\includegraphics[width=0.35\textwidth]{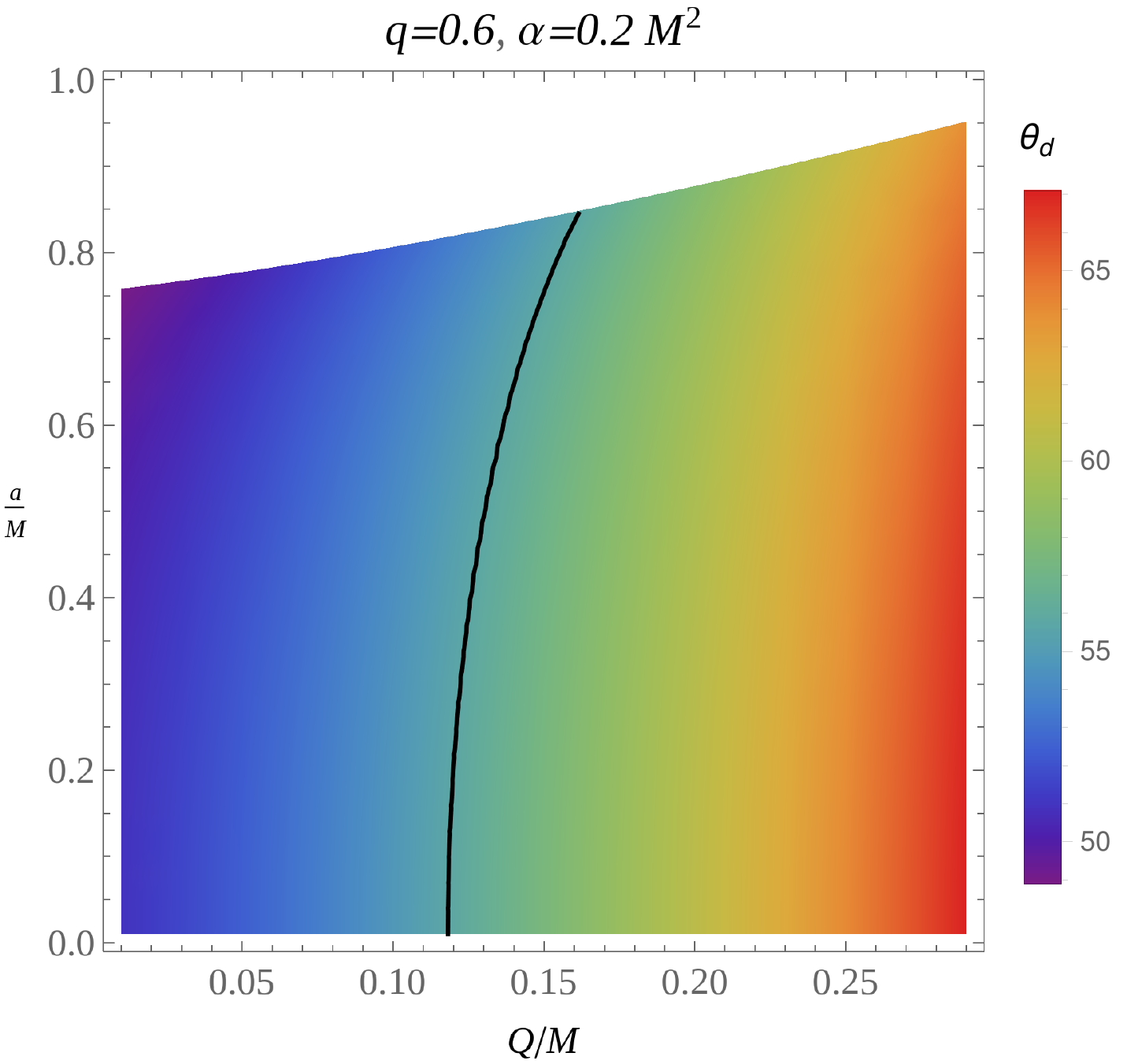}\\
\includegraphics[width=0.35\textwidth]{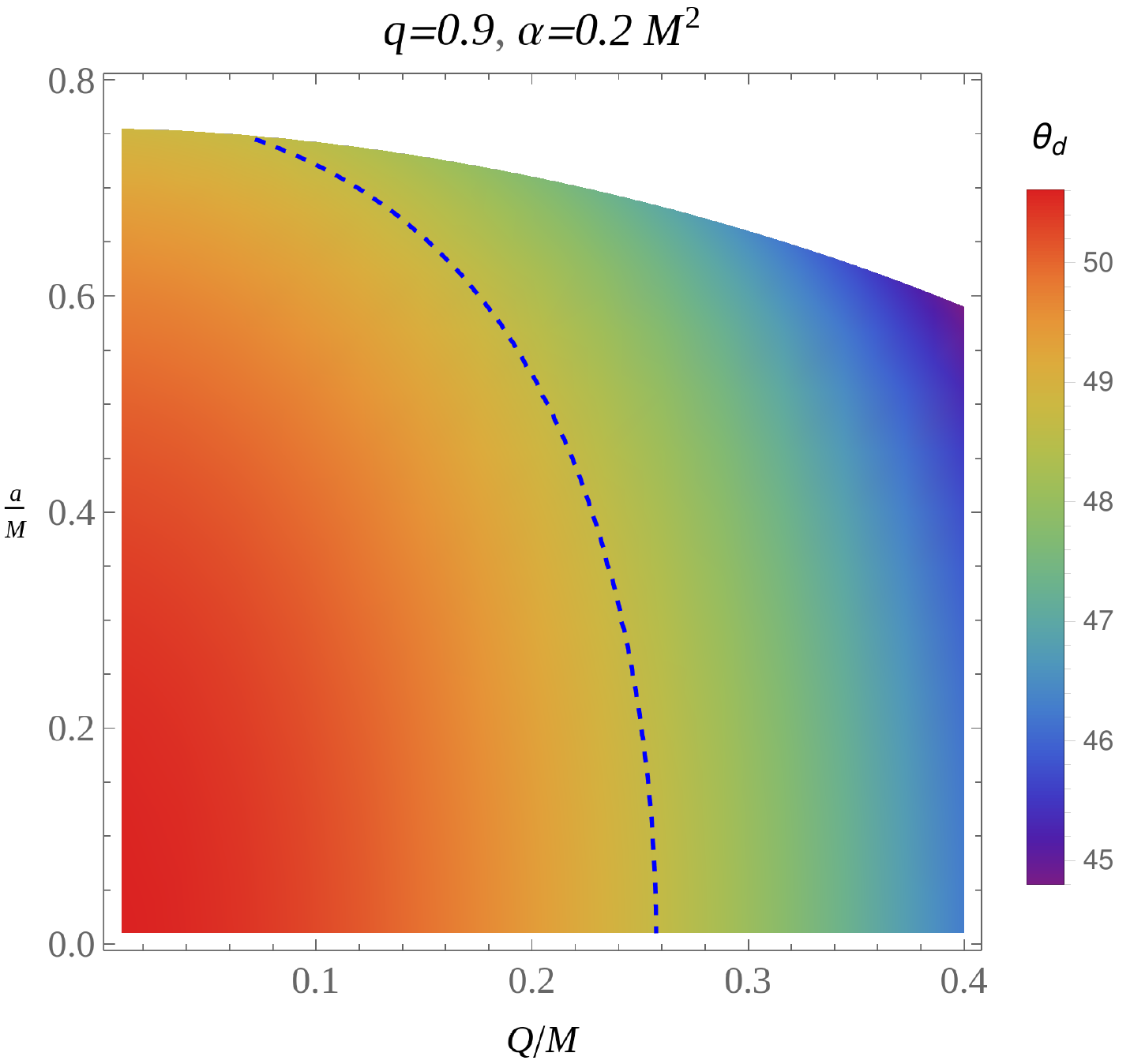}
\includegraphics[width=0.35\textwidth]{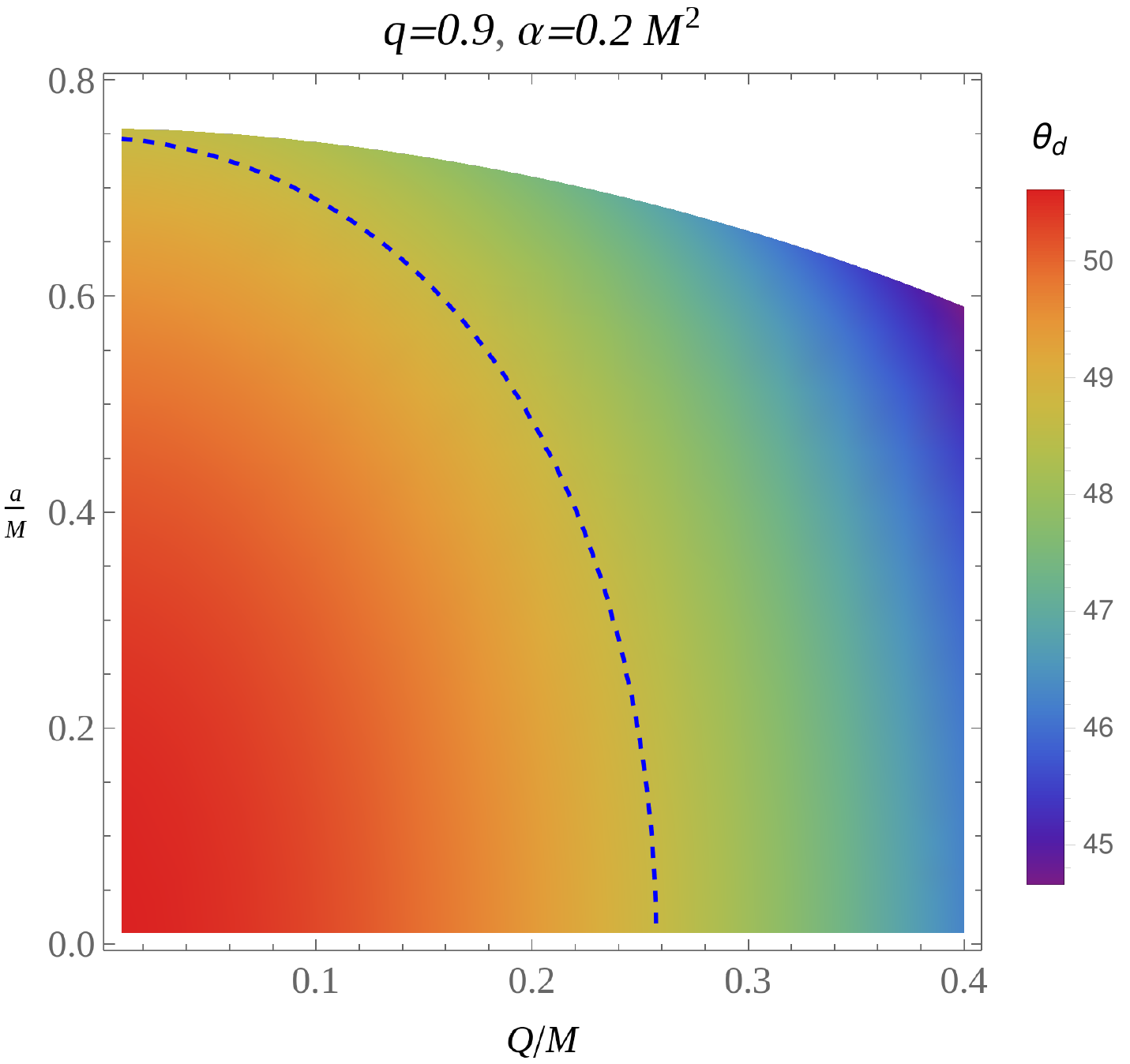}
\caption{Angular diameter observable $\theta_d$ for the BH shadows as a function of parameters $a/M$ and $Q/M$ at inclinations $90^\circ$ (left panels) and $50^\circ$ (right panels). The black curves correspond to $\theta_d=55.7\mu as$ within the measured angular diameter, $\theta_d=48.7\pm7\mu as$ of the Sgr. A* BH reported by the EHT. The blue dashed curves correspond to $48.7\mu as$.} \label{constraintsS1}
\end{figure*}
\begin{figure*}\centering
\includegraphics[width=0.35\textwidth]{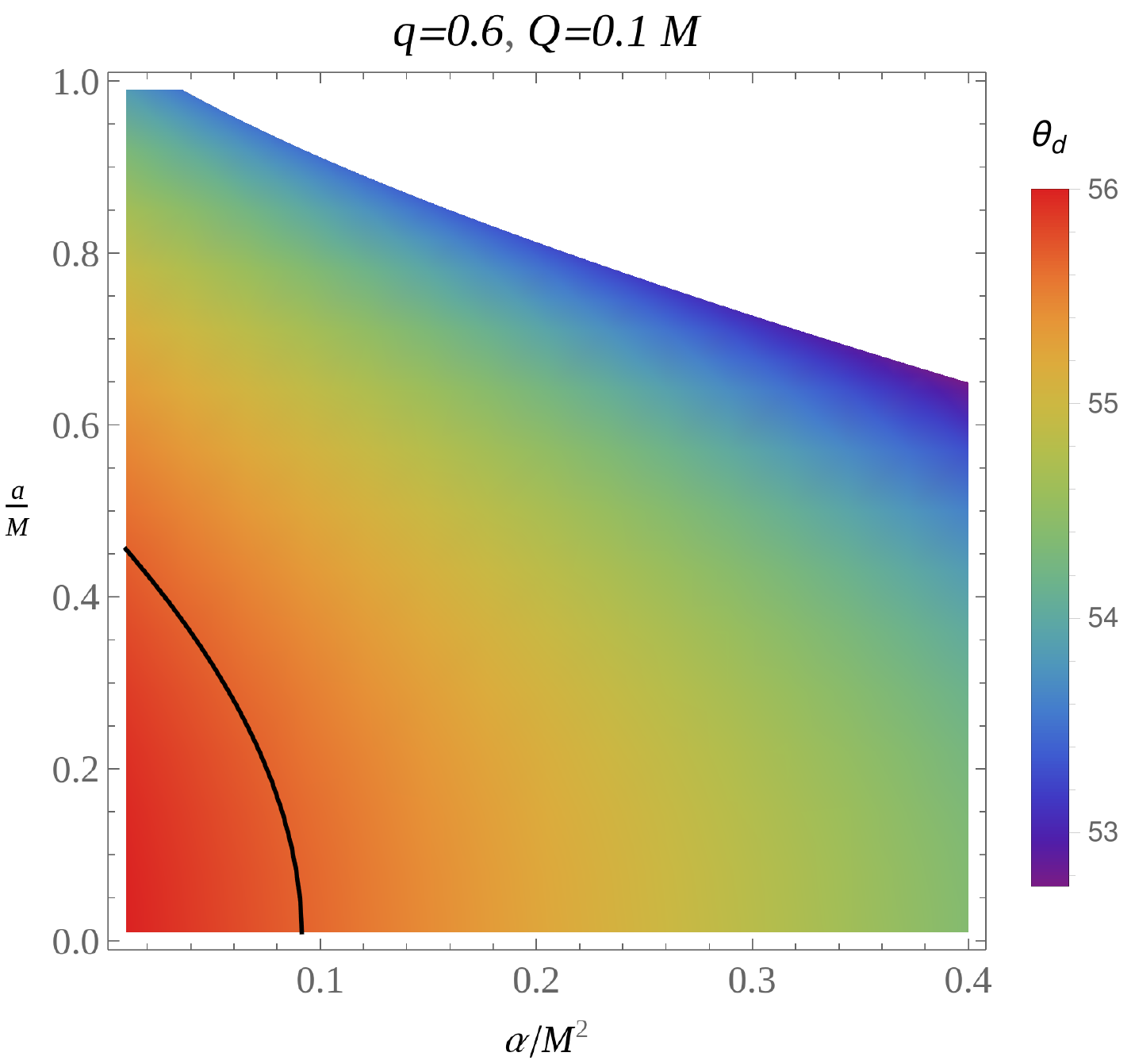}
\includegraphics[width=0.35\textwidth]{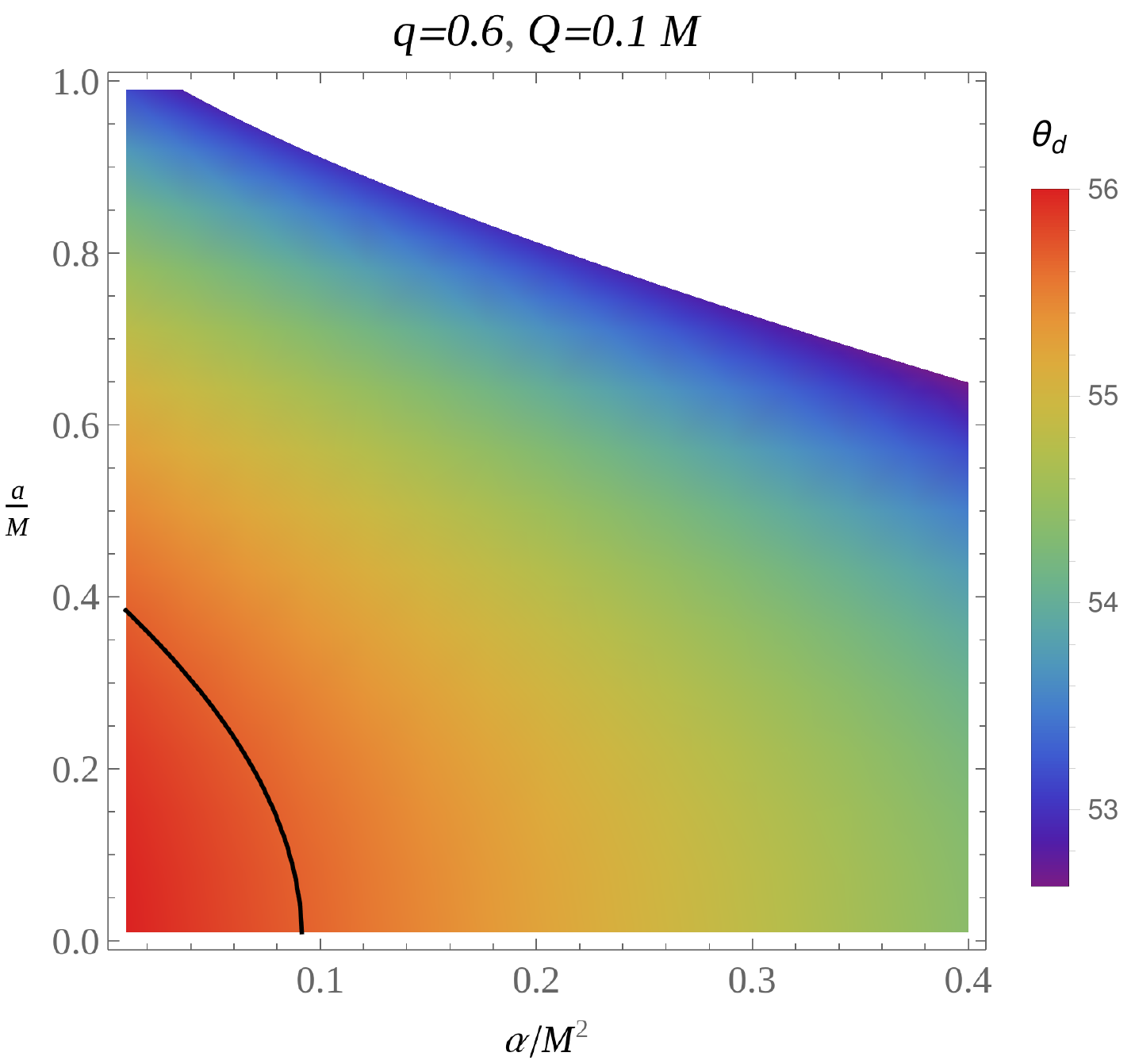}\\
\includegraphics[width=0.35\textwidth]{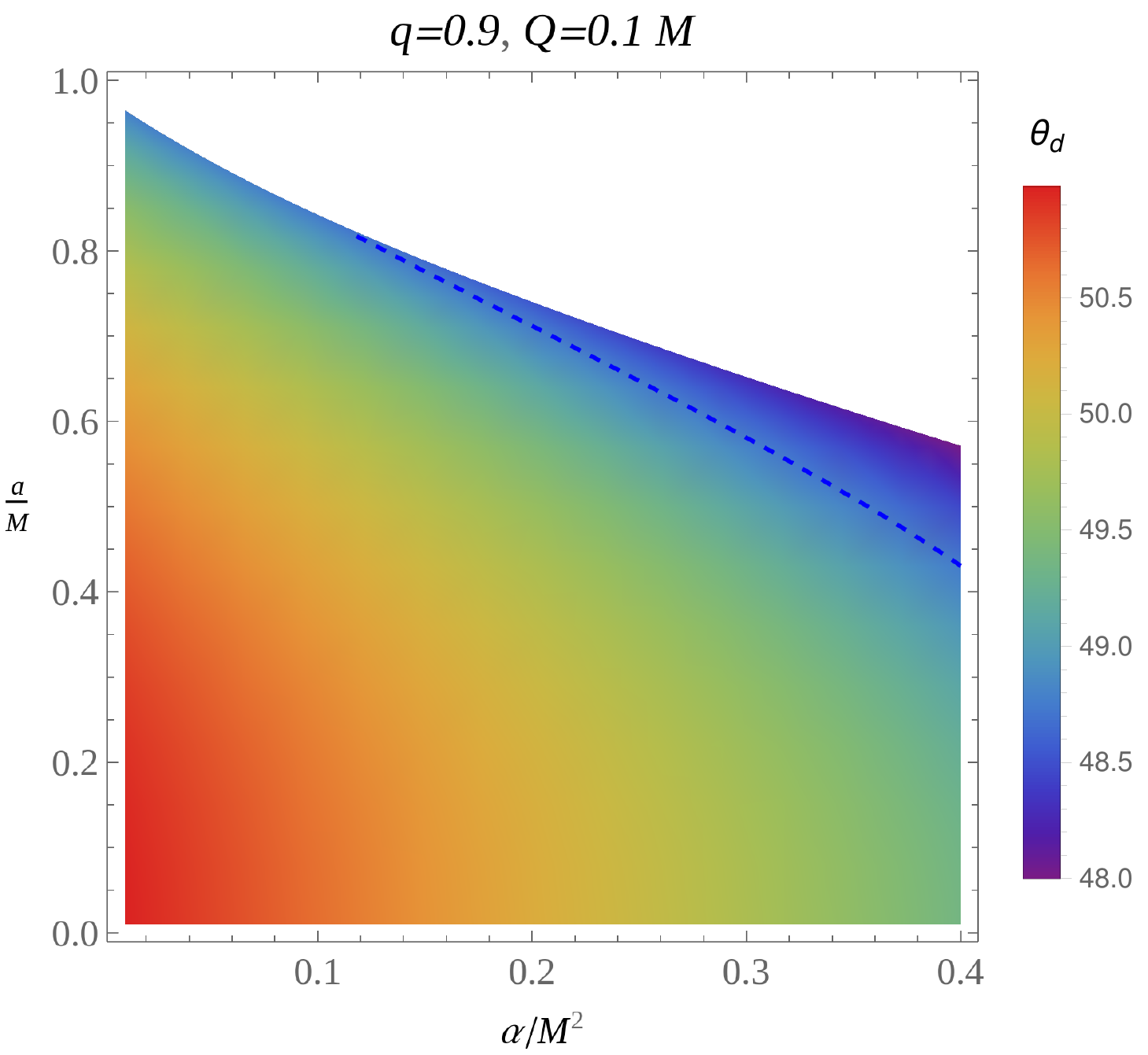}
\includegraphics[width=0.35\textwidth]{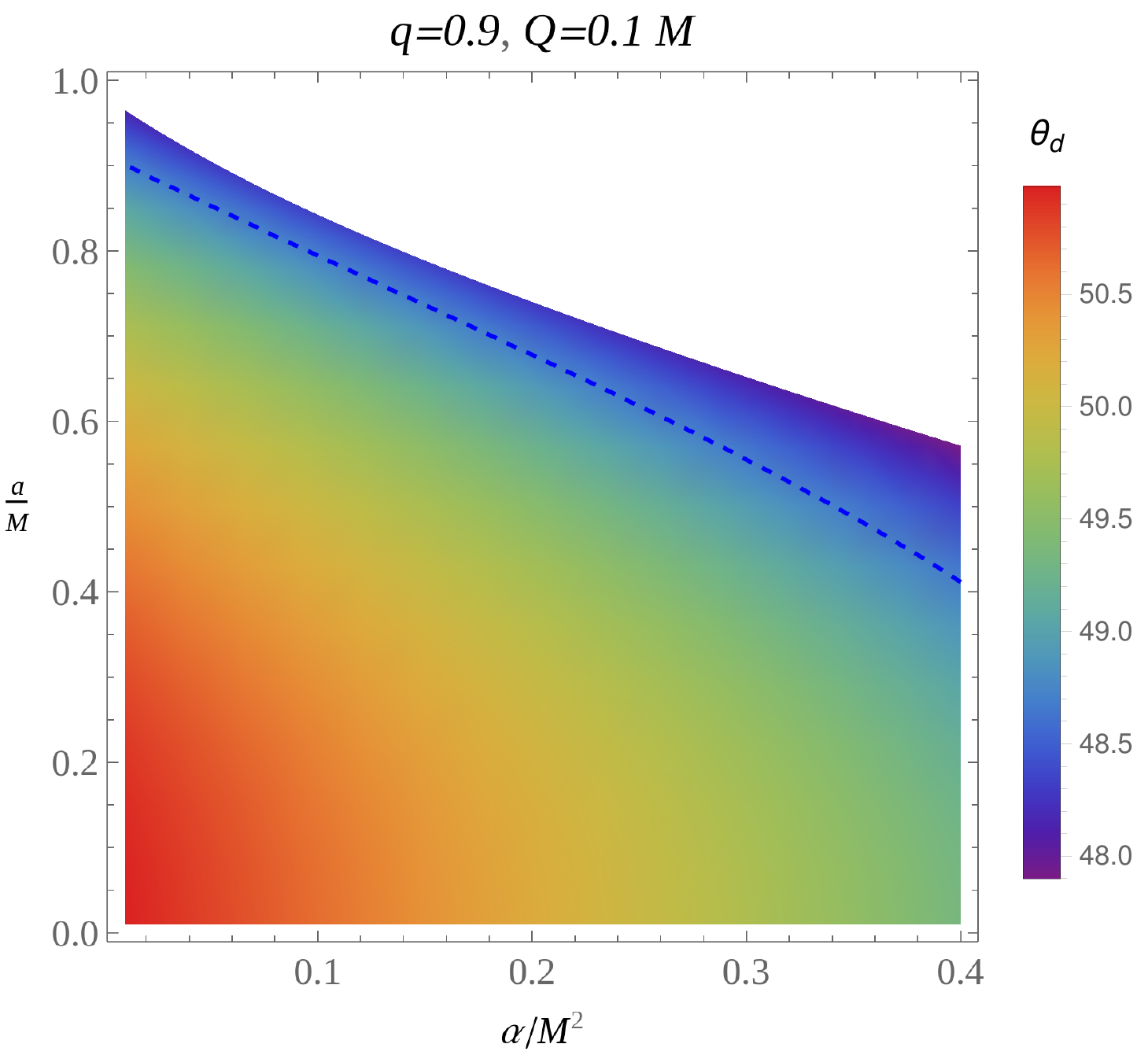}
\caption{Angular diameter observable $\theta_d$ for the BH shadows as a function of parameters $a/M$ and $\alpha/M^2$ at inclinations $90^\circ$ (left panels) and $50^\circ$ (right panels). The black curves correspond to $\theta_d=55.7\mu as$ within the measured angular diameter, $\theta_d=48.7\pm7\mu as$, of the Sgr. A* BH reported by the EHT. The blue dashed curves correspond to $48.7\mu as$.} \label{constraintsS2}
\end{figure*}
Similarly, in Figs. \ref{constraintsS1} and \ref{constraintsS2} we obtain limits for the spin and charge parameters for Sgr. A*. These plots describe the angular diameter observable from BH parameters at inclinations $90^\circ$ (left panels) and $50^\circ$ (right panels).
Since the shadow observables depend differently on each parameter of the BH, therefore we will get different plots for various values of the BH parameters. In the top panel of Fig. \ref{constraintsS1}, we obtain upper limits for the spin and charge parameters for $q=0.6$ and $\alpha=0.2M^2$. However, for $q=0.9$, the upper limits shift up and there are no upper and lower bounds for the spin and charge parameters. All possible values of image size that the BH parameters provide, lie in the range of observational data due to less accuracy of the observations. In the top panel of Fig. \ref{constraintsS2}, we obtain lower limits for the spin and GB parameters of the BH Sgr. A*, at $q=0.6$ and $Q=0.1M$. However, there are no limits for the parameters when $q=0.9$ as it is obtained in Fig. \ref{constraintsS1}. We can conclude that these constraints may be useful in estimating the BH parameters using astrophysical observations of BH shadow and may help to find out the range of possible values of its parameters. The shadow of the BH that we are studying can be compared with the shadow images of M$87$* and Sgr. A*. 

\section{Energy Emission Rate}
In Classical Mechanics, if something enters a BH then it is lost forever in contradiction with the Quantum Mechanical view. The quantum fluctuations are responsible for the particle creation and annihilation inside the BH, near its horizon. Under the effect of quantum tunnelling, the particles may possess the energy $E>0$ and escape by crossing the horizon. This emitted energy helps the BH to evaporate. The absorption process is measured in terms of absorption gravitational cross-section probability. At infinity, the high energy of the absorption cross-section ($\sigma_{lim}$) is related to the shadow of the BH. At high energies, a constant value $\sigma_{lim}$ determines the absorption cross-section of the BH. The value of $\sigma_{lim}\approx A_G$, where $A_G$ is the geometric area of the photon region. As we know, the innermost layer of the photon sphere is the outer boundary of the shadow, thus, it is related to the absorption cross-section \cite{2013JCAP...11..063W,1973PhRvD...7.2807M,2011PhRvD..83d4032D} as
\begin{equation}
\sigma_{lim}\approx\pi R_{sh}^2. \label{66}
\end{equation}
The energy emission rate has identical forms as
\begin{equation}
\mathcal{E}_{\omega t}:=\frac{d^2\mathcal{E}(\omega)}{d\omega dt} =\frac{2\pi^2\sigma_{lim}\omega^3}{e^{\omega/T_H}-1}\approx\frac{2\pi^3R_{sh}^2\omega^3}{e^{\omega/T_H}-1}, \label{67}
\end{equation}
where $\omega$ denotes the angular frequency, $T_H=\kappa/2\pi$ is the Hawking temperature and
\begin{equation}
\kappa=\frac{\Delta'(r)}{2(a^2+r^2)}\bigg|_{r=r_+} \label{68}
\end{equation}
is the surface gravity at the outer horizon. The surface gravity for the non-rotating case becomes
\begin{equation}
\kappa=\frac{1}{2}f'(r)\bigg|_{r=r_+}. \label{69}
\end{equation}
\begin{figure}[t]
\begin{center}
\subfigure{
\includegraphics[height=5cm,width=7.28cm]{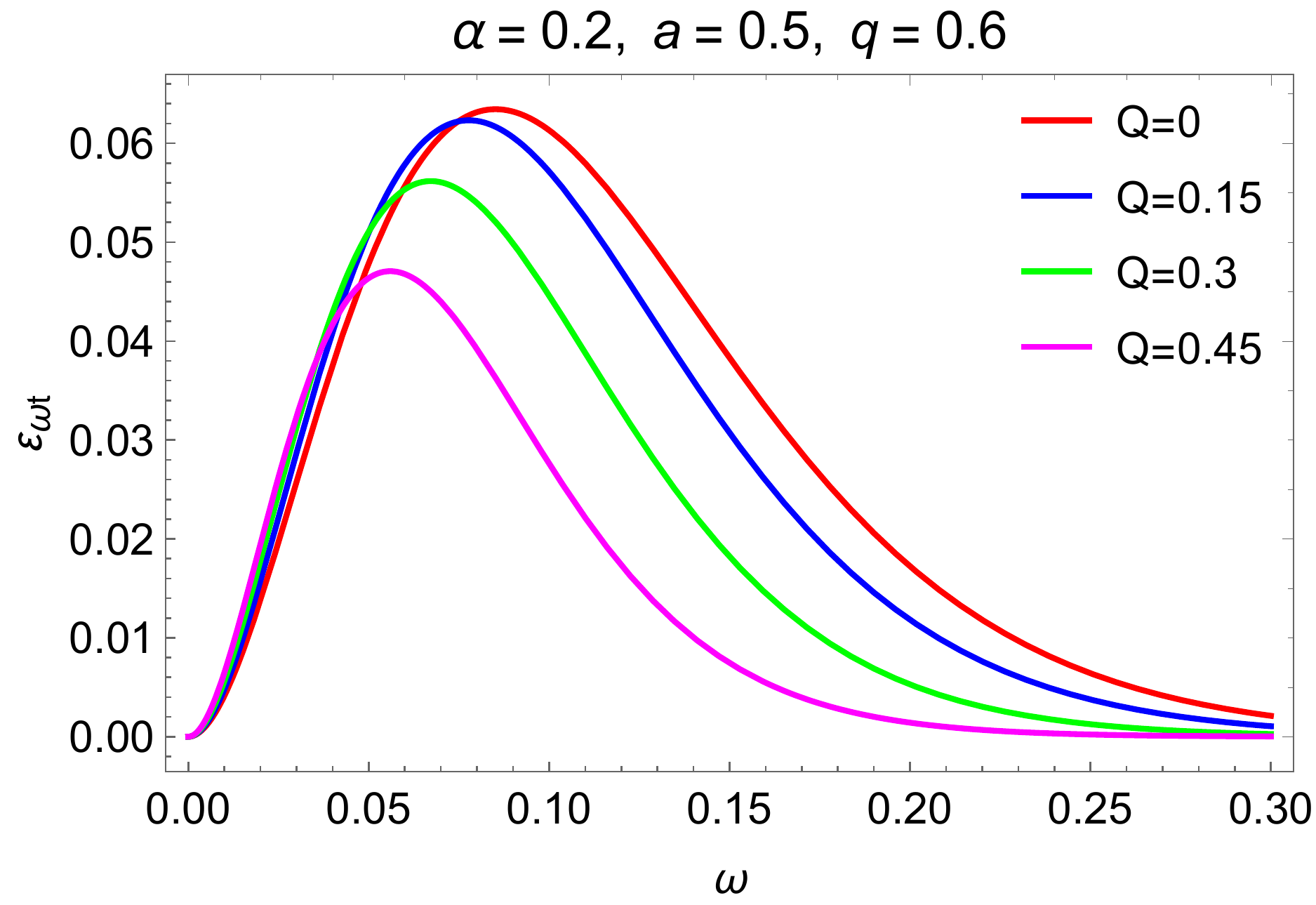}}
~~~~
\subfigure{
\includegraphics[height=5cm,width=7.28cm]{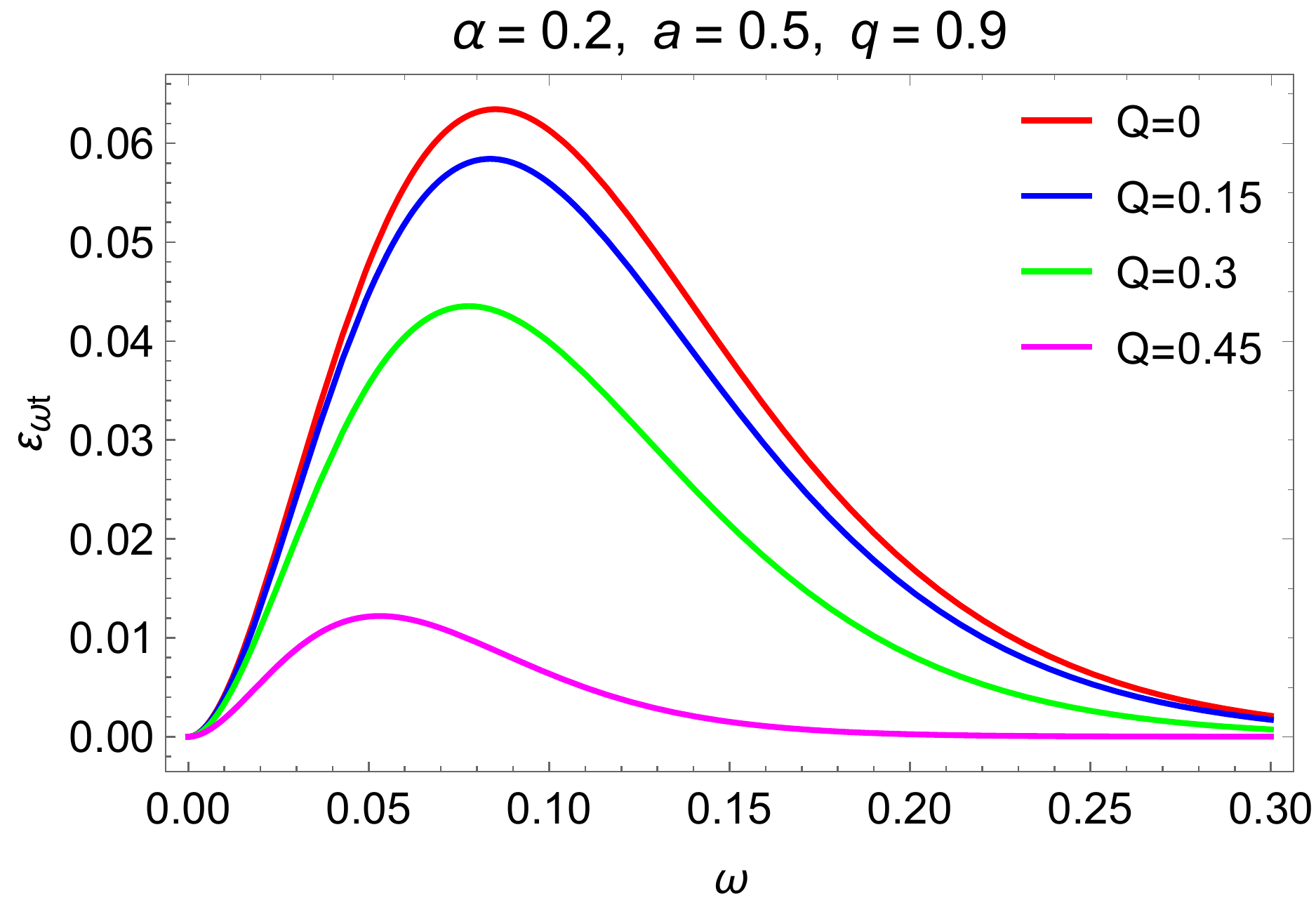}}
\subfigure{
\includegraphics[height=5cm,width=7.28cm]{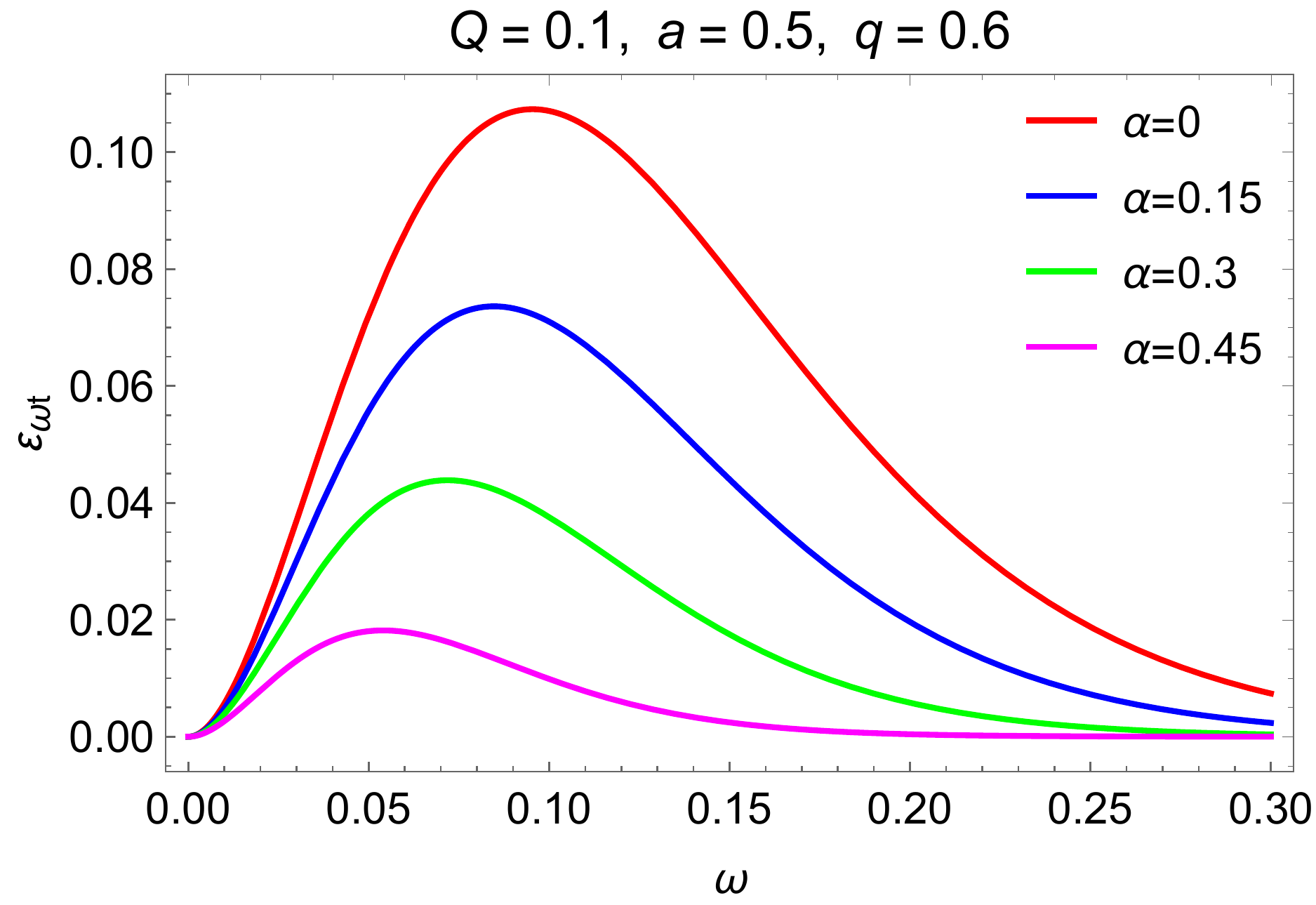}}
~~~~
\subfigure{
\includegraphics[height=5cm,width=7.28cm]{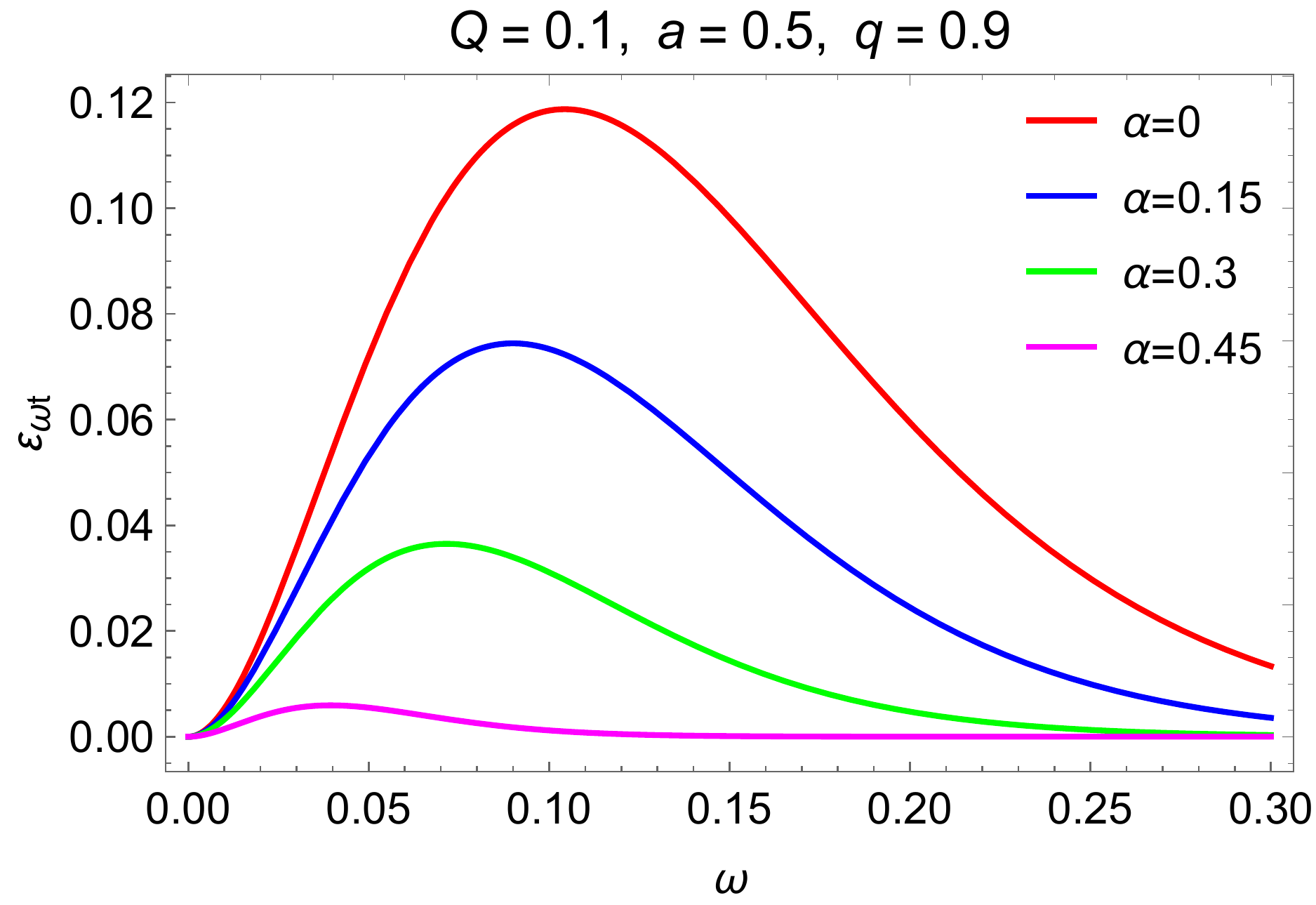}}
\subfigure{
\includegraphics[height=5cm,width=7.2cm]{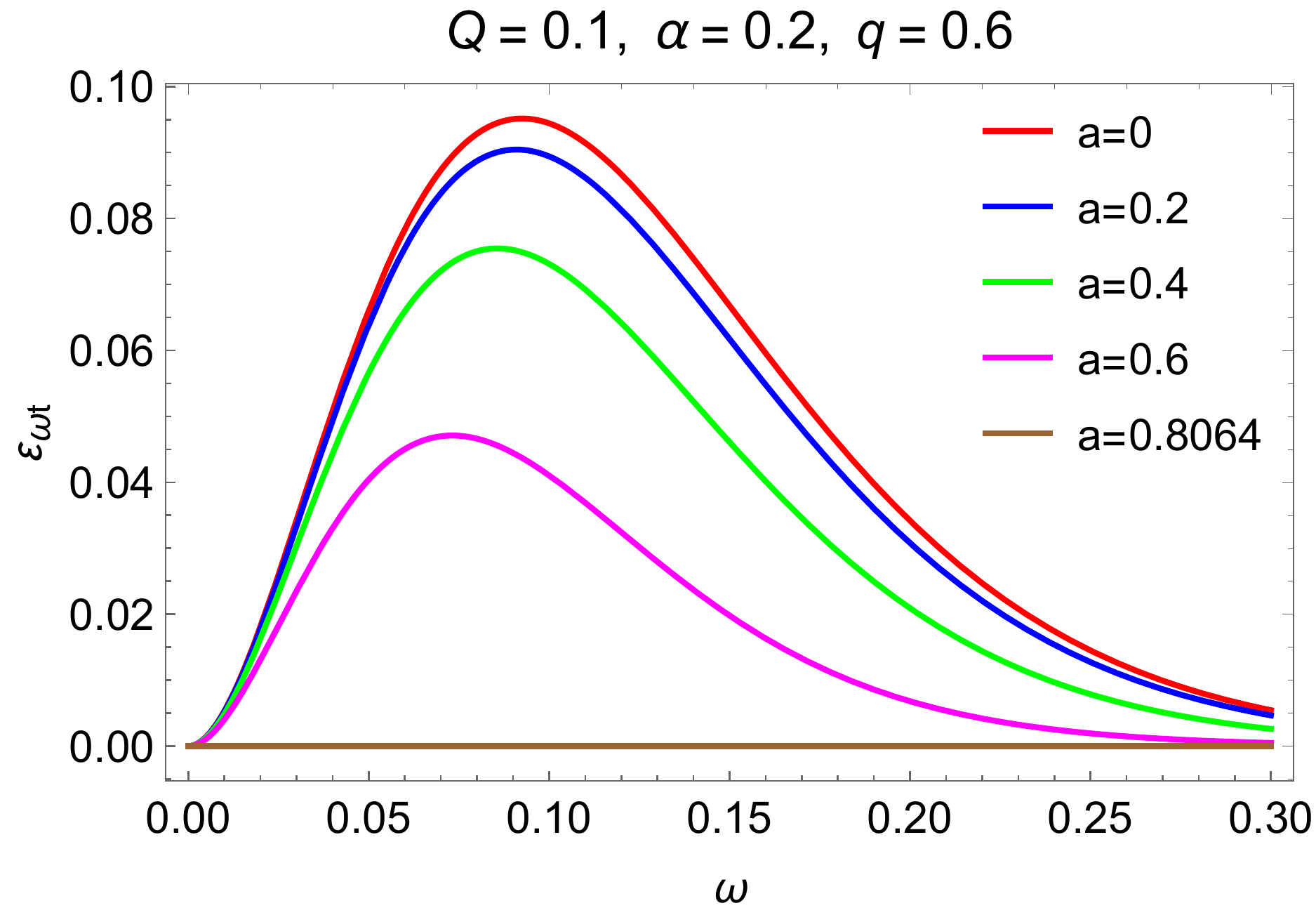}}
~~~~
\subfigure{
\includegraphics[height=5cm,width=7.28cm]{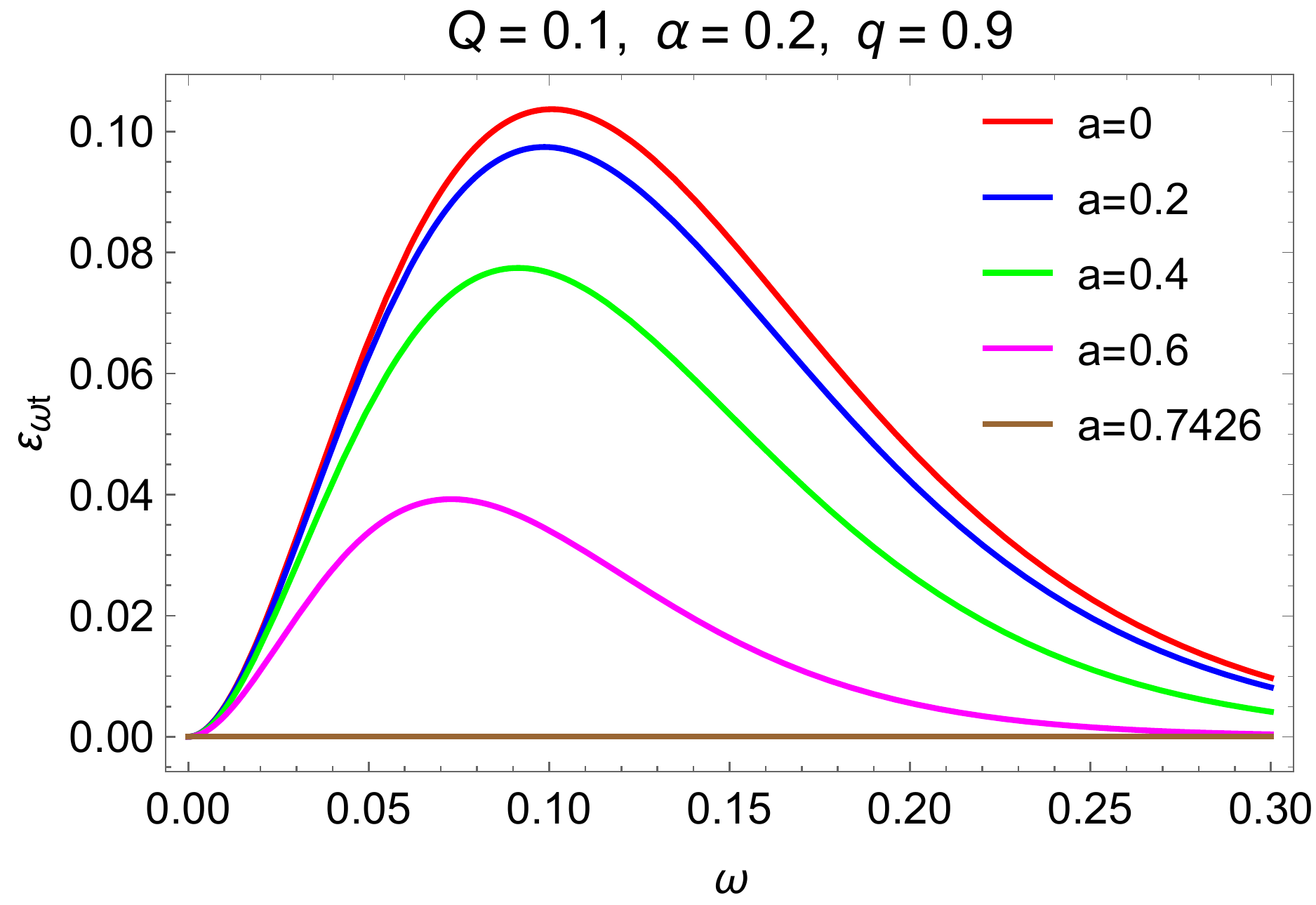}}
\subfigure{
\includegraphics[height=5cm,width=7.28cm]{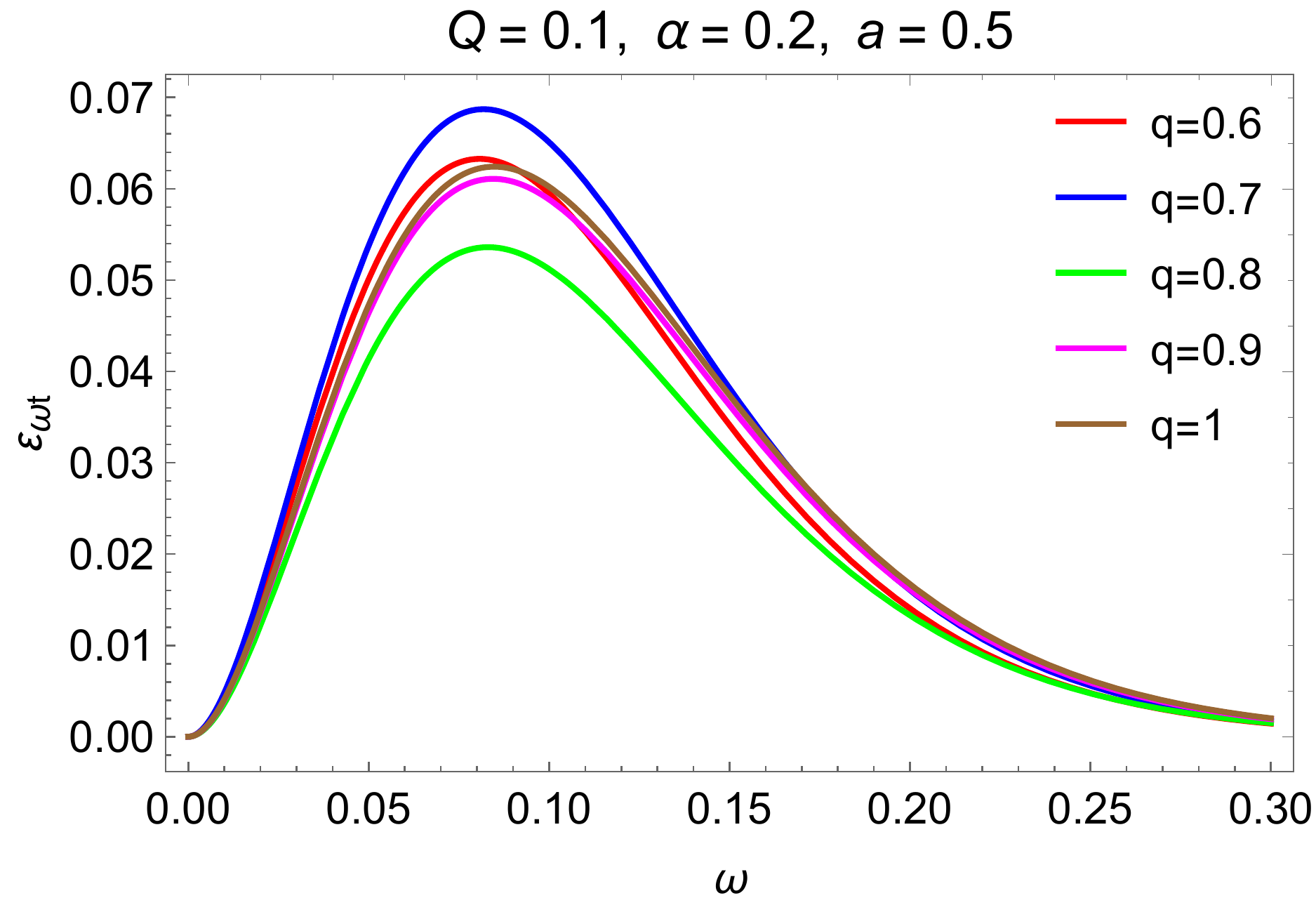}}
\end{center}
\caption{Plots for the energy emission rate w.r.t the angular frequency for different values of $a$, $q$, $Q$ and $\alpha$.} \label{EERfig}
\end{figure}
The rate of energy emission is plotted in Fig. \ref{EERfig}. All of these plots correspond to the relevant shadow plots in Fig. \ref{shadowfig}. The left and right plots in the top panel show that by increasing the value of $Q$, the crest of the energy emission curve drops which corresponds to the slow evaporation rate of the BH. However, in the right plot, the crest of the emission curve drops drastically w.r.t $Q$ as compared to that in the left plot. In the seconds panel, the peaks in both plots drop with almost the same behaviour w.r.t $\alpha$ thus causing the slow BH evaporation. The difference in both plots is merely the height of the peak of the curves. In the third panel, the peaks get dropped w.r.t $a$ causing the slow evaporation of the BH. Moreover, for extremal values of $a$, there is no BH evaporation because the extreme BHs are non-radiating and the Hawking temperature is zero Kelvin. As the evaporation rate of the BH decreases for all of the above cases, however, the evaporation rate increases w.r.t $q$ for both intervals.

\section{Conclusion}
In this paper, we aimed to understand the effect of YM charge and its power, together with spin and GB parameters on different BH properties and related observables. The modified NJA is applied to a static metric in $4D$ EGB gravity with a PYM field in order to derive a rotating BH solution. The size of the photon sphere is explored for the static BH and then the horizon radius is investigated for the rotating BH. In order to derive the analytical scheme for the shadows, the HJ formulation is employed and then by introducing Bardeen's procedure for an observer at infinity, the impact parameters, shadows and distortion are presented. The shadows are compared with the EHT data for Sgr. A* and M$87$* BHs and the parametric constraints are obtained. Finally, the rate of energy emission is discussed. The results presented in previous sections are summarized below:
\begin{itemize}
\item In Fig. \ref{phsphere}, it can be found that the size of the photon sphere is decreased with an increase in $\alpha$. When $Q$ is increased, the size of the photon sphere is increased for $q<0.75$ and is decreased for $q>0.75$. It shows that the effect of $Q$ is dependent upon $q$ as well. A discontinuity is observed for the radius of photon sphere when $q\rightarrow0.75$;
\item  Figure \ref{horizonrad} shows the behaviour of horizon radius for rotating BH. It can be seen that with the increase in spin up to its extremal value, the CH is increased and EH is decreased for all cases. As mentioned above, the effect of $Q$ is altered by varying $q$. Therefore, as $Q$ is increased, we measured an increase in EH for $q<0.75$ and a decrease in EH for $q>0.75$, whereas, the curves converge to a fixed value of CH for $q<0.75$ and for $q>0.75$, the CH increases. The curves are shifted inwards w.r.t $\alpha$ thus causing the EH to decrease and CH to increase. The curves are expanded w.r.t at both intervals of $q$. However, $q$ is more sensitive for the EH as compared to the CH. The central singularity is not identified in all cases.
\item From the plots for effective potential, we perceive that the unstable circular orbits get enlarged for $q<0.75$ and shrunk for $q>0.75$ w.r.t increase in $Q$. This is because $Q$ is dependent upon the power $q$. Moreover, as we see that $\alpha$ and $a$ are independent of $q$, therefore, the unstable circular orbits are reduced in size w.r.t $\alpha$ and $a$ $\forall$ $q$. Furthermore, the unstable orbits are enlarged w.r.t both intervals of $q$.
\item Shadows are presented in Fig. \ref{shadowfig} and is found that the shadow size is increased for $q<0.75$ and is decreased for $q>0.75$ w.r.t the increase in $Q$. The shadow size decreases and the flatness increases w.r.t increase in $\alpha$ $\forall$ $q$. The GB parameter appears as a less sensitive parameter because the shadow size is decreased by a small amount as $\alpha$ is increased. With respect to spin $a$, the shadow size is also not altered significantly, however, the loops are shifted towards the right with an increased flatness. Therefore, $a$ cannot be considered a sensitive parameter. For the extreme value of $a$, the shadow of the BH is Kerr-like.  Moreover, by increasing $q$ in both intervals, the shadow size also increases.
\item The quantity of flatness in the shadows known as distortion is plotted in Fig. \ref{distrfig}. The distortion is increased with an increase in $a$ and $\alpha$ $\forall$ $q$. Whereas, for $q<0.75$, the distortion is decreased linearly and is increased for $q>0.75$ w.r.t increase in $Q$. In both intervals for $q$, the distortion decreases.
\item {The constraint values for the spin and BH charge together with the GB parameters using EHT observations of SMBHs Sgr. A* and M$87$* for $q=0.6$ and $0.9$ are obtained. It is found that there are upper and lower values for the charge and spin of the BH M$87$* at $q=0.6$, while only upper bounds have been obtained for the charge and spin for Sgr. A* due to bigger error in shadow observations of the Sgr. A*. However, there is only an upper bound in the $a$-$\alpha$ space of M$87$* at $q=0.9$, while there are no limits for the case of Sgr. A*.}
\item The energy emission rate is shown in Fig. \ref{EERfig}. It can be seen that the emission peak is dropped with the increase in $Q$, $a$ and $\alpha$ $\forall$ $q$ causing the BH to evaporate slowly. However, the extreme BH does not evaporate at all due to Hawking temperature being zero Kelvin. The BH evaporation becomes faster w.r.t $q$ in both intervals.
\end{itemize}
It is concluded that the spin and GB parameters have a usual effect on the radius of photon sphere, horizon radius, shadow and related observables i.e. the spin caused a distortion in the shadows and the GB parameter reduces the size of the photon sphere, the EH and shadows. The YM magnetic charge behaves somewhat identically with the electric charge for $q>0.75$ as the field strength reduces the shadow size with an increase in $Q$. Whereas, for $q<0.75$, the YM charge behaves in contrast with the electric charge as the YM field causes the shadow size to increase with an increase in $Q$.

\end{document}